\documentclass[amsmath,showpacs,nofootinbib,11pt,superscriptaddress]{revtex4-2}
\usepackage{graphicx}
\usepackage{dcolumn}
\usepackage{bm}
\usepackage{subcaption}
\usepackage{color} 
\usepackage{slashed}
\usepackage{float}
\usepackage{amsfonts}
\usepackage{multirow}
\begin{document}
\newcommand{\hs}{\hspace*{0.5cm}}
\newcommand{\vs}{\vspace*{0.5cm}}
\newcommand{\be}{\begin{equation}}
\newcommand{\ee}{\end{equation}}
\newcommand{\bea}{\begin{eqnarray}}
\newcommand{\eea}{\end{eqnarray}}
\newcommand{\ben}{\begin{enumerate}}
\newcommand{\een}{\end{enumerate}}
\newcommand{\bde}{\begin{widetext}}
\newcommand{\ede}{\end{widetext}}
\newcommand{\nn}{\nonumber}
\newcommand{\crn}{\nonumber \\}
\newcommand{\Tr}{\mathrm{Tr}}
\newcommand{\non}{\nonumber}
\newcommand{\noi}{\noindent}
\newcommand{\al}{\alpha}
\newcommand{\la}{\lambda}
\newcommand{\bet}{\beta}
\newcommand{\ga}{\gamma}
\newcommand{\va}{\varphi}
\newcommand{\om}{\omega}
\newcommand{\pa}{\partial}
\newcommand{\+}{\dagger}
\newcommand{\fr}{\frac}
\newcommand{\bc}{\begin{center}}
\newcommand{\ec}{\end{center}}
\newcommand{\Ga}{\Gamma}
\newcommand{\de}{\delta}
\newcommand{\De}{\Delta}
\newcommand{\ep}{\epsilon}
\newcommand{\varep}{\varepsilon}
\newcommand{\ka}{\kappa}
\newcommand{\La}{\Lambda}
\newcommand{\si}{\sigma}
\newcommand{\Si}{\Sigma}
\newcommand{\ta}{\tau}
\newcommand{\up}{\upsilon}
\newcommand{\Up}{\Upsilon}
\newcommand{\ze}{\zeta}
\newcommand{\ps}{\psi}
\newcommand{\Ps}{\Psi}
\newcommand{\ph}{\phi}
\newcommand{\vph}{\varphi}
\newcommand{\Ph}{\Phi}
\newcommand{\Om}{\Omega}
\newcommand{\AdrHEPC}{Phenikaa Institute for Advanced Study, Phenikaa University, Yen Nghia, Ha Dong, Hanoi 12116, Vietnam}
\newcommand{\AdrIOP}{Institute of Physics, Vietnam Academy of Science and Technology, 10 Dao Tan, Ba Dinh, Hanoi 100000, Vietnam}

\title{Tri-hypercharge versus tri-darkcharge} 

\author{Duong Van Loi}
\email{loi.duongvan@phenikaa-uni.edu.vn (corresponding author)}
\affiliation{\AdrHEPC} 
\author{A.E. C\'{a}rcamo Hern\'{a}ndez}
\email{antonio.carcamo@usm.cl}
\affiliation{Universidad T\'{e}cnica Federico Santa Mar\'{\i}a, Casilla 110-V, Valpara%
\'{\i}so, Chile}
\affiliation{Centro Cient\'ifico-Tecnol\'ogico de Valpara\'iso, Casilla 110-V, Valpara%
\'{\i}so, Chile}
\affiliation{Millennium Institute for Subatomic Physics at high energy frontier - SAPHIR, Fernandez Concha 700, Santiago, Chile}
\author{Van Que Tran}
\email{que.tranvan@phenikaa-uni.edu.vn}
\affiliation{Physics Division, National Center for Theoretical Sciences, Taipei, Taiwan 10617, R.O.C.}
\affiliation{\AdrHEPC}
\author{N. T. Duy}
\email{ntduy@iop.vast.vn}
\affiliation{\AdrIOP}

\date{\today}

\begin{abstract}
We propose a minimal, ultraviolet-complete, and renormalizable extension of the Standard Model, in which the three generations of ordinary fermions are distinguished by family-dependent hypercharges, while three right-handed neutrinos are separated by a dark gauge symmetry that is trivial for all Standard Model fields. This setup yields a fully flipped inert doublet model. The model naturally realizes a hybrid scotoseesaw mechanism that accounts for the smallness of neutrino masses and the largeness of lepton mixing. Simultaneously, it explains the stability and relic abundance of dark matter through a residual dark parity and addresses the hierarchies of charged fermion masses and the suppression of quark mixing via higher-dimensional operators involving high-scale scalar singlets and vector-like fermions. We explore the phenomenological implications of the model and derive constraints from electroweak precision tests, collider searches, flavor-changing processes, and observations of dark matter.  
\end{abstract}

\maketitle

\section{Introduction}
Two major puzzles of the Standard Model (SM) are the discovery of neutrino oscillations and the existence of dark matter (DM). The SM predicts that neutrinos are massless and that flavor lepton numbers are conserved. However, experimental evidence for neutrino oscillations indicates that neutrinos have mass and that lepton flavor is violated \cite{Kajita:2016cak, McDonald:2016ixn}. Furthermore, the SM particle content lacks any viable candidate for DM, which constitutes most of the mass in galaxies and galaxy clusters \cite{Planck:2018vyg}. Another longstanding issue is the flavor problem: in the SM, all three fermion generations are identical under gauge symmetry. Consequently, the theory does not explain why there are precisely three fermion generations, nor the large hierarchies observed in charged fermion masses and mixings \cite{ParticleDataGroup:2024cfk}.

Various mechanisms have been proposed to explain the origin of neutrino masses. Among the most well-known are the seesaw mechanism \cite{Minkowski:1977sc,Gell-Mann:1979vob,Yanagida:1979as,Glashow:1979nm,Mohapatra:1979ia,Mohapatra:1980yp,Lazarides:1980nt,Schechter:1980gr,Schechter:1981cv,VanLoi:2023kgl} and the scotogenic mechanism \cite{Zee:1980ai,Zee:1985id,Babu:1988ki,Krauss:2002px,Ma:2006km,VanLoi:2023utt,VanDong:2023xmf}. The smallness of neutrino masses may also be explained via a hybrid scenario in which both mechanisms contribute—the so-called scotoseesaw mechanism \cite{Kubo:2006rm,Rojas:2018wym,Barreiros:2020gxu,Mandal:2021yph,Ganguly:2022qxj,Kumar:2023moh,VanDong:2023xbd,VanLoi:2024ptt,VanDong:2024lry}.

Regarding DM, the most straightforward approach involves introducing a real, neutral singlet scalar with a $\mathbb{Z}_2$ symmetry into the SM, where the added scalar is odd while all other fields are even \cite{Silveira:1985rk}. Another familiar candidate is a neutral singlet fermion, often introduced to cancel gauge anomalies. The $\mathbb{Z}_2$ symmetry may be included {\it ad hoc}, or it may arise as a residual gauge symmetry \cite{Krauss:1988zc,Martin:1997ns,Kubo:2006rm}.

The issue of fermion generation number can be approached using anomaly cancellation and QCD asymptotic freedom arguments \cite{Frampton:1992wt,VanDong:2022cin,VanLoi:2024ptt}, while charged fermion mass and mixing hierarchies may be explained through flavor-deconstructed models \cite{Froggatt:1978nt,Rajpoot:1980ib,Li:1981nk,Barbieri:1994cx,Carone:1995ge}.

In this work, we construct a simple extension of the SM by “fully flipping” all fermion generations, including three right-handed neutrino singlets—the counterparts of the usual left-handed neutrinos—resulting in what we call the fully flipped inert doublet model. Concretely, we decompose the SM hypercharge symmetry into three generation-specific hypercharge symmetries,
\be U(1)_Y\to U(1)_{Y_1}\otimes U(1)_{Y_2}\otimes U(1)_{Y_3},\ee
 such that each ordinary fermion generation $a$ is charged only under the corresponding $U(1)_{Y_a}$, similar to the decomposition of lepton number into generation lepton numbers and in the spirit of recent works \cite{FernandezNavarro:2023rhv,FernandezNavarro:2024hnv,Davighi:2023evx}. Since the right-handed neutrino singlets carry zero hypercharge, they remain uncharged under all $U(1)_{Y_a}$. To distinguish them, we introduce an additional dark gauge symmetry, $U(1)_\mathsf{D}$, under which the right-handed neutrinos are charged, while all SM fermions are neutral (i.e., $\mathsf{D} = 0$), as suggested in \cite{Lindner:2013awa,VanDong:2023xbd,VanDong:2024lry}.

This setup leads to several interesting phenomenological consequences. First, if the SM Higgs doublet carries only the third-generation hypercharge, then only third-generation charged fermion masses are generated at the renormalizable level. The masses of the first- and second-generation charged fermions and the CKM mixings arise from higher-dimensional operators involving high-scale scalar singlets—flavons—which spontaneously break the three $U(1)_{Y_a}$ symmetries down to the SM hypercharge. This explains both the heaviness of the third generation and the smallness of the CKM mixing angles.

Second, assigning distinct dark charges $\delta_{1,2,3}$ to the right-handed neutrinos $\nu_{1,2,3R}$, anomaly cancellation for $[\mathrm{Gravity}]^2 U(1)_\mathsf{D}$ and $U(1)_\mathsf{D}^3$ requires
\be \delta_1+\delta_2+\delta_3=0,\hs \delta^3_1+\delta^3_2+\delta^3_3=0. \ee
A minimal nontrivial solution is $\delta_1 = -\delta_2 \neq 0$ and $\delta_3 = 0$. Normalizing $\delta_1 = 1$, we obtain $\mathsf{D}_{\nu{1,2,3R}} = 1, -1, 0$, consistent with \cite{Lindner:2013awa,VanDong:2023xbd,VanDong:2024lry}.

Third, a residual dark parity emerges, $P_\mathsf{D} = (-1)^\mathsf{D}$, under which only $\nu_{1R}$ and $\nu_{2R}$ are odd ($P_\mathsf{D} = -1$), while all other fields are even. This structure naturally realizes a scotoseesaw mechanism: the seesaw contribution arises via $\nu_{3R}$, generating one light active neutrino, and the scotogenic contribution—mediated by $P_\mathsf{D}$-odd scalars and $\nu_{1,2R}$—generates a second light active neutrino.

Fourth, the model provides a viable DM candidate in the form of a dark Majorana neutrino, which is stabilized by the conserved dark parity and realized as the lightest $P_\mathsf{D}$-odd field. Finally, since each generation hypercharge is assigned to only one fermion generation, the number of $U(1)_{Y_a}$ symmetries equals the number of fermion generations, naturally explaining the origin of three generations.

We would like to emphasize that the main novelty of this work, compared with the recent studies \cite{FernandezNavarro:2023rhv, FernandezNavarro:2024hnv, Davighi:2023evx}, lies in simultaneously addressing the origin of neutrino masses and DM. Whereas in the models of Refs \cite{FernandezNavarro:2023rhv, FernandezNavarro:2024hnv} tiny active neutrino masses are generated through a type-I seesaw mechanism, in our model these masses arise from an interplay of tree-level type-I and one-loop radiative seesaw mechanism. Furthermore, our model provides an explanation for the hierarchy between the atmospheric and solar neutrino mass-squared splittings, since the former arises at tree level whereas the later originates at one loop. In our model, the radiative nature of the seesaw mechanism that yields the solar neutrino mass-squared difference is guaranteed by a residual parity $P_\mathsf{D}$ symmetry, resulting from the spontaneous breaking of the dark $U(1)_\mathsf{D}$ gauge symmetry. This residual parity $P_\mathsf{D}$ symmetry also ensures the stability of the DM candidates. In our framework, the successful implementation of the seesaw mechanisms that generate tiny active neutrino masses, together with the requirement of an anomaly-free $U(1)_\mathsf{D}$ dark gauge symmetry, naturally implies the existence of exactly three right-handed neutrinos and two vector-like neutral leptons. The parity-even and parity-odd neutral leptonic fields mediate a hybrid scotoseesaw mechanism that produces the tiny neutrino masses, while the lightest parity-odd neutral lepton serves as a DM candidate stabilized by the residual parity $P_\mathsf{D}$ symmetry. All SM fermions are neutral under $U(1)_\mathsf{D}$, whereas the three right-handed neutrinos carry charges $\mathsf{D}=1,-1,0$. As a result, the $U(1)_\mathsf{D}$ gauge boson does not couple directly to SM fermions. Distinctive collider signatures of the present model, compared with scenarios without $U(1)_\mathsf{D}$, thus arise mainly from the production of $P_\mathsf{D}$-odd right-handed neutrinos and scalars, leading to characteristic final states with leptons and missing energy.

The remainder of this paper is organized as follows. In \textbf{Section II}, we present the fundamental aspects of the model, including the gauge symmetry, particle content, residual symmetries, Yukawa interactions, and scalar potential. \textbf{Section III} is devoted to the scalar and gauge boson mass spectra, while the fermion mass spectrum is analyzed in \textbf{Section IV}. Constraints from electroweak precision observables, collider searches, and flavor physics are discussed in \textbf{Sections V, VI, and VII}, respectively. \textbf{Section VIII} examines dark matter phenomenology, and our conclusions are summarized in \textbf{Section IX}. Mathematical details concerning the general hypercharge decomposition are given in Appendix \ref{decom}. The vector and axial-vector couplings are discussed in Appendix \ref{AppB}. Appendix \ref{AppC} provides a detailed discussion of the one-loop Landau poles for the Abelian factors $U(1)_{Y_{1}}\otimes U(1)_{Y_{2}}\otimes U(1)_{Y_{3}}\otimes U(1)_\mathsf{D}$.

\section{\label{model}The fully flipped inert doublet model}
As mentioned, this work considers a fully flipped inert doublet model where the scalar sector is enlarged by the inclusion of several gauge singlet scalars and an inert scalar doublet, while the fermionic content is added by charged vector like fermions and right-handed Majorana neutrinos, whose inclusion is necessary for the implementation of seesaw mechanisms that yields the small masses of the first and second generation of SM charged femions as well as tree level type I and one loop level radiative seesaw mechanisms that produces the tiny masses of the active neutrinos. The tree-level type I seesaw mechanism generates the atmospheric neutrino mass squared splitting, whereas the solar neutrino mass squared difference arises from a radiative seesaw mechanism at one-loop level. Our theory is based on the gauge symmetry 
\be SU(3)_C\otimes SU(2)_L\otimes U(1)_{Y_1}\otimes U(1)_{Y_2}\otimes U(1)_{Y_3}\otimes U(1)_\mathsf{D},\label{gaugesymmetry}\ee
 where the electric charge operator is embedded in that symmetry, so that it is given by 
\be Q=T_3+Y_1+Y_2+Y_3.\ee
In our proposed framework, the local $U(1)_\mathsf{D}$ gauge symmetry is assumed to be spontaneously broken down to a preserved discrete symmetry, referred to as dark parity $P_\mathsf{D}$. This residual symmetry plays a crucial role in ensuring the radiative origin of the solar neutrino mass-squared splitting, as it forbids tree-level contributions to certain neutrino masses. Moreover, the presence of $P_\mathsf{D}$ guarantees the stability of DM candidates in the model. The fermionic content of the model and their transformations properties under the extended gauge symmetry group $SU(3)_C\otimes SU(2)_L\otimes U(1)_{Y_1}\otimes U(1)_{Y_2}\otimes U(1)_{Y_3}\otimes U(1)_\mathsf{D}$,  as well as under the residual dark parity $P_\mathsf{D}$ (defined as $P_\mathsf{D} = (-1)^\mathsf{D}$), are summarized in Table \ref{tab1}. Here, $q_{aL} = (u_{aL}, d_{aL})^T$ and $l_{aL} = (\nu_{aL}, e_{aL})^T$ denote the left-handed quark and lepton doublets for the three generations $a = 1, 2, 3$.
 
\begin{table}[h!]
\bc
\begin{tabular}{l|cccccccc}
\hline\hline
Field  & $SU(3)_C\otimes SU(2)_L$ & $U(1)_{Y_1}$ & $U(1)_{Y_2}$ & $U(1)_{Y_3}$ & $U(1)_\mathsf{D}$ & $P_\mathsf{D}$\\ \hline
$q_{1L}$ & $({\bf 3,2})$ & $1/6$ & $0$ & $0$ & $0$ & $+$\\
$u_{1R}$ & $({\bf 3,1})$ & $2/3$ & $0$ & $0$ & $0$ & $+$\\
$d_{1R}$ & $({\bf 3,1})$ & $-1/3$ & $0$ & $0$ & $0$ & $+$\\
$l_{1L}$ & $({\bf 1,2})$ & $-1/2$ & $0$ & $0$ & $0$ & $+$\\
$e_{1R}$ & $({\bf 1,1})$ & $-1$ & $0$ & $0$ & $0$ & $+$\\
$q_{2L}$ & $({\bf 3,2})$ & $0$ & $1/6$ & $0$ & $0$ & $+$\\
$u_{2R}$ & $({\bf 3,1})$ & $0$ & $2/3$ & $0$ & $0$ & $+$\\
$d_{2R}$ & $({\bf 3,1})$ & $0$ & $-1/3$ & $0$ & $0$ & $+$\\
$l_{2L}$ & $({\bf 1,2})$ & $0$ & $-1/2$ & $0$ & $0$ & $+$\\
$e_{2R}$ & $({\bf 1,1})$ & $0$ & $-1$ & $0$ & $0$ & $+$\\
$q_{3L}$ & $({\bf 3,2})$ & $0$ & $0$ & $1/6$ & $0$ & $+$\\
$u_{3R}$ & $({\bf 3,1})$ & $0$ & $0$ & $2/3$ & $0$ & $+$\\
$d_{3R}$ & $({\bf 3,1})$ & $0$ & $0$ & $-1/3$ & $0$ & $+$\\
$l_{3L}$ & $({\bf 1,2})$ & $0$ & $0$ & $-1/2$ & $0$ & $+$\\
$e_{3R}$ & $({\bf 1,1})$ & $0$ & $0$ & $-1$ & $0$ & $+$\\
$\nu_{1R}$ & $({\bf 1,1})$ & $0$ & $0$ & $0$ & $1$ & $-$\\
$\nu_{2R}$ & $({\bf 1,1})$ & $0$ & $0$ & $0$ & $-1$ & $-$\\
$\nu_{3R}$ & $({\bf 1,1})$ & $0$ & $0$ & $0$ & $0$ & $+$\\
\hline\hline
\end{tabular}
\caption[]{\label{tab1}Fermion fields and their quantum numbers. $P_\mathsf{D}$ is residual dark parity of $U(1)_\mathsf{D}$.}
\ec
\end{table}

To break the extended gauge symmetry and generate the correct mass spectrum for the particles, we introduce several scalar fields in addition to the SM Higgs doublet $H=(H^+\, H^0)^T$. First, a scalar singlet $\Phi$ is included to spontaneously break the $U(1)_\mathsf{D}$ symmetry down to the residual dark parity $P_\mathsf{D}$, while simultaneously generating Majorana masses for the right-handed neutrinos $\nu_{1,2R}$. Furthermore, four electrically neutral scalar singlets—$\varphi_{12}$, $\varphi_{23}$, $\phi_{12}$, and $\phi_{23}$—are introduced. These scalars, referred to as flavons, carry non-zero generation hypercharges (associated with $U(1)_{Y_{1,2,3}}$), arranged such that the total hypercharge remains zero. Their roles are twofold: (i) to break the generation hypercharge groups spontaneously down to the SM hypercharge $U(1)_Y$, and (ii) to explain the observed mass hierarchies and the small mixing angles in the quark sector. Each of these scalar fields acquires a vacuum expectation value (VEV), given by
\bea \langle H\rangle &=&\frac{1}{\sqrt2}\begin{pmatrix} 0\\v \end{pmatrix}, \\
 \langle\Phi\rangle &=&\frac{1}{\sqrt2}\La,\hs
 \langle\varphi_{23}\rangle\approx\langle\phi_{23}\rangle\approx\frac{1}{\sqrt2}v_{23}, \hs \langle\varphi_{12}\rangle\approx\langle\phi_{12}\rangle\approx\frac{1}{\sqrt2}v_{12},\label{vevs}
 \eea
where $v=246$ GeV denotes the electroweak VEV. The other symmetry-breaking scales are assumed to lie well above the Fermi scale, i.e., $v_{12,23},\La\gg v$. This hierarchy ensures that the couplings of the $126$ GeV SM-like Higgs boson remain very close to their SM values, preserving compatibility with current experimental constraints. To account for the observed mass hierarchies among the SM charged fermions and the smallness of quark mixing angles, we further assume a hierarchy among the flavon VEVs, specifically $v_{12}\gg v_{23}$. In addition to the aforementioned scalar fields, we introduce an inert scalar doublet $\eta=(\eta^0\,\eta^-)^T$ and a complex scalar singlet $\rho$. Both of these are odd under the preserved dark parity $P_\mathsf{D}$ and, as such, must have vanishing VEVs to maintain $P_\mathsf{D}$ conservation. The presence of these inert scalars is crucial for realizing the scotogenic seesaw mechanism, in which the light neutrino mass matrix receives an additional radiative contribution at the one-loop level. The quantum numbers of all scalar fields under the gauge symmetry group in Eq. (\ref{gaugesymmetry}) and under $P_\mathsf{D}$ are summarized in Table \ref{tab2}. Notably, the fields $\phi_{12(23)}$ and $\varphi_{12(23)}^*$ transform identically under the full symmetry, i.e., $\phi_{12(23)} \sim \varphi_{12(23)}^{*3}$. Therefore, in principle, one could eliminate two scalar flavons (e.g., $\phi_{12}$ and $\phi_{23}$) and still retain the same symmetry-breaking pattern and phenomenology. This would yield a more economical scalar sector and a simpler scalar potential. However, such a minimal setup would complicate the ultraviolet completion of the model, especially if one aims to generate the effective Yukawa couplings via heavy vector-like fermions, rendering the ultraviolet theory less minimal and more involved.

\begin{table}[h!]
\bc
\begin{tabular}{l|cccccccl}
\hline\hline
\multicolumn{1}{c|}{Field} & 
\multicolumn{1}{c}{$SU(3)_C\otimes SU(2)_L$} & 
\multicolumn{1}{c}{$U(1)_{Y_1}$} & 
\multicolumn{1}{c}{$U(1)_{Y_2}$} & 
\multicolumn{1}{c}{$U(1)_{Y_3}$} & 
\multicolumn{1}{c}{$U(1)_\mathsf{D}$} & 
\multicolumn{1}{c}{$P_\mathsf{D}$} & 
\multicolumn{1}{c}{VEV [GeV]} & 
\multicolumn{1}{c}{Roles/Purposes} \\ \hline
$H$ & $({\bf 1,2})$ & $0$ & $0$ & $1/2$ & $0$ & $+$ & $v=246$ & \multirow{1}{*}{\begin{tabular}{@{}l@{}}
$\left.\vphantom{\begin{array}{c}x\end{array}}\right\} \begin{array}{l}
\text{Breaks }SU(2)_L\otimes U(1)_Y
\end{array}$
\end{tabular}}\\[3pt]
$\varphi_{23}$ & $({\bf 1,1})$ & $0$ & $1/6$ & $-1/6$ & $0$ & $+$ & $v_{23}\sim 10^4$ & \multirow{2}{*}{\begin{tabular}{@{}l@{}}
$\left.\vphantom{\begin{array}{c}x\\x\end{array}}\right\} \begin{array}{l}
\text{Breaks }U(1)_{Y_{12}}\otimes U(1)_{Y_3};\\
\text{flavon for Yukawa hierarchy}
\end{array}$
\end{tabular}}\\[3pt]
$\phi_{23}$ & $({\bf 1,1})$ & $0$ & $-1/2$ & $1/2$ & $0$ & $+$ & $v_{23}\sim 10^4$ & \\
$\varphi_{12}$ & $({\bf 1,1})$ & $1/6$ & $-1/6$ & $0$ & $0$ & $+$ & $v_{12}\sim 10^6$ & \multirow{2}{*}{\begin{tabular}{@{}l@{}}
$\left.\vphantom{\begin{array}{c}x\\x\end{array}}\right\} \begin{array}{l}
\text{Breaks }U(1)_{Y_1}\otimes U(1)_{Y_2};\\
\text{flavon for Yukawa hierarchy}
\end{array}$
\end{tabular}}\\[2pt]
$\phi_{12}$ & $({\bf 1,1})$ & $-1/2$ & $1/2$ & $0$ & $0$ & $+$ & $v_{12}\sim 10^6$ & \\
$\Phi$ & $({\bf 1,1})$ & $0$ & $0$ & $0$ & $2$ & $+$ & $\La\sim 10^3$ & \multirow{1}{*}{\begin{tabular}{@{}l@{}}
$\left.\vphantom{\begin{array}{c}x\end{array}}\right\} \begin{array}{l}
\text{Breaks }U(1)_\mathsf{D}\to R_\mathsf{D}
\end{array}$
\end{tabular}}\\[3pt]
$\eta$ & $({\bf 1,2})$ & $0$ & $0$ & $-1/2$ & $-1$ & $-$ & $0$ & \multirow{2}{*}{\begin{tabular}{@{}l@{}}
$\left.\vphantom{\begin{array}{c}x\\x\end{array}}\right\} \begin{array}{l}
\text{Scotogenic neutrino mass}\\
\text{generation; DM candidate}
\end{array}$
\end{tabular}}\\[3pt]
$\rho$ & $({\bf 1,1})$ & $0$ & $0$ & $0$ & $1$ & $-$ & $0$ & \\
\hline\hline
\end{tabular}
\caption[]{\label{tab2}Scalar fields, their quantum numbers, VEV, and roles/purposes in the model.}
\ec
\end{table}

The spontaneous gauge symmetry breaking is implemented through the following way, $SU(3)_C\otimes SU(2)_L\otimes U(1)_{Y_1}\otimes U(1)_{Y_2}\otimes U(1)_{Y_3}\otimes U(1)_\mathsf{D}\stackrel{v_{12}}\longrightarrow SU(3)_C\otimes SU(2)_L\otimes U(1)_{Y_{12}\equiv Y_1+Y_2}\otimes U(1)_{Y_3}\otimes U(1)_\mathsf{D} \stackrel{v_{23}}\longrightarrow SU(3)_C\otimes SU(2)_L\otimes U(1)_Y\otimes U(1)_\mathsf{D}\stackrel{\La}\longrightarrow SU(3)_C\otimes SU(2)_L\otimes U(1)_Y\otimes R_\mathsf{D} \stackrel{v}\longrightarrow SU(3)_C\otimes U(1)_Q\otimes R_\mathsf{D}$, in which we have assumed $\La<v_{23}$ for the potential discovery of new physics at the LHC. Here $U(1)_Q$ is the electromagnetic symmetry and $R_\mathsf{D}$ is a residual symmetry of $U(1)_\mathsf{D}$ that conserves all the VEVs, $R_\mathsf{D}=(-1)^{k\mathsf{D}}$ for $k$ integer. This residual symmetry is automorphic to a discrete group, such as $R_\mathsf{D}\cong\mathbb{Z}_2=\{1,(-1)^\mathsf{D}\}$, for which $P_\mathsf{D}=(-1)^\mathsf{D}$ is called a residual dark parity of $U(1)_\mathsf{D}$. Under $P_\mathsf{D}$, the SM fields, $\nu_{3R}$, $\Phi$, $\varphi_{12,23}$, and $\phi_{12,23}$ transform trivially, i.e., $P_\mathsf{D}=1$, whereas $\nu_{1,2R}$, $\eta$, and $\rho$ transform nontrivially, i.e., $P_\mathsf{D}=-1$, as presented in Tables \ref{tab1} and \ref{tab2}. 

With the above scalar content, the scalar potential can be split in two parts, such as $V=V_1+V_2$, where $V_1$ contains terms involving $P_\mathsf{D}$-even scalars, $H,\Phi,\phi_{12,23}$, and $\varphi_{12,23}$, while $V_2$ contains terms of $P_\mathsf{D}$-odd scalars, $\eta$ and $\rho$, and mixing terms between these two kinds, i.e.,
\bea V_1 &=& \mu_1^2H^\dag H + \la_1(H^\dag H)^2 + \mu_2^2\Phi^*\Phi + \la_2(\Phi^*\Phi)^2 + \la_3(\Phi^*\Phi)(H^\dag H) \crn
&&+ \mu_3^2\varphi_{23}^*\varphi_{23} + \mu_4^2\phi_{23}^*\phi_{23} + \la_4(\varphi_{23}^*\varphi_{23})^2 + \la_5(\phi_{23}^*\phi_{23})^2 + \la_6(\varphi_{23}^*\varphi_{23})(\phi_{23}^*\phi_{23})\crn
&&+ (\varphi_{23}^*\varphi_{23})(\la_7\Phi^*\Phi + \la_8H^\dag H) + (\phi_{23}^*\phi_{23})(\la_9\Phi^*\Phi + \la_{10}H^\dag H)\crn
&&+ \mu_5^2\varphi_{12}^*\varphi_{12} +\mu_6^2\phi_{12}^*\phi_{12} + \la_{11}(\varphi_{12}^*\varphi_{12})^2 + \la_{12}(\phi_{12}^*\phi_{12})^2 + \la_{13}(\varphi_{12}^*\varphi_{12})(\phi_{12}^*\phi_{12})\crn
&&+ (\varphi_{12}^*\varphi_{12})(\la_{14}\varphi_{23}^*\varphi_{23} + \la_{15}\phi_{23}^*\phi_{23} + \la_{16}\Phi^*\Phi + \la_{17}H^\dag H)\crn
&&+ (\phi_{12}^*\phi_{12})(\la_{18}\varphi_{23}^*\varphi_{23} + \la_{19}\phi_{23}^*\phi_{23} + \la_{20}\Phi^*\Phi + \la_{21}H^\dag H)\crn
&&+ (\la_{22}\varphi_{12}^3\phi_{12} + \la_{23}\varphi_{23}^3\phi_{23} + \mathrm{H.c.}),\\
V_2 &=& \mu_7^2\eta^\dag\eta + \mu_8^2\rho^\dag\rho + \la_{24}(\eta^\dag\eta)^2 + \la_{25}(\rho^\dag\rho)^2 + \la_{26}(\eta^\dag\eta)(\rho^\dag\rho) \crn
&&+ (\eta^\dag\eta)(\la_{27}\varphi_{12}^*\varphi_{12} + \la_{28}\phi_{12}^*\phi_{12} +\la_{29}\varphi_{23}^*\varphi_{23} + \la_{30}\phi_{23}^*\phi_{23} + \la_{31}\Phi^*\Phi + \la_{32}H^\dag H)\crn
&&+ (\rho^\dag\rho)(\la_{33}\varphi_{12}^*\varphi_{12} + \la_{34}\phi_{12}^*\phi_{12} +\la_{35}\varphi_{23}^*\varphi_{23} + \la_{36}\phi_{23}^*\phi_{23} + \la_{37}\Phi^*\Phi + \la_{38}H^\dag H)\crn
&&+ \la_{39}(H^\dag\eta)(\eta^\dag H) + (\mu_9H\eta\rho + \mu_{10}\rho\rho\Phi^* +\mathrm{H.c.}), \label{potendark}\eea
where the parameters $\la$’s are dimensionless, while $\mu$’s have a mass dimension. Hermiticity of the scalar potential requires all parameters to be real, except for $\la_{22,23}$ and $\mu_{9,10}$, which may, in general, be complex. However, any complex phases in $\la_{22,23}$ and $\mu_{9,10}$ can be removed by suitable phase redefinitions of the scalar fields $\varphi_{12,23},\eta$, and $\rho$. Therefore, without loss of generality, we assume that all parameters in the scalar potential are real throughout our analysis. Furthermore, the quartic scalar couplings of the form $\varphi_{12}^3\phi_{12}$ and $\varphi_{23}^3\phi_{23}$ explicitly break the global $U(1)$ phase symmetries associated with the four flavon fields. This breaking ensures that no physical Nambu–Goldstone bosons appear in the spectrum, rendering the model free from unwanted massless scalar degrees of freedom.

Since the SM Higgs doublet carries only the third-generation hypercharge, only the third-generation Yukawa couplings are allowed at the renormalizable level. In contrast, the masses of the first and second generations of SM charged fermions, as well as the fermionic mixing angles, arise solely from nonrenormalizable Yukawa operators. Adopting an effective field theory (EFT) framework, we can express the relevant Yukawa interactions as follows:
\bea \mathcal{L} &\supset& \begin{pmatrix}
\bar{q}_{1L} & \bar{q}_{2L} & \bar{q}_{3L}
\end{pmatrix}\begin{pmatrix}
y^u_{11}\frac{\phi_{12}}{\La_{12}}\frac{\phi_{23}}{\La_{23}} & y^u_{12}\frac{\varphi_{12}}{\La_{12}}\frac{\phi_{23}}{\La_{23}} & y^u_{13}\frac{\varphi_{12}}{\La_{12}}\frac{\varphi_{23}}{\La_{23}}\\
y^u_{21}\frac{\phi_{12}}{\La_{12}}\frac{\varphi^*_{12}}{\La_{12}}\frac{\phi_{23}}{\La_{23}} & y^u_{22}\frac{\phi_{23}}{\La_{23}} & y^u_{23}\frac{\varphi_{23}}{\La_{23}}\\
y^u_{31}\frac{\phi_{12}}{\La_{12}}\frac{\varphi^*_{12}}{\La_{12}}\frac{\phi_{23}}{\La_{23}}\frac{\varphi^*_{23}}{\La_{23}} & y^u_{32}\frac{\phi_{23}}{\La_{23}}\frac{\varphi^*_{23}}{\La_{23}} & y^u_{33}
\end{pmatrix} \begin{pmatrix}
u_{1R} \\ u_{2R} \\ u_{3R}
\end{pmatrix}\tilde{H}\crn
&&+\begin{pmatrix}
\bar{q}_{1L} & \bar{q}_{2L} & \bar{q}_{3L}
\end{pmatrix}\begin{pmatrix}
y^d_{11}\frac{\phi^*_{12}}{\La_{12}}\frac{\phi^*_{23}}{\La_{23}} & y^d_{12}\frac{\varphi_{12}}{\La_{12}}\frac{\phi^*_{23}}{\La_{23}} & y^d_{13}\frac{\varphi_{12}}{\La_{12}}\frac{\varphi_{23}}{\La_{23}}\\
y^d_{21}\frac{\phi^*_{12}}{\La_{12}}\frac{\varphi^*_{12}}{\La_{12}}\frac{\phi^*_{23}}{\La_{23}} & y^d_{22}\frac{\phi^*_{23}}{\La_{23}} & y^d_{23}\frac{\varphi_{23}}{\La_{23}}\\
y^d_{31}\frac{\varphi^2_{12}}{\La^2_{12}}\frac{\varphi^2_{23}}{\La^2_{23}} & y^d_{32}\frac{\varphi^2_{23}}{\La^2_{23}} & y^d_{33}
\end{pmatrix} \begin{pmatrix}
d_{1R} \\ d_{2R} \\ d_{3R}
\end{pmatrix}H \crn
&&+\begin{pmatrix}
\bar{l}_{1L} & \bar{l}_{2L} & \bar{l}_{3L}
\end{pmatrix}\begin{pmatrix}
y^e_{11}\frac{\phi^*_{12}}{\La_{12}}\frac{\phi^*_{23}}{\La_{23}} & y^e_{12}\frac{\phi_{12}}{\La_{12}}\frac{\phi^*_{23}}{\La_{23}} & y^e_{13}\frac{\phi_{12}}{\La_{12}}\frac{\phi_{23}}{\La_{23}}\\
y^e_{21}\frac{\phi^{*2}_{12}}{\La^2_{12}}\frac{\phi^*_{23}}{\La_{23}} & y^e_{22}\frac{\phi^*_{23}}{\La_{23}} & y^e_{23}\frac{\phi_{23}}{\La_{23}}\\
y^e_{31}\frac{\phi^{*2}_{12}}{\La^2_{12}}\frac{\phi^{*2}_{23}}{\La^2_{23}} & y^e_{32}\frac{\phi^{*2}_{23}}{\La^2_{23}} & y^e_{33}
\end{pmatrix} \begin{pmatrix}
e_{1R} \\ e_{2R} \\ e_{3R}
\end{pmatrix}H  + \mathrm{H.c.},\eea 
where the coefficients $y$'s are dimensionless, $\La_{12,23}$ denote the EFT cutoff scales, and $\tilde{H}=i\sigma_2 H^*$ with $\sigma_2$ being the second Pauli matrix. To obtain a complete and renormalizable model, we include heavy fermionic messenger fields whose masses correspond to the heavy scales $\La_{12,23}$. Indeed, we add to the fermionic spectrum three heavy vector-like $SU(2)_L$ singlet fermions for each charged sector, namely $u_{12,13,23}$, $d_{12,13,23}$ and $e_{12,13,23}$ for the up-type quark, down-type quark and charged lepton sectors, respectively. Their quantum number assignments under the model symmetries are presented in Table \ref{tab3}. Accordingly, we construct a set of renormalizable Yukawa interactions and bare mass terms for each charged fermion sector, given by:
\bea \mathcal{L}_\mathrm{Y}^u &=& y_1^u \bar{q}_{1L}\tilde{H}u_{13R} + y_2^u \bar{u}_{13L}\phi_{23}u_{12R} + y_3^u \bar{u}_{12L}\phi_{12}u_{1R} + y_4^u \bar{u}_{13L}\varphi_{12}u_{23R}\crn 
&& + y_5^u \bar{u}_{23L}\phi_{23}u_{2R} + y_6^u \bar{u}_{23L}\varphi_{23}u_{3R} + y_7^u \bar{q}_{2L}\tilde{H}u_{23R} + y_8^u \bar{u}_{23L}\varphi_{12}^*u_{13R} \crn
&&+ y_{33}^u \bar{q}_{3L}\tilde{H}u_{3R} + m_{u_{12}}\bar{u}_{12L}u_{12R} + m_{u_{13}}\bar{u}_{13L}u_{13R} + m_{u_{23}}\bar{u}_{23L}u_{23R} + \mathrm{H.c.},\\
\mathcal{L}_\mathrm{Y}^d &=& \mathcal{L}_\mathrm{Y}^u (y^u\to y^d, u\to d, \tilde{H}\to H, \phi\to\phi^*, m_u\to m_d),\\
\mathcal{L}_\mathrm{Y}^e &=& \mathcal{L}_\mathrm{Y}^u (y^u\to y^e, q\to l, u\to e, \tilde{H}\to H, \phi\to\phi^*, \varphi\to\phi, m_u\to m_e), \eea
where the coefficients $y$'s are dimensionless, whereas $m$'s have mass dimension.

\begin{table}[h]
\bc
\begin{tabular}{l|ccccccl}
\hline\hline
\multicolumn{1}{c|}{Field} & 
\multicolumn{1}{c}{$SU(3)_C\otimes SU(2)_L$} & 
\multicolumn{1}{c}{$U(1)_{Y_1}$} & 
\multicolumn{1}{c}{$U(1)_{Y_2}$} & 
\multicolumn{1}{c}{$U(1)_{Y_3}$} & 
\multicolumn{1}{c}{$U(1)_\mathsf{D}$} & 
\multicolumn{1}{c}{$P_\mathsf{D}$} & 
\multicolumn{1}{c}{Roles/Purposes} \\ \hline
$u_{12}$ & $({\bf 3,1})$ & $1/6$ & $1/2$ & $0$ & $0$ & $+$ & \multirow{3}{*}{\begin{tabular}{@{}l@{}}
$\left.\vphantom{\begin{array}{c}x\\x\\x\end{array}}\right\} \begin{array}{l}
\text{Messengers for up-type} \\
\text{quark mass generation} 
\end{array}$
\end{tabular}}\\[3pt]
$u_{13}$ & $({\bf 3,1})$ & $1/6$ & $0$ & $1/2$ & $0$ & $+$\\
$u_{23}$ & $({\bf 3,1})$ & $0$ & $1/6$ & $1/2$ & $0$ & $+$\\
$d_{12}$ & $({\bf 3,1})$ & $1/6$ & $-1/2$ & $0$ & $0$ & $+$ & \multirow{3}{*}{\begin{tabular}{@{}l@{}}
$\left.\vphantom{\begin{array}{c}x\\x\\x\end{array}}\right\} \begin{array}{l}
\text{Messengers for down-type} \\
\text{quark mass generation} 
\end{array}$
\end{tabular}}\\[3pt]
$d_{13}$ & $({\bf 3,1})$ & $1/6$ & $0$ & $-1/2$ & $0$ & $+$\\
$d_{23}$ & $({\bf 3,1})$ & $0$ & $1/6$ & $-1/2$ & $0$ & $+$\\
$e_{12}$ & $({\bf 1,1})$ & $-1/2$ & $-1/2$ & $0$ & $0$ & $+$ & \multirow{3}{*}{\begin{tabular}{@{}l@{}}
$\left.\vphantom{\begin{array}{c}x\\x\\x\end{array}}\right\} \begin{array}{l}
\text{Messengers for charged} \\
\text{lepton mass generation}
\end{array}$
\end{tabular}}\\[3pt]
$e_{13}$ & $({\bf 1,1})$ & $-1/2$ & $0$ & $-1/2$ & $0$ & $+$\\
$e_{23}$ & $({\bf 1,1})$ & $0$ & $-1/2$ & $-1/2$ & $0$ & $+$\\
$\nu_{13}$ & $({\bf 1,1})$ & $-1/2$ & $0$ & $1/2$ & $0$ & $+$ & \multirow{2}{*}{\begin{tabular}{@{}l@{}}
$\left.\vphantom{\begin{array}{c}x\\x\end{array}}\right\} \begin{array}{l}
\text{Messengers for seesaw} \\
\text{neutrino mass generation} 
\end{array}$
\end{tabular}}\\[3pt]
$\nu_{23}$ & $({\bf 1,1})$ & $0$ & $-1/2$ & $1/2$ & $0$ & $+$\\
\hline\hline
\end{tabular}
\caption[]{\label{tab3}Vector-like fermions, their quantum numbers, and roles/purposes in the model.}
\ec
\end{table}

We also introduce two vector-like neutrinos, $\nu_{13,23}$, which are complete singlets under the SM gauge group. These fields enable the generation of light neutrino masses and mixings at tree level via the seesaw mechanism. Their quantum numbers under the gauge symmetry of Eq. (\ref{gaugesymmetry}) are listed in the last two rows of Table \ref{tab3}. The corresponding renormalizable Yukawa interactions relevant for neutrino mass generation are given by:
\bea \mathcal{L}^\nu_\mathrm{Y} &=& y_1\bar{l}_{1L}\tilde{H}\nu_{13R} + y_2\bar{l}_{2L}\tilde{H}\nu_{23R} + y_3\bar{l}_{3L}\tilde{H}\nu_{3R} + x_1\bar{\nu}_{13L}\phi_{12}\nu_{23R}+ x_2\bar{\nu}_{23L}\phi_{12}^*\nu_{13R}\crn
&& + z_1\bar{\nu}_{23L}\phi_{23}\nu_{3R} + z_2\bar{\nu}_{3R}^c\phi_{23}^*\nu_{23R} - m_{\nu_{13}}\bar{\nu}_{13L}\nu_{13R} - m_{\nu_{23}}\bar{\nu}_{23L}\nu_{23R}  - \fr 1 2 M_3\bar{\nu}_{3R}^c\nu_{3R} \crn
&& + \kappa\bar{l}_{3L}\eta\nu_{1R}+ \fr 1 2 f_1\bar{\nu}_{1R}^c\Phi^*\nu_{1R} + \fr 1 2 f_2\bar{\nu}_{2R}^c\Phi\nu_{2R} - m_{12}\bar{\nu}_{1R}^c\nu_{2R}+ \mathrm{H.c.},\label{yuneu} \eea
where the coefficients $y_{1,2,3}$, $x_{1,2}$, $z_{1,2}$, $\kappa$ and $f_{1,2}$ are dimensionless, whereas $m$'s and $M_3$ have mass dimension. 

\section{\label{gausca}Gauge and scalar sectors}
\subsection{Gauge sector}
In our model the covariant derivative takes the form $D_\mu=\partial_\mu+ig_st_pG_{p\mu}+igT_nA_{n\mu}+ig_1Y_1B_{1\mu}+ig_2Y_2B_{2\mu}+ig_3Y_3B_{3\mu}+ig_\mathsf{D}\mathsf{D} C_\mu$, where $(g_s,g,g_1,g_2,g_3,g_\mathsf{D})$, $(t_p,T_n,Y_1,Y_2,Y_3,\mathsf{D})$, and $(G_{p\mu},A_{n\mu},B_{1\mu},B_{2\mu},B_{3\mu},C_\mu)$ are coupling constants, generators, and gauge bosons of the $(SU(3)_C$, $SU(2)_L,U(1)_{Y_1},U(1)_{Y_2},U(1)_{Y_3},U(1)_\mathsf{D})$ groups, respectively. Substituting them into the scalar kinetic term $\sum_S(D^\mu\langle S\rangle)^\dag(D_\mu\langle S\rangle)$ for $S=H,\Phi,\phi_{12,23},\varphi_{12,23}$, we find\footnote{In this work, we only discuss the mass mixing among the gauge bosons. In contrast, kinetic mixing effects associated with the $U(1)$ gauge fields are assumed to be negligible and thus suppressed for simplicity.}
\be\mathcal{L}\supset\frac{g^2v^2}{4}W^{\mu +}W_\mu^- + 2 g_\mathsf{D}^2\La^2 \mathsf{Z}^\mu \mathsf{Z}_\mu +\fr 1 2 (A^\mu_3\, B^\mu_3\, B^\mu_2\, B^\mu_1)M_0^2(A_{3\mu}\, B_{3\mu}\, B_{2\mu}\, B_{1\mu})^T,\ee
where $W^\pm_\mu=(A_{1\mu}\mp iA_{2\mu})/\sqrt2$ and $\mathsf{Z}_\mu\equiv C_\mu$ are physical fields by themselves, which are respectively identified with the SM charged and $U(1)_\mathsf{D}$ gauge bosons with their masses given by $m^2_W=g^2v^2/4$ and $m^2_\mathsf{Z}=4g^2_\mathsf{D}\La^2$. For the last term, the mixing matrix
\be M_0^2=\begin{pmatrix}
\fr{g^2v^2}{4} & -\frac{gg_3v^2}{4} & 0 & 0\\
-\frac{gg_3v^2}{4} & \frac{g_3^2(10v_{23}^2+9v^2)}{36} & -\frac{5g_2g_3v_{23}^2}{18} & 0\\
0 & -\frac{5g_2g_3v_{23}^2}{18} & \frac{5g_2^2(v_{12}^2+v_{23}^2)}{18} & -\frac{5g_1g_2v_{12}^2}{18}\\
0 & 0 & -\frac{5g_1g_2v_{12}^2}{18}& \frac{5g_1^2v_{12}^2}{18}
\end{pmatrix}\ee
always provides a zero eigenvalue with the corresponding eigenstate (photon field)
\be A_\mu = s_WA_3+c_W[s_{23}B_{3\mu}+c_{23}(s_{12}B_{2\mu}+c_{12}B_{1\mu})], \ee
where the Weinberg angle $\theta_W$ and other mixing angles $\theta_{12,23}$ are given by \be s_W =\fr{g_Y}{\sqrt{g_Y^2+g^2}},\hs
s_{12}=\frac{g_1}{\sqrt{g_1^2+g_2^2}},\hs s_{23}=\frac{g_{12}}{\sqrt{g_{12}^2+g_3^2}},\ee
for $g_Y=g_{12}g_3/\sqrt{g_{12}^2+g_3^2}$ and $g_{12}=g_1g_2/\sqrt{g_1^2+g_2^2}$. Here and throughout this work, we use a type of notation as $s_x\equiv\sin x$, $c_x\equiv\cos x$, and $t_x\equiv\tan x$ for any mixing angle either $\theta_x$ or $x$. Hence, we determine the SM $Z$ boson and two new neutral gauge bosons as follows
\bea Z_\mu &=& c_WA_3-s_W[s_{23}B_{3\mu}+c_{23}(s_{12}B_{2\mu}+c_{12}B_{1\mu})],\\
 Z_{1\mu} &=& c_{23}B_{3\mu}-s_{23}(s_{12}B_{2\mu}+c_{12}B_{1\mu}),\\
 Z_{2\mu} &=& c_{12}B_{2\mu}-s_{12}B_{1\mu}.
 \eea
 
In the $(A_\mu,Z_\mu,Z_{1\mu},Z_{2\mu})$ basis, the photon field is decoupled as a physical massless field, whereas the other states mix among themselves via the following squared mass matrix: 
\be \mathcal{M}_0^2=\begin{pmatrix}
\fr{g^2v^2}{4c_W^2} & -\fr{g_3g^2t_Wv^2}{4g_{12}c_W} & 0\\
 -\fr{g_3g^2t_Wv^2}{4g_{12}c_W} &  \fr{g_3^2(9g^4t_W^4v^2 + 10g_{12}^4v_{23}^2)}{36g^2g_{12}^2t_W^2} &  -\fr{5g_1g_2g_3c_{12}^2v_{23}^2}{18gt_W}\\
0 & -\fr{5g_1g_2g_3c_{12}^2v_{23}^2}{18gt_W} & \fr{5[(g_1^2+g_2^2)v_{12}^2+g_2^2c_{12}^2v_{23}^2]}{18}
\end{pmatrix}. \ee
This matrix can be approximately diagonalized by using the usual type I seesaw formula. Introducing a new basis as $(\mathcal{Z}_\mu, \mathcal{Z}_{1\mu},\mathcal{Z}_{2\mu})$ for which the light $\mathcal{Z}_\mu$ boson is separated from the heavy bosons $\mathcal{Z}_{1\mu}$ and $\mathcal{Z}_{2\mu}$, we obtain $\mathcal{Z}_\mu\simeq Z_\mu-\varepsilon_1 Z_{1\mu}-\varepsilon_2 Z_{2\mu}$, $\mathcal{Z}_{1\mu}\simeq \varepsilon_1 Z_\mu+Z_{1\mu}$, $\mathcal{Z}_{2\mu}\simeq \varepsilon_2 Z_\mu+Z_{2\mu}$, and the mass of $\mathcal{Z}_\mu$ to be
\be m^2_{\mathcal{Z}}\simeq \frac{g^2v^2}{4c_W^2}\left(1+\fr{\varepsilon_1 g_3s_W}{g_{12}}\right),\ee 
where the mixing parameters $\varepsilon_{1,2}$ are strongly suppressed by $(v/v_{23})^2$ and $(v/v_{12})^2$, respectively,
\be \varepsilon_1\simeq-\fr{9g^4t_W^3v^2}{10g_{12}^3g_3c_Wv_{23}^2}, \hs \varepsilon_2\simeq-\fr{9g^3c_{12}^2t_W^2v^2}{10g_{12}g_1g_2c_Wv_{12}^2}.\ee The $\mathcal{Z}_\mu$ boson that has a mass at the weak scale is identical to the SM $Z$ boson.

The $\mathcal{Z}_{1\mu}$ and $\mathcal{Z}_{2\mu}$ bosons mix by themselves via a mass matrix which approximates the bottom-right $2\times 2$ submatrix of $\mathcal{M}_0^2$. Diagonalizing this mass matrix, we obtain the corresponding physical states,
\be Z_{23\mu}=c_\zeta \mathcal{Z}_{1\mu}- s_\zeta\mathcal{Z}_{2\mu},\hs Z_{12\mu}=s_\zeta \mathcal{Z}_{1\mu}+ c_\zeta\mathcal{Z}_{2\mu}, \ee
with their respective masses
\be m^2_{Z_{23}}\simeq \fr{5g_{12}^2g_3^2v_{23}^2}{18g^2t_W^2}, \hs m^2_{Z_{12}}\simeq \fr{5[(g_1^2+g_2^2)v_{12}^2+g_2^2c_{12}^2v_{23}^2]}{18}, \ee
to be very heavy at the $v_{23}$ and $v_{12}$ scales, respectively.
The $\zeta$ mixing angle is given by \be t_{2\zeta}\simeq -\frac{2gg_1g_2g_3t_Wc_{12}^2v_{23}^2}{g^2(g_1^2+g_2^2)t_W^2v_{12}^2+(g^2g_2^2t_W^2c_{12}^2-g_{12}^2g_3^2)v_{23}^2},\ee
strongly suppressed by $(v_{23}/v_{12})^2$.

\subsection{Scalar sector}
To obtain the physical scalar spectrum, we first expand the scalar fields around their VEVs, such as
\bea
H &=&\begin{pmatrix}
H^+ \\ \frac{1}{\sqrt2}(v+S_1+iA_1)
\end{pmatrix}, \hs \Phi =\frac{1}{\sqrt2}(\La+S_2+iA_2),\\
\phi_{23}&=&\frac{1}{\sqrt2}(v_{23}+S_3+iA_3),\hs \varphi_{23}=\frac{1}{\sqrt2}(v_{23}+S_4+iA_4),\\
\phi_{12}&=&\frac{1}{\sqrt2}(v_{12}+S_5+iA_5),\hs \varphi_{12}=\frac{1}{\sqrt2}(v_{12}+S_6+iA_6),\\
\eta &=&\begin{pmatrix}
\frac{1}{\sqrt2}(R_\eta+i I_\eta)\\ \eta^-
\end{pmatrix}, \hs \rho =\frac{1}{\sqrt2}(R_\rho+i I_\rho),
\eea
and then substitute them into the scalar potential. Hence, the scalar potential minimization conditions 
are given by
\bea 2\mu_1^2+2\la_1v^2+\la_3\La^2+(\la_8+\la_{10})v_{23}^2+(\la_{17}+\la_{21})v_{12}^2 &=& 0,\\
2\mu_2^2+\la_3v^2+2\la_2\La^2+(\la_7+\la_9)v_{23}^2+(\la_{16}+\la_{20})v_{12}^2 &=& 0,\\
2\mu_3^2+\la_8v^2+\la_7\La^2+(2\la_4+\la_6+3\la_{23})v_{23}^2+(\la_{14}+\la_{18})v_{12}^2 &=& 0,\\
2\mu_4^2+\la_{10}v^2+\la_9\La^2+(2\la_5+\la_6+\la_{23})v_{23}^2+(\la_{15}+\la_{19})v_{12}^2 &=& 0,\\
2\mu_5^2+\la_{17}v^2+\la_{16}\La^2+(\la_{14}+\la_{15})v_{23}^2+(2\la_{11}+\la_{13}+3\la_{22})v_{12}^2 &=& 0,\\
2\mu_6^2+\la_{21}v^2+\la_{20}\La^2+(\la_{18}+\la_{19})v_{23}^2+(2\la_{12}+\la_{13}+\la_{22})v_{12}^2 &=& 0. \eea

For the CP-odd and $P_\mathsf{D}$-even scalar sector, $A_{1,2,3,4,5,6}$, we find two physical heavy mass eigenstates, $\mathcal{A}_1=-(A_3+3A_4)/\sqrt{10}$ and $\mathcal{A}_2=-(A_5+3A_6)/\sqrt{10}$, with corresponding masses, $m_{\mathcal{A}_1}=-5\la_{23}v_{23}^2$ and $m_{\mathcal{A}_2}=-5\la_{22}v_{12}^2$, then implying that 
the parameters $\la_{22,23}$ have be negative. Additionally, we also obtain four massless eigenstates, $G_Z=A_1$, $G_\mathsf{Z}=A_2$, $G_{Z_1}=(3A_3-A_4)/\sqrt{10}$, and $G_{Z_2}=(3A_5-A_6)/\sqrt{10}$, which are the Goldstone bosons eaten by the longitudinal components of the neutral gauge bosons, $Z$, $\mathsf{Z}$, $Z_1$, and $Z_2$, respectively.

For the CP-even and $P_\mathsf{D}$-even scalar sector, $S_{1,2,3,4,5,6}$, we find the following squared mass matrix 
\be M^2_S=\begin{pmatrix}
2\la_1v^2 & \la_3\La v & \la_{10}v_{23}v & \la_8 v_{23}v & \la_{21}v_{12}v & \la_{17}v_{12}v \\
\la_3\La v & 2\la_2\La^2 & \la_9\La v_{23} & \la_7\La v_{23} & \la_{20}\La v_{12} & \la_{16}\La v_{12} \\
\la_{10}v_{23}v & \la_9\La v_{23} & \fr{(4\la_5-\la_{23})v_{23}^2}{2} & \fr{(2\la_6+3\la_{23})v_{23}^2}{2} & \la_{19} v_{12}v_{23} & \la_{15}v_{12}v_{23} \\
\la_8v_{23}v & \la_7\La v_{23} & \fr{(2\la_6+3\la_{23})v_{23}^2}{2} & \fr{(4\la_4+3\la_{23})v_{23}^2}{2} & \la_{18} v_{12}v_{23} & \la_{14}v_{12}v_{23} \\
\la_{21}v_{12}v & \la_{20} v_{12}\La & \la_{19} v_{12}v_{23} & \la_{18}v_{12}v_{23} & \fr{(4\la_{12}-\la_{22})v_{12}^2}{2} & \fr{(2\la_{13}+3\la_{22})v_{12}^2}{2}\\
\la_{17}v_{12}v & \la_{16} v_{12}\La & \la_{15} v_{12}v_{23} & \la_{14}v_{12}v_{23} & \fr{(2\la_{13}+3\la_{22})v_{12}^2}{2} & \fr{(4\la_{11}+3\la_{22})v_{12}^2}{2}
\end{pmatrix}. \ee
This matrix can be approximately diagonalized by using the hierarchies $v\ll \La,v_{23}\ll v_{12}$. Indeed, neglecting the mixing between $S_{5,6}$ and $S_{1,2,3,4}$, the bottom-right $2\times 2$ submatrix provides two physical eigenstates $\mathcal{H}_{5,6}$ with their masses to be very heavy at the $v_{12}$ scale. For the rest, we use the seesaw approximation to separate the light state $S_1$ from the heavy states $S_{2,3,4}$. In a new basis denoted $(h,H_2,H_3,H_4)$ such that $h$ is decoupled as a physical field, we get
\be h \simeq S_1 -\epsilon_2 S_2-\epsilon_3 S_3-\epsilon_4 S_4\ee
with its mass to be
\be m^2_h\simeq 2\la_1v^2 -[\epsilon_2\la_3\La+(\epsilon_3\la_{10}+\epsilon_4\la_8)v_{23}]v, \ee 
where the mixing parameters are suppressed as $\epsilon_2\sim v/\La$ and $\epsilon_{3,4}\sim v/v_{23}$. The remaining states $H_2\simeq\epsilon_2 S_1+ S_2$, $H_3 \simeq\epsilon_3 S_1+ S_3$, and $H_4\simeq\epsilon_4 S_1+ S_4$ mix by themselves via a $3\times 3$ submatrix, which provide three physical eigenstates $\mathsf{H}$ and $\mathcal{H}_{3,4}$ with their masses to be $m_{\mathsf{H}}$ heavy at the $\La$ scale and $m_{\mathcal{H}_{3,4}}$ heavy at the $v_{23}$ scale. Hence, the $h$ boson with a mass in the weak scale is identified with the SM Higgs boson.

For the $P_\mathsf{D}$-odd scalars, $R_{1,2}$ and $I_{1,2}$, we find two mass matrices where they mix in each pair, namely
\bea V &\supset& \fr 1 2 \begin{pmatrix} R_\eta & R_\rho \end{pmatrix}
\begin{pmatrix} M^2_\eta & -\fr{\mu_9 v}{\sqrt{2}}\\
-\fr{\mu_9 v}{\sqrt{2}} & M^2_\rho + \sqrt2 \mu_{10}\La \end{pmatrix}
\begin{pmatrix} R_\eta \\ R_\rho\end{pmatrix}\crn
&& +\fr 1 2  \begin{pmatrix} I_\eta & I_\rho \end{pmatrix}
\begin{pmatrix} M^2_\eta & \fr{\mu_9 v}{\sqrt{2}}\\
\fr{\mu_9 v}{\sqrt{2}} & M^2_\rho - \sqrt2 \mu_{10}\La \end{pmatrix}
\begin{pmatrix} I_\eta \\ I_\rho\end{pmatrix}, \eea 
where we have denoted $M^2_\eta=[2\mu^2_7+(\la_{27}+\la_{28})v_{12}^2+(\la_{29}+\la_{30})v_{23}^2+\la_{31}\La^2+\la_{32}v^2]/2$ and $M^2_\rho=[2\mu^2_8+(\la_{33}+\la_{34})v_{12}^2+(\la_{35}+\la_{36})v_{23}^2+\la_{37}\La^2+\la_{38}v^2]/2$. Defining two mixing angles $\theta_{R,I}$ via the tangent function as \be t_{2R,2I}=\fr{\mp\sqrt{2}\mu_9 v}{M^2_{\rho}- M^2_\eta\pm\sqrt{2}\mu_{10} \La},\ee
we obtain $P_D$-odd physical mass eigenstates $R_1=c_R R_\eta -s_R R_\rho$, $R_2=s_R R_\eta +c_R R_\rho$, $I_1=c_I I_\eta -s_I I_\rho$, and $I_2=s_I I_\eta +c_I I_\rho$, with respective masses,
\bea m^2_{R_1,I_1} &\simeq& M^2_\eta - \fr{\mu^2_9 v^2/2}{M^2_\rho-M^2_\eta\pm\sqrt{2}\mu_{10} \La},\\
m^2_{R_2,I_2} &\simeq& M^2_\rho \pm\sqrt{2} \mu_{10} \La + \fr{\mu^2_9 v^2/2}{M^2_\rho - M^2_\eta \pm\sqrt{2} \mu_{10} \La},\eea 
in which the approximations come from $|\theta_{R,I}|\ll 1$. Indeed, since the last two terms associated with $\mu_{9,10}$ in Eq. (\ref{potendark}) are not suppressed by any existing symmetry, $\mu_{9,10}$ may be as large as the highest scale, i.e. $\mu_{9,10}\sim v_{12}$. In such case, we still have $|\theta_{R,I}|\sim v/v_{12}\ll 1$. Also, it is straightforward to derive $(m^2_{R_1}-m^2_{I_1})/m^2_{R_1,I_1}\sim (v^2 /v_{12}^2)(\La/v_{12}) \ll 1$ and $(m^2_{R_2}-m^2_{I_2})/m^2_{R_2,I_2}\sim \La/v_{12} \ll 1$.

For the charged scalars, we directly obtain a massless eigenstate, $G_{W^\pm}=H^\pm$, which is the Goldstone boson eaten by the $W^\pm$ charged gauge boson. Additionally, the charged dark scalar $\eta^\pm$ is a physical field by itself with mass in the $v_{12}$ scale, i.e. $m^2_{\eta^\pm}=M^2_\eta +\la_{39}v^2/2$.

For completeness, we have analyzed the perturbativity of the extra Abelian gauge interactions in Appendix~\ref{AppC}, where the one-loop Landau poles of the four factors $U(1)_{Y_1}\otimes U(1)_{Y_2}\otimes U(1)_{Y_3}\otimes U(1)_\mathsf{D}$ are derived. The most restrictive case corresponds to $U(1)_{Y_3}$, where the corresponding Landau pole is found at the scale $\mu_{\text{LP},Y_3} \approx 10^{18}$ TeV, i.e. well above both the GUT and Planck scales. From this analysis we also extract upper bounds on the gauge couplings ensuring perturbativity up to given reference scales, namely $g_3\lesssim 0.846$ at the typical seesaw scale ($M_3\sim 10^{11}$ TeV), $g_3\lesssim 0.733$ at the GUT scale, and $g_3\lesssim 0.655$ at the Planck scale. These results justify the parameter ranges adopted in the phenomenological study of flavor observables, collider signatures, and DM in the following sections.

\section{\label{fermion}Fermion mass and mixing}

In this section, we study the generation of fermion masses and the structure of mixings. Since the model introduces several new energy scales, it is useful to begin with a global overview. Table~\ref{tab:scales} summarizes the characteristic scales and their typical values, highlighting the hierarchy that later appears in the effective operators. This overview provides a clear reference for the scale ratios relevant to the generation of active neutrino and SM charged fermion masses, as well as to the fermionic mixing pattern.

\begin{table}[h]
\centering
\renewcommand{\arraystretch}{1.15}
\begin{tabular}{lc}
\hline\hline
New scale(s) & Typical value \\
\hline
$\Lambda,\,M_{1,2}$ & $\mathcal{O}(1)\,\text{TeV}$ \\
$v_{23}$ & $\mathcal{O}(10)\,\text{TeV}$ \\
$v_{12},\,m_{u_{23}},\,m_{\nu_{13,23}},\,m_{R_{1,2}},\,m_{I_{1,2}}$ & $\mathcal{O}(10^{3})\,\text{TeV}$ \\
$m_{u_{13}},\,m_{d_{13,23}},\,m_{e_{23}}$ & $\mathcal{O}(10^{4})\,\text{TeV}$ \\
$m_{u_{12}},\,m_{d_{12}},\,m_{e_{12,13}}$ & $\mathcal{O}(10^{5})\,\text{TeV}$ \\
$M_{3}$ & $\mathcal{O}(10^{11})\,\text{TeV}$ \\
\hline\hline
\end{tabular}
\caption{\label{tab:scales}New energy scales appearing in the model together with their typical values.}
\end{table}

\subsection{Charged fermion sector}
The Feynman diagrams responsible for the mass generation and mixing of the up-type quarks are shown in Fig. \ref{plotmassude}. The corresponding diagrams for the down-type quarks (charged leptons) can be obtained by making the substitutions: $u\to d$, $\tilde{H}\to H$, and $\phi\to\phi^*$ ($q\to l$, $u\to e$, $\tilde{H}\to H$, $\phi\to\phi^*$, and $\varphi\to\phi$). After integrating out the heavy messenger fields, we obtain effective Yukawa interactions for the SM chiral charged fermions. Once the flavon fields acquire VEVs, these effective interactions generate the mass matrices for the SM charged fermions, which take the form:
\begin{figure}[h]
\centering
\includegraphics[scale=1]{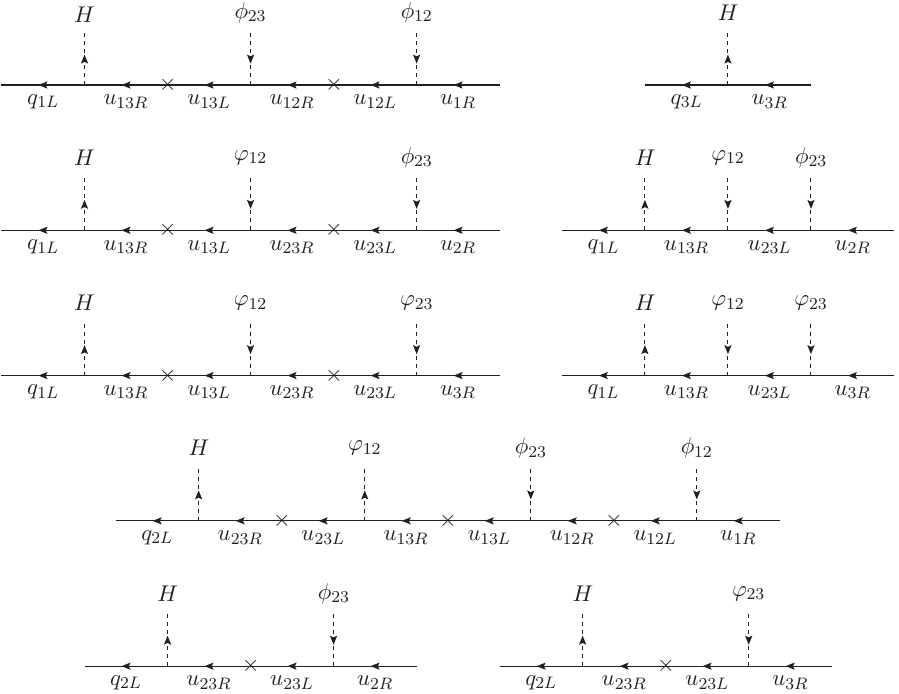}
\caption[]{\label{plotmassude}Diagrams relate to the generation of up-quark masses and mixing.}
\end{figure}

\bea \mathcal{M}_u &=& \frac{v}{\sqrt2}\begin{pmatrix}
\frac{y_1^uy_2^uy_3^u\langle\phi_{12}\rangle\langle\phi_{23}\rangle}{m_{u_{12}}m_{u_{13}}} & \frac{y_1^uy_5^u\langle\phi_{23}\rangle}{y_{8}^u\langle\varphi^*_{12}\rangle}+\frac{y_1^uy_4^uy_5^u\langle\varphi_{12}\rangle\langle\phi_{23}\rangle}{m_{u_{13}}m_{u_{23}}} & \frac{y_1^uy_6^u\langle\varphi_{23}\rangle}{y_{8}^u\langle\varphi^*_{12}\rangle}+\frac{y_1^uy_4^uy_6^u\langle\varphi_{12}\rangle\langle\varphi_{23}\rangle}{m_{u_{13}}m_{u_{23}}}\\
\frac{y_2^uy_3^uy_7^uy_8^u\langle\varphi_{12}^*\rangle\langle\phi_{12}\rangle\langle\phi_{23}\rangle}{m_{u_{12}}m_{u_{13}}m_{u_{23}}} & \frac{y_5^uy_7^u\langle\phi_{23}\rangle}{m_{u_{23}}} & \frac{y_6^uy_7^u\langle\varphi_{23}\rangle}{m_{u_{23}}} \\
0 & 0 & y_{33}^u \end{pmatrix},\\
\mathcal{M}_d &=& \frac{v}{\sqrt2}\begin{pmatrix}
\frac{y_1^dy_2^dy_3^d\langle\phi^*_{12}\rangle\langle\phi^*_{23}\rangle}{m_{d_{12}}m_{d_{13}}} & \frac{y_1^dy_5^d\langle\phi^*_{23}\rangle}{y_{8}^d\langle\varphi^*_{12}\rangle}+\frac{y_1^dy_4^dy_5^d\langle\varphi_{12}\rangle\langle\phi^*_{23}\rangle}{m_{d_{13}}m_{d_{23}}} & \frac{y_1^dy_6^d\langle\varphi_{23}\rangle}{y_{8}^d\langle\varphi^*_{12}\rangle}+\frac{y_1^dy_4^dy_6^d\langle\varphi_{12}\rangle\langle\varphi_{23}\rangle}{m_{d_{13}}m_{d_{23}}}\\
\frac{y_2^dy_3^dy_7^dy_8^d\langle\varphi_{12}^*\rangle\langle\phi^*_{12}\rangle\langle\phi^*_{23}\rangle}{m_{d_{12}}m_{d_{13}}m_{d_{23}}} & \frac{y_5^dy_7^d\langle\phi^*_{23}\rangle}{m_{d_{23}}} & \frac{y_6^dy_7^d\langle\varphi_{23}\rangle}{m_{d_{23}}} \\
0 & 0 & y_{33}^d \end{pmatrix},\\
\mathcal{M}_e &=& \frac{v}{\sqrt2}\begin{pmatrix}
\frac{y_1^ey_2^ey_3^e\langle\phi^*_{12}\rangle\langle\phi^*_{23}\rangle}{m_{e_{12}}m_{e_{13}}} & \frac{y_1^ey_5^e\langle\phi^*_{23}\rangle}{y_{8}^e\langle\phi^*_{12}\rangle}+\frac{y_1^ey_4^ey_5^e\langle\phi_{12}\rangle\langle\phi^*_{23}\rangle}{m_{e_{13}}m_{e_{23}}} & \frac{y_1^ey_6^e\langle\phi_{23}\rangle}{y_{8}^u\langle\phi^*_{12}\rangle}+\frac{y_1^ey_4^ey_6^e\langle\phi_{12}\rangle\langle\phi_{23}\rangle}{m_{e_{13}}m_{e_{23}}}\\
\frac{y_2^ey_3^ey_7^ey_8^e\langle\phi_{12}^*\rangle\langle\phi^*_{12}\rangle\langle\phi^*_{23}\rangle}{m_{e_{12}}m_{e_{13}}m_{e_{23}}} & \frac{y_5^ey_7^e\langle\phi^*_{23}\rangle}{m_{e_{23}}} & \frac{y_6^ey_7^e\langle\phi_{23}\rangle}{m_{e_{23}}} \\
0 & 0 & y_{33}^e \end{pmatrix}.
\eea
It is precisely that the $(3,1)$ and $(3,2)$ elements are not be generated at tree-level, while other components are naturally small with respect to the $(3,3)$ element, as suppressed by the ratios of flavon VEVs over messenger masses as well as the assumption $v_{23}\ll v_{12}$. The large amount of parametric freedom allows us to parametrize the low-energy SM charged fermion mass matrices as follows: 
\bea \mathcal{M}_u &=& \fr{v}{\sqrt2}\begin{pmatrix}
y^u_{11}\la^7 & y^u_{12}\la^6 & y^u_{13}\la^5\\
y^u_{21}\la^8 & y^u_{22}\la^4 & y^u_{23}\la^3 \\
0 & 0 & y_{33}^u \end{pmatrix},\hs \mathcal{M}_d = \fr{v}{\sqrt2}\begin{pmatrix}
y_{11}^d\la^7 & y_{12}^d\la^7 & y_{13}^d\la^6\\
y_{21}^d\la^9 & y_{22}^d\la^5 & y_{23}^d\la^4 \\
0 & 0 & y_{33}^d \end{pmatrix},\\
\mathcal{M}_e &=& \fr{v}{\sqrt2}\begin{pmatrix}
y_{11}^e\la^6 & y_{12}^e\la^6 & y_{13}^e\la^6\\
y_{21}^e\la^7 & y_{22}^e\la^5 & y_{23}^e\la^5 \\
0 & 0 & y_{33}^e \end{pmatrix}, \eea
where the coefficients $y$'s are dimensionless and $\la$ is the Wolfenstein parameter, $\la=0.22501(68)$ \cite{UTfit:2022hsi}.

By applying biunitary transformations, we can diagonalize the $\mathcal{M}_{u,d,e}$ mass matrices separately, and then get the realistic masses of the up quarks $u,c,t$, the down quarks $d,s,b$, as well as the charged leptons $e,\mu,\tau$, such as 
\bea V_{u_L}^\dag \mathcal{M}_u V_{u_R} &=& \text{diag}(m_u,m_c,m_t),\\
V_{d_L}^\dag \mathcal{M}_d V_{d_R} &=& \text{diag}(m_d,m_s,m_b),\\
V_{e_L}^\dag \mathcal{M}_e V_{e_R} &=& \text{diag}(m_e,m_\mu,m_\tau), \eea
where $V_{u_{L,R}}$, $V_{d_{L,R}}$ and $V_{e_{L,R}}$ are unitary matrices, linking gauge states, $u_a=(u_1,u_2,u_3)$, $d_a=(d_1,d_2,d_3)$, $e_a=(e_1,e_2,e_3)$, to mass eigenstates, $u_i=(u,c,t)$, $d_i=(d,s,b)$, $e_i=(e,\mu,\tau)$, respectively, namely
\be u_{aL,R} = [V_{u_{L,R}}]_{ai}u_{iL,R}, \hs d_{aL,R} = [V_{d_{L,R}}]_{ai}d_{iL,R}, \hs e_{aL,R} = [V_{e_{L,R}}]_{ai}e_{iL,R}.\ee
Here, $a(i)=1,2,3$ labels mass (gauge) eigenstates. Then, the Cabibbo-Kobayashi-Maskawa (CKM) matrix is defined by $V=V_{u_L}^\dag V_{d_L}$. 

\subsection{\label{neu}Neutral fermion sector}
From the terms of the first two rows of Eq. (\ref{yuneu}), we obtain a full neutrino mass matrix in the basis $(\nu_{aL}, \nu_{13L}, \nu_{23L}, \nu_{13R}^c, \nu_{23R}^c, \nu_{3R}^c)$, which has the form
\be \begin{pmatrix}
0 & M_D \\ M_D^T & M_M
\end{pmatrix}, \ee
where the submatrices are given by
\bea M_D &=& -\frac{v}{\sqrt2} \begin{pmatrix}
0 & 0 & y_1 & 0 & 0\\ 0 & 0 & 0 & y_2 & 0\\ 0 & 0 & 0 & 0 & y_3
\end{pmatrix},\\
M_M &=& -\frac{1}{\sqrt2} \begin{pmatrix}
0 & 0 & m_{\nu_{13}} & x_1v_{12} & 0\\ 0 & 0 & x_2v_{12} & m_{\nu_{23}} & z_1v_{23}\\ m_{\nu_{13}} & x_2v_{12} & 0 & 0 & 0 \\ x_1v_{12} & m_{\nu_{23}} & 0 & 0 & z_2v_{23} \\ 0 & z_1v_{23} & 0 & z_2v_{23} & \sqrt2 M_3
\end{pmatrix}. \eea
With the aid of the hierarchy $M_3,m_{\nu_{13,23}},v_{12,23}\gg v$, one finds $M_M\gg M_D$, so that the seesaw formula can be applied to extract a $3\times 3$ submatrix for the light neutrino sector. The resulting neutrino mass matrix takes a factorized form and can be written as the outer product of two vectors, i.e., 
\bea M_\nu^\mathrm{tree} &\simeq& -M_DM_M^{-1}M_D^T\crn
&=& \begin{pmatrix} -\fr{v^2 v_{12}^2 v_{23}^2 x_1^2 y_1^2 z_1^2}{2 F_1 F_2}\\ \fr{m_{\nu_{13}} v^2 v_{12} v_{23}^2 x_1 y_1 y_2 z_1^2}{2 F_1 F_2}\\ -\fr{v^2 v_{12} v_{23} x_1 y_1 y_3 z_1}{2 F_2} \end{pmatrix}\begin{pmatrix} 1 & -\fr{m_{\nu_{13}} y_2}{v_{12} x_1 y_1} & \fr{F_1 y_3}{v_{12} v_{23} x_1 y_1 z_1} \end{pmatrix},\eea
where $F_1 = m_{\nu_{13}}m_{\nu_{23}} - x_1 x_2 v_{12}^2$ and $F_2 = M_3 F_1 - \sqrt{2} m_{\nu_{13}} z_1 z_2 v_{23}^2$. Further, assuming $M_3\gg m_{\nu_{13}}\simeq m_{\nu_{23}}\simeq v_{12}$ and taking $z_1=z_2=1$, we obtain
\be M_\nu^\mathrm{tree}\simeq \frac{v^2}{2M_3}\begin{pmatrix}
\frac{y_1^2x_1^2v_{23}^2}{(x_1x_2-1)^2v_{12}^2} & -\frac{y_1y_2x_1v_{23}^2}{(x_1x_2-1)^2v_{12}^2} & \frac{y_1y_3x_1v_{23}}{(x_1x_2-1)v_{12}} \\
-\frac{y_1y_2x_1v_{23}^2}{(x_1x_2-1)^2v_{12}^2} & \frac{y_2^2v_{23}^2}{(x_1x_2-1)^2v_{12}^2} & -\frac{y_2y_3v_{23}}{(x_1x_2-1)v_{12}}\\
\frac{y_1y_3x_1v_{23}}{(x_1x_2-1)v_{12}} & -\frac{y_2y_3v_{23}}{(x_1x_2-1)v_{12}} & y_3^2
\end{pmatrix}. \ee
Notably, although most entries of $M_\nu^{\mathrm{tree}}$ are suppressed by the ratio $v_{23}/v_{12}\sim 10^{-2}$, this is compensated by the small denominator $(x_1x_2-1)\sim 10^{-2}$, so that all entries of the mass matrix can be of the same order. Consequently, the texture naturally yields large neutrino mixing angles without requiring fine-tuning of small Yukawa couplings. For representative values obtained in the numerical analysis below, $x_1\simeq 1.27$, $x_2\simeq 0.74$, $y_1\simeq 1.10$, $y_2\simeq 0.24$, $y_3\simeq 0.24$, the Yukawa couplings remain $\mathcal{O}(1)$, as typically expected in flavor models. Furthermore, the observed neutrino mass scale $m_\nu \sim 0.1$ eV is reproduced for a heavy mass scale $M_3\sim 10^{11}$ TeV, while the gauge symmetry-breaking scales $v_{12}$ and $v_{23}$ can lie in the multi-TeV range.

\begin{figure}[h]
\centering
\includegraphics[scale=1]{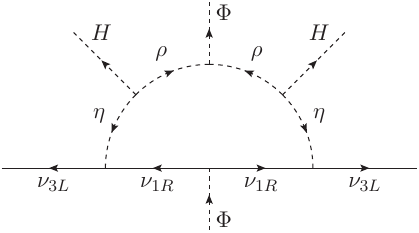}
\caption[]{\label{plotmassnu}Scotogenic neutrino mass generation governed by residual dark parity.}
\end{figure}

It is worth emphasizing that the tree-level neutrino mass matrix $M_\nu^\mathrm{tree}$ generated via the seesaw mechanism, takes the factorized form discussed above and is therefore manifestly of rank~1. Consequently, it predicts only one massive active neutrino, while the other two remain massless, in conflict with current neutrino oscillation data \cite{ParticleDataGroup:2024cfk}. However, the interactions of the third row in Eq. (\ref{yuneu}), together with the last two scalar couplings in Eq. (\ref{potendark}), induce a one-loop radiative correction to the $(3,3)$ entry of the neutrino mass matrix, as illustrated in Fig. \ref{plotmassnu}. As a result, in our model the atmospheric neutrino mass-squared splitting is generated at tree level, whereas the solar neutrino mass-squared difference arises at one loop. In the mass eigenstate basis, the interactions relevant to this radiative contribution are given by:
\be \mathcal{L}\supset \frac{\kappa V_{1\al}}{\sqrt2}\bar{\nu}_{3L}(c_R R_1 +s_R R_2+ic_I I_1 +i s_I I_2 ) N_{\al R} + \mathrm{H.c.}, \ee 
where $V$ is a $2\times 2$ rotation matrix,\be V=\begin{pmatrix}
c_\xi & s_\xi \\ -s_\xi & c_\xi
\end{pmatrix}, \ee relating $\nu_{1,2R}$ to their two mass eigenstates $N_{\al R}$ for $\al=1,2$. The respective mass eigenvalues and mixing angle are given by $M_{1,2}=\fr 1 2 [m_{\nu_{1R}}+m_{\nu_{2R}}\mp\sqrt{(m_{\nu_{1R}}-m_{\nu_{2R}})^2+4m_{12}^2}]$ and $t_{2\xi}=2m_{12}/(m_{\nu_{1R}}-m_{\nu_{2R}})$ with $m_{\nu_{1,2R}}=-f_{1,2}\La/\sqrt2$. Hence, the radiative contribution is defined by
\be [M_\nu^\mathrm{rad}]_{33} = \frac{(\kappa V_{1\al})^2M_\al}{32\pi^2}\left(\frac{c^2_Rm^2_{R_1}\ln\frac{M_\al^2}{m^2_{R_1}}}{M_\al^2-m^2_{R_1}}-\frac{c^2_I m^2_{I_1}\ln\frac{M_\al^2}{m^2_{I_1}}}{M_\al^2-m^2_{I_1}}+\frac{s^2_R m^2_{R_2}\ln\frac{M_\al^2}{m^2_{R_2}}}{M_\al^2-m^2_{R_2}}-\frac{s^2_Im^2_{I_2}\ln\frac{M_\al^2}{m^2_{I_2}}}{M_\al^2-m^2_{I_2}}\right).\ee
Because the dark scalar mixings and mass splittings are significantly suppressed, i.e., $|\theta_{R,I}|\sim v/v_{12}\ll 1$, $(m^2_{R_1}-m^2_{I_1})/m^2_{R_1,I_1}\sim (v^2 /v_{12}^2)(\La/v_{12}) \ll 1$, $(m^2_{R_2}-m^2_{I_2})/m^2_{R_2,I_2}\sim \La/v_{12} \ll 1$, and the physical $P_\mathsf{D}$-odd right-handed neutrinos are much lighter than the dark neutral scalars, i.e., $M_\al/ m_{R_1,R_2,I_1,I_2}\sim\La/v_{12}\ll 1$, the radiative contribution is proportional to $(\kappa^2 V_{1\al}^2/32\pi^2)(v^2 /v_{12})(\La^2/v_{12}^2)\sim 0.1$ eV, taking $\kappa\sim\mathcal{O}(1)$, $\La\sim\mathcal{O}(1)$ TeV, and $v_{12}\sim\mathcal{O}(10^3)$ TeV, as expected.

Finally, it is worth noting that the total neutrino mass matrix, given by \be M_\nu^\mathrm{tot} = M_\nu^\mathrm{tree}+M_\nu^\mathrm{rad},\ee
has rank 2. This structure yields two massive active neutrinos and one massless neutrino, which is sufficient to accommodate current neutrino oscillation data \cite{ParticleDataGroup:2024cfk}. This outcome is reminiscent of what one obtains in a minimal type-I seesaw/scotogenic/scotoseesaw extension of the SM with only two right-handed neutrinos \cite{Schechter:1980gr,King:1999mb,VanDong:2023thb,Ma:2006km,Rojas:2018wym}. The total neutrino mass matrix can be diagonalized via a unitary transformation: $V_{\nu_L}^T M_\nu^\mathrm{tot}V_{\nu_L}=\text{diag}(0,m_2,m_3)$, where $m_{2,3}$ are the physical neutrino masses, and $V_{\nu_L}$ is a unitary matrix that relates the gauge basis neutrino states $\nu_{aL}=(\nu_{1L},\nu_{2L},\nu_{3L})^T$ to the physical flavor eigenstates $\nu_{iL}=(\nu_{e L},\nu_{\mu L},\nu_{\tau L})^T$ via $\nu_{aL}=[V_{\nu_L}]_{ai}\nu_{iL}$. Consequently, the Pontecorvo–Maki–Nakagawa–Sakata (PMNS) matrix, which governs neutrino mixing in charged-current interactions, is given by $V_\text{PMNS}=V_{\nu_L}^\dag V_{e_L}$, where $V_{e_L}$ is the unitary matrix that diagonalizes the charged lepton mass matrix. 

\subsection{Number analysis of fermion spectrum}

\begin{figure}[h]
\centering
\includegraphics[scale=0.426]{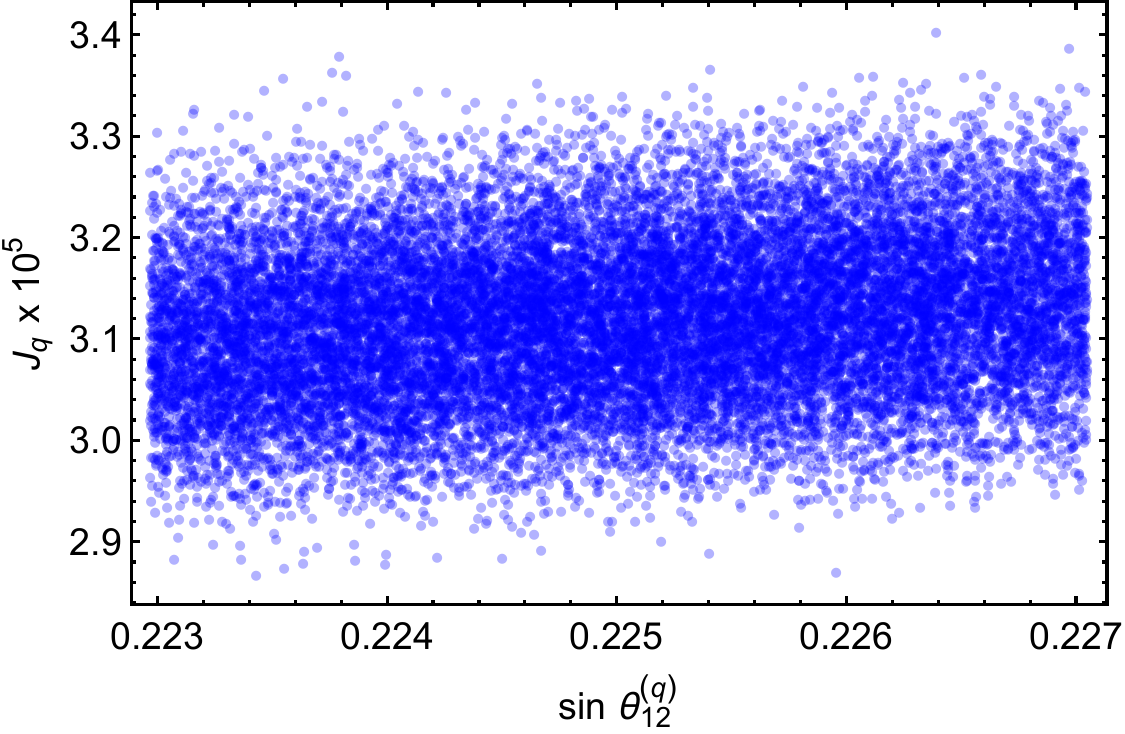}
\includegraphics[scale=0.42]{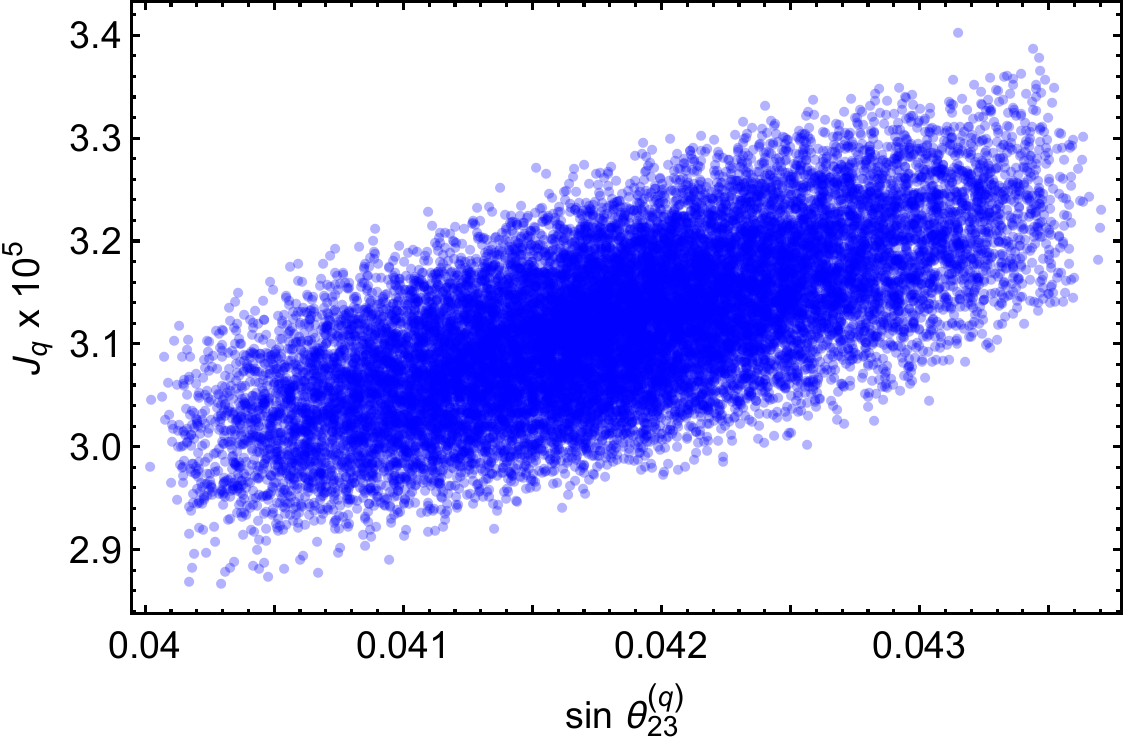}
\includegraphics[scale=0.42]{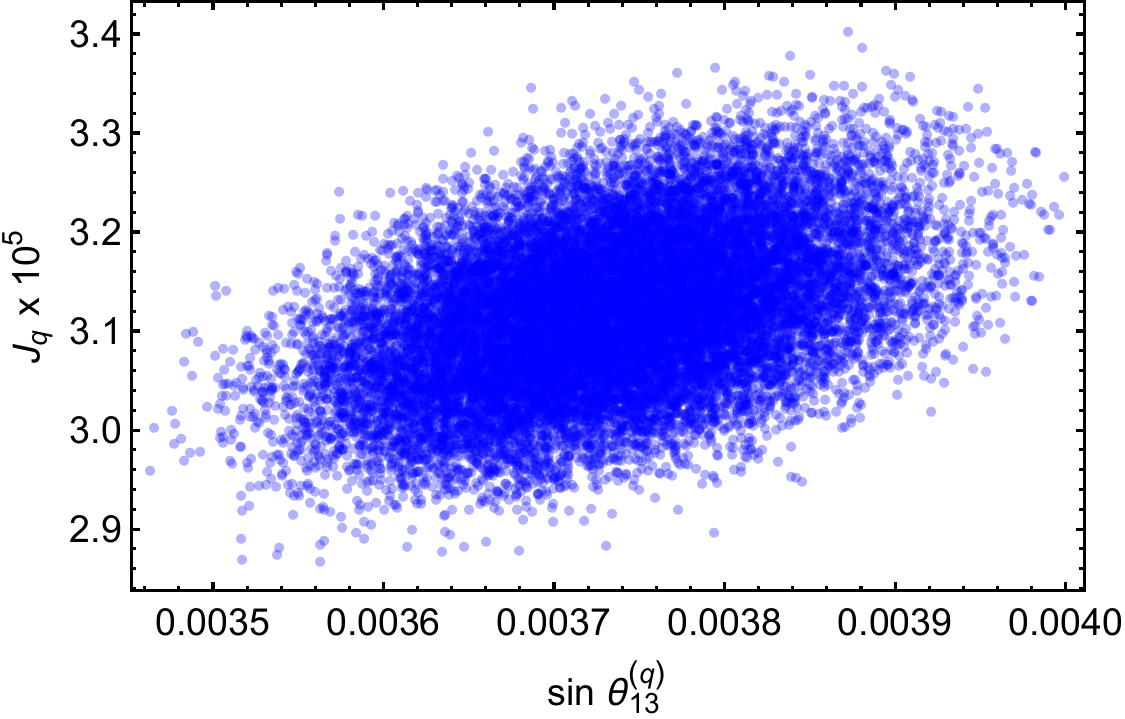}
\caption[]{\label{plotsvsJq}Correlation between the quark mixing angles and the Jarlskog invariant.}
\end{figure}

\begin{table}[
]
\begin{center}
\begin{tabular}{|c|c|c|}
\hline
Observable & Experimental value & Model value \\ \hline
$m_{u}$ [MeV] &  $1.24\pm 0.22$ & $1.24901$ \\ \hline
$m_{c}$ [GeV] & $0.62\pm 0.02$ & $0.61993$ \\ \hline
$m_{t}$ [GeV] & $172.9\pm 0.4$ & $172.901$ \\
\hline
$m_{d}$ [MeV] & $2.69\pm 0.19$ & $2.68948$ \\ \hline
$m_{s}$ [MeV] & $53.5\pm 4.6$ & $53.4558$ \\ \hline
$m_{b}$ [GeV] & $2.86\pm 0.03$ & $2.86009$ \\ \hline
$\sin \theta^{(q)} _{12}$ & $0.22501\pm 0.00068$ & $0.225011$\\
\hline
$\sin \theta^{(q)} _{23}$ & $0.04183_{-0.00069}^{+0.00079}$ & $0.041829$\\
\hline
$\sin \theta^{(q)} _{13}$ & $0.003732^{+0.000090}_{-0.000085}$ & $0.003732 $\\
\hline
$J_q$ & $\left(3.12^{+0.13}_{-0.12}\right) \times 10^{-5}$ & $3.11993 \times 10^{-5}$ \\ \hline
\end{tabular}
\end{center}
\caption{Experimental values of the SM quark masses \cite{Xing:2020ijf} and CKM parameters \cite{ParticleDataGroup:2024cfk} along with the model values obtained for the best fit solution corresponding to $\chi^2\simeq 2\times 10^{-4}$.}
\label{tabquark}
\end{table}
To determine the best-fit parameters that reproduce the observed fermion masses and mixing, we minimize the $\chi^2$ function, defined as 
\be \chi^2=\sum_i\left(\frac{O_i^\text{calc}-O_i^\text{exp}}{\Delta O_i}\right)^2, \label{chi}\ee
where $O_i^\text{calc}$ and $O_i^\text{exp}$ denote the model prediction and the experimental central value of the $i$-th observable, respectively, and $\Delta O_i$ represents its associated uncertainty. In the quark sector, the summation runs over the six quark masses, the CKM mixing angles, and the Jarlskog invariant, with the uncertainties $\Delta O_i$ taken as the $3\sigma$ experimental errors. As summarized in Table \ref{tabquark}, our model successfully accommodates the observed quark mass spectrum, CKM mixing angles, and Jarlskog invariant, taking 
\bea y_{11}^u &\simeq & -0.274808, \hs y_{12}^u \simeq -2.8602, \hs y_{13}^u \simeq 6.23966, \hs y_{21}^u \simeq 1.17024, \\
y_{22}^u &\simeq & 1.38277, \hs y_{23}^u \simeq 0.283248, \hs y_{33}^u \simeq 0.993968, \hs \text{arg}~(y_{13}^u) \simeq 108.115^\circ,\\
y_{11}^d &\simeq & -0.52976, \hs y_{12}^d \simeq -1.28349, \hs y_{13}^d \simeq 0.883176, \hs y_{21}^d \simeq 0.598669,\\
y_{22}^d &\simeq & -0.52933, \hs y_{23}^d \simeq -0.286541, \hs y_{33}^d \simeq -0.0164254. \eea
These results indicate that the Yukawa couplings $y^{u,d}_{1,2,\cdots,8}$ are all of order $\mathcal{O}(1)$. In the analysis above, all parameters are assumed to be real, except for $y_{13}^u$, which is taken to be complex and is solely responsible for generating the CP-violating phase in the quark sector. Furthermore, the correlation between the quark mixing angles and the Jarlskog invariant is illustrated in Fig. \ref{plotsvsJq}. The figure demonstrates that the Jarlskog invariant exhibits strong sensitivity to variations in the mixing angles $\theta_{13}^{(q)}$ and $\theta_{23}^{(q)}$, while being largely insensitive to changes in $\theta_{12}^{(q)}$. This highlights the crucial role of the former two angles in controlling CP violation in the quark sector.

\begin{figure}[h!]
	\centering
	\includegraphics[width=6.5cm, height=4.5cm]{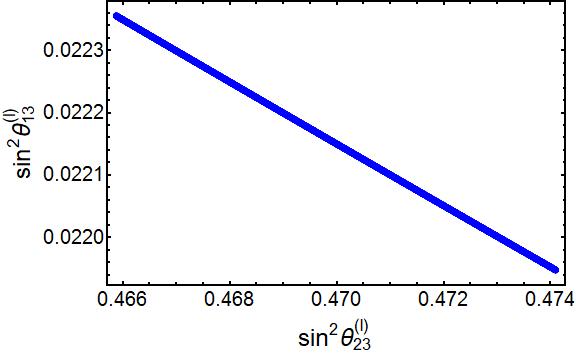}
	\includegraphics[width=6.5cm, height=4.5cm]{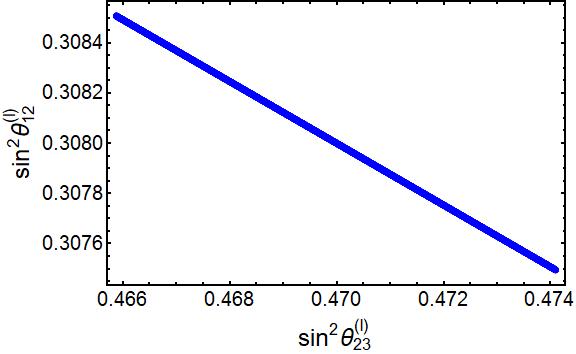}\newline
	\includegraphics[width=6.5cm, height=4.5cm]{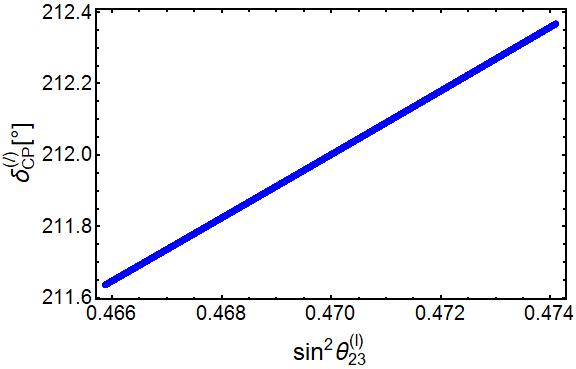}
	\caption[]{\label{leptonobs}Correlation among some lepton sector observables.}
\end{figure}

\begin{table}[h]
\begin{center}
\begin{tabular}{|c|c|c|c|c|c|c|}
\hline
\multirow{2}{*}{Observable} & \multicolumn{2}{c|}{Experimental value} & \multirow{2}{*}{Model value}\\
\cline{2-3}
& $1\sigma$ range & $3\sigma$ range &\\ \hline
$m_e$ [MeV] & $0.4883266(17)$ & $0.4883266(51)$ & 0.488341\\ \hline
$m_\mu$ [MeV] & $102.87267(21)$ & $102.87267(63)$ & 102.873\\ \hline
$m_\tau$ [MeV] & $1747.43(12)$ & $1747.43(36)$ & 1747.43\\ \hline
$\Delta m_{21}^{2}$ $[10^{-5}$eV$^{2}]$ & $7.49_{-0.19}^{+0.19}$ & 6.92--8.05 & 7.49004 \\ \hline
$\Delta m_{31}^{2}$ $[10^{-3}$eV$^{2}]$ & $2.513_{-0.019}^{+0.021}$ & 2.451--2.578 & 2.513\\ \hline
$\sin^2\theta^{(\ell)}_{12}/10^{-1}$ & $3.08_{-0.11}^{+0.12}$ & 2.75--3.45 & 3.08 \\ \hline
$\sin^2\theta^{(\ell)}_{23}/10^{-1}$ & $4.70_{-0.13}^{+0.17}$ & 4.35--5.85 & 4.69999\\ \hline
$\sin^2\theta^{(\ell)}_{13}/10^{-2}$ & $2.215^{+0.056}_{-0.058}$ & 2.030--2.388 & 2.215\\ \hline
$\delta^{(\ell)}_{\text{CP}}$ $[^\circ]$ & $212_{-41}^{+26}$ & 124--364 & 211.995\\ \hline
\end{tabular}
\end{center}
\caption{Experimental values of the SM charged lepton masses \cite{Xing:2020ijf} and neutrino mass squared differences, leptonic mixing parameters, and CP-violating phase for the scenario of normal order neutrino mass \cite{Esteban:2024eli} along with the model values obtained for the best fit solution corresponding to $\chi^2=1.46\times 10^{-9}$.}
\label{tablepton}
\end{table}
For the lepton sector, the summation in the $\chi^2$ function defined in Eq. (\ref{chi}) includes the charged lepton masses, the experimentally measured neutrino mass-squared differences, the lepton mixing angles, and the leptonic Dirac CP-violating phase. The neutrino masses are fitted under the assumption that the symmetry breaking scales are $v_{23}\sim \mathcal{O}(10)$ TeV, $v_{12}\sim \mathcal{O}(10^3)$ TeV, and $M_3\sim \mathcal{O}(10^{11})$ TeV. As presented in Table \ref{tablepton}, the model successfully reproduces the experimental data in the neutrino sector under the normal mass ordering scenario. By solving the eigenvalue problem for the charged lepton and active neutrino mass matrices, we identify a benchmark solution that yields the correct charged lepton masses, neutrino mass-squared differences, leptonic mixing angles, and the Dirac CP phase, all of which are consistent with experimental observations. This solution leads to the following forms for the mass matrices of the SM charged leptons and light active neutrinos:
\begin{eqnarray}
M_e&=&\left(
\begin{array}{ccc}
 -50.4475-69.6146 i & 55.1413 -3.22028 i & 256.03 +20.9526 i \\
 25.2226+4.60226 i & -11.4663+11.4492 i & -374.481+673.853 i \\
 0 & 0 & -84.572+1544.43 i \\
\end{array}
\right)\mathrm{MeV},\\
M_{\nu}&=&\left(
\begin{array}{ccc}
 -19.0017+3.14894 i & -1.61237+2.90527 i & 1.48751 +18.0746 i \\
 -1.61237+2.90527 i & 0.219646 +0.529448 i & 2.60182 +1.73744 i \\
 1.48751+18.0746 i & 2.60182 +1.73744 i & 38.9502 +0. i \\
\end{array}
\right)\mathrm{meV}.
\end{eqnarray}
The correlation among several observables in the lepton sector is illustrated in Fig. \ref{leptonobs}. It is evident that $\sin^2\theta_{23}^{(\ell)}$ exhibits an inverse correlation with both $\sin^2\theta_{13}^{(\ell)}$ and $\sin^2\theta_{12}^{(\ell)}$ — that is, as $\sin^2\theta_{23}^{(\ell)}$ increases, both $\sin^2\theta_{13}^{(\ell)}$ and $\sin^2\theta_{12}^{(\ell)}$ tend to decrease, and vice versa. Moreover, a positive correlation is observed between $\sin^2\theta_{23}^{(\ell)}$ and the leptonic Dirac CP-violating phase $\delta^{(\ell)}_\text{CP}$, implying that larger values of $\sin^2\theta_{23}^{(\ell)}$ correspond to an increase in $\delta^{(\ell)}_\text{CP}$. 

\section{Electroweak precision test\label{EWtest}}
\subsection{$\rho$ parameter}
The model under consideration predicts a tree-level mixing between the SM $Z$ boson and the new neutral gauge bosons. This mixing leads to a reduction in the physical mass of the $\mathcal{Z}$ boson relative to the SM expectation, whereas the mass of the SM $W$ boson remains unaffected. As a result, the $\rho$ parameter receives a tree-level correction, which is given by
\be\Delta\rho = \frac{m^2_W}{c_W^2m_\mathcal{Z}^2}-1\simeq \frac{9v^2}{10v_{23}^2}\left(1-\frac{g_Y^2}{g_3^2}\right)^2. \ee
Using the global fit value, $\rho=1.00031\pm 0.00019$ \cite{ParticleDataGroup:2024cfk}, we derive a low bound on the new physics scale $v_{23}$, shown as the red curve in Fig. \ref{plotsv23vsg3}. In this analysis, we have taken $v=246$ GeV, $s_W^2=0.231$, $g=0.652$, and imposed $\Delta\rho=0.0005$. The shaded region corresponds to the excluded parameter space. The constraint is most stringent for $g_3\gtrsim 0.79$. For reference, the dashed blue line represents a natural gauge unification scenario in which the three gauge couplings are of similar magnitude, i.e., $g_1=g_2=g_3=\sqrt3 g_Y$. In this case, we obtain the lower bounds $v_{23} \gtrsim 7.0$ TeV and $m_{Z_{23}}\gtrsim 2.8$ TeV.

\subsection{$\varepsilon_{1,2}$ mixing parameters}
Due to the mixing between the SM $Z$ boson and the new neutral gauge bosons $Z_{1,2}$, the well-measured vector and axial-vector couplings of the SM $Z$ boson to fermions are modified by terms proportional to the mixing parameters $\varepsilon_{1,2}$, namely, 
\be g_{V,A}^\mathcal{Z}(f)= g_{V,A}^Z(f)+\mathcal{O}(\varepsilon_{1,2}), \ee 
as explicitly shown in Table \ref{Tab1} in Appendix \ref{AppB}. According to the electroweak precision data \cite{ParticleDataGroup:2024cfk}, the new physics effects remain consistent with observations if $|\varepsilon_{1,2}|\lesssim 10^{-3}$ \cite{ALEPH:2005ab,Erler:2009jh}. Hence, we take the bound $|\varepsilon_1|=10^{-3}$ into account and then obtain the allowed parameter space as determined by the brown curve in Fig. \ref{plotsv23vsg3}. In the gauge unification scenario $g_1=g_2=g_3=\sqrt3 g_Y$,  this translates to the bounds $v_{23}\gtrsim 6.0$ TeV and $m_{Z_{23}}\gtrsim 2.4$ TeV.

\subsection{Total $\mathcal{Z}$ decay width}
\begin{figure}[h]
\centering
\includegraphics[scale=0.42]{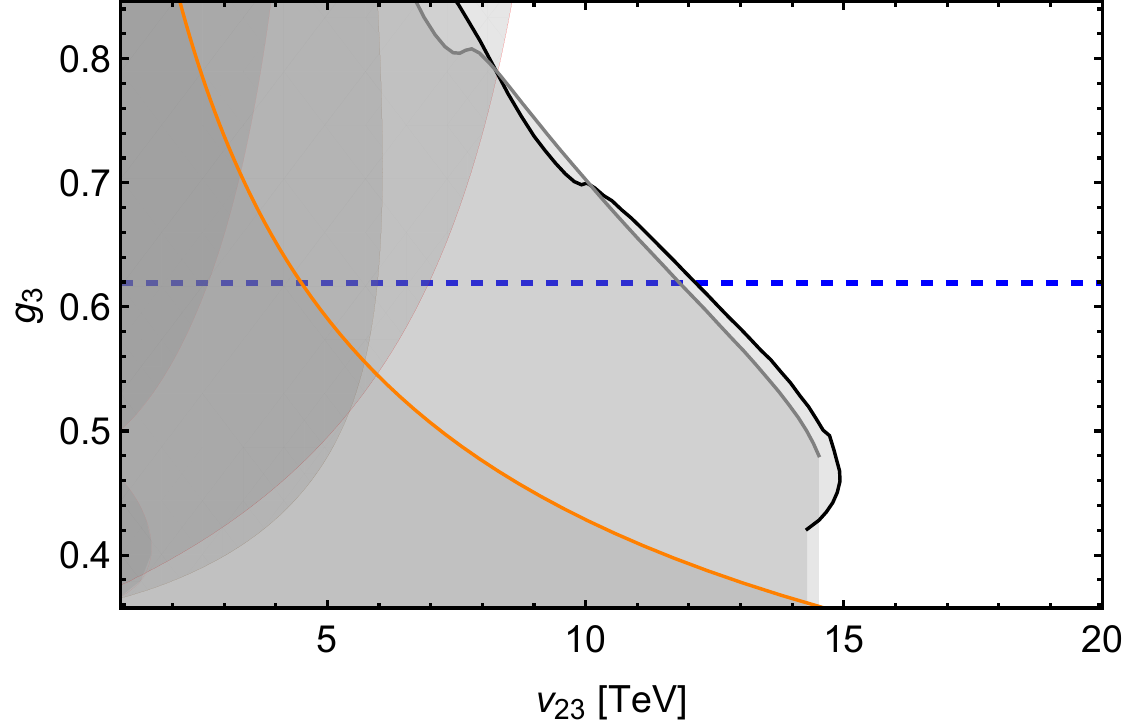}
\includegraphics[scale=0.413]{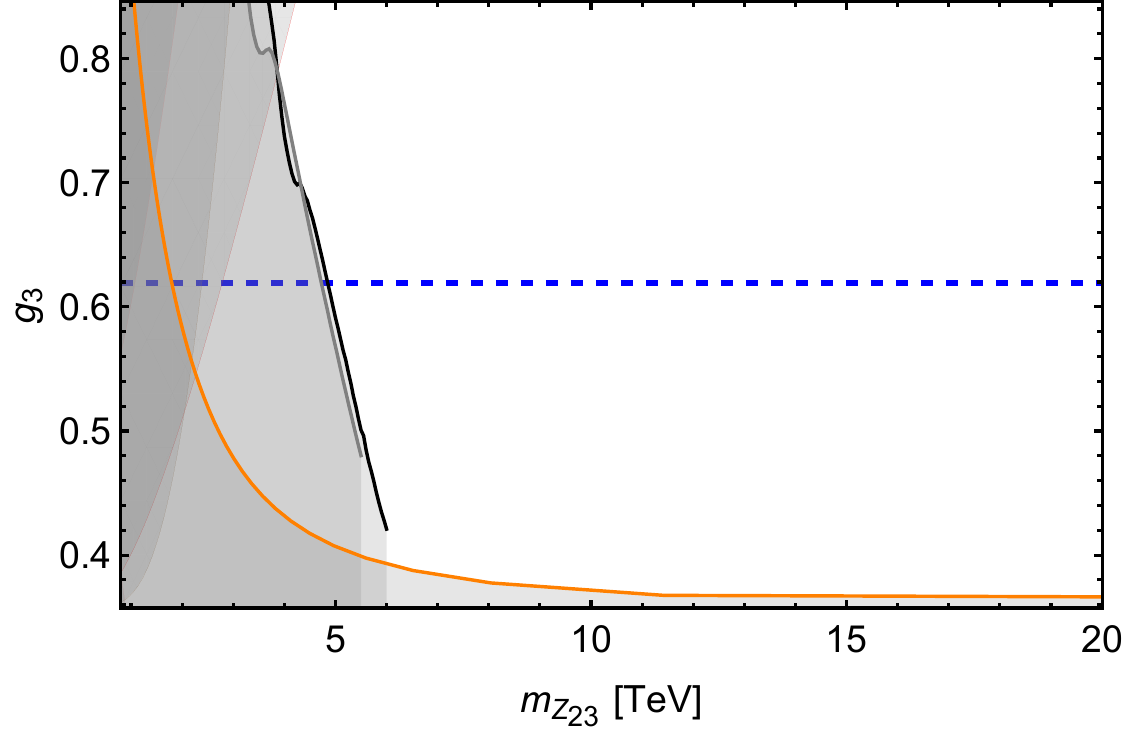}
\caption[]{\label{plotsv23vsg3}Red, brown, pink, orange, gray, and black curves denote the exclusion bounds derived from the $\rho$ parameter \cite{ParticleDataGroup:2024cfk}, $\varepsilon_{1,2}$ mixing parameters \cite{ALEPH:2005ab, Erler:2009jh}, total $\mathcal{Z}$ decay width \cite{ParticleDataGroup:2024cfk}, LEPII \cite{ALEPH:2013dgf}, ATLAS \cite{ATLAS:2019erb} and CMS \cite{CMS:2021ctt}, respectively. The shaded regions indicate the excluded parameter space. The dashed blue line corresponds to the gauge unification condition $g_1=g_2=g_3=\sqrt3 g_Y$.}
\end{figure}

The precision measurement of the total decay width of the $\mathcal{Z}$ boson allows us to place constraints on the free parameters of the model. The relevant interaction between $\mathcal{Z}$ and the SM fermions is described by the Lagrangian: 
\be \mathcal{L} \supset - \frac{g}{2c_W}\left\{\bar{\nu}_{iL}\ga^\mu\tilde{G}_L^\mathcal{Z}(\nu_{iL})(1+\delta^{\nu_{iL}})\nu_{iL} + \bar{f}\ga^\mu [\tilde{G}_{V}^\mathcal{Z}(f)(1+\delta_V^{f})-\tilde{G}_{A}^\mathcal{Z}(f)(1+\delta_A^{f})\ga_5]f\right\} \mathcal{Z}_\mu,\ee
where $f$ runs over all the SM charged fermions, and
\bea \tilde{G}_L^\mathcal{Z}(\nu_{iL}) &=& \fr{1}{2}\sum_{k=1}^3(V^*_{\nu L})_{ki}(V_{\nu L})_{ki},\\
\tilde{G}_{V,A}^\mathcal{Z}(e_i) &=& \fr{1}{2}\sum_{k=1}^3\left[-c_{2W}(V^*_{eL})_{ki}(V_{eL})_{ki}\pm (1-c_{2W})(V^*_{eR})_{ki}(V_{eR})_{ki}\right],\\
\tilde{G}_{V,A}^\mathcal{Z}(u_i) &=& \fr{1}{6}\sum_{k=1}^3\left[(1+2c_{2W})(V^*_{uL})_{ki}(V_{uL})_{ki}\pm 2(c_{2W}-1)(V^*_{uR})_{ki}(V_{uR})_{ki}\right],\\
\tilde{G}_{V,A}^\mathcal{Z}(d_i) &=& \fr{1}{6}\sum_{k=1}^3\left[-(2+c_{2W})(V^*_{dL})_{ki}(V_{dL})_{ki}\pm (1-c_{2W})(V^*_{dR})_{ki}(V_{dR})_{ki}\right],\eea
and $\delta^{\nu_{iL}}=\tilde{g}^\mathcal{Z}_L(\nu_{iL})/\tilde{G}_L^\mathcal{Z}(\nu_{iL})-1$, $\delta_{V,A}^{f}=\tilde{g}^\mathcal{Z}_{V,A}(f)/G_{V,A}^\mathcal{Z}(f)-1$, in which the couplings in mass eigenstates $\tilde{g}^\mathcal{Z}_L(\nu_{iL})$ and $\tilde{g}^\mathcal{Z}_{V,A}(f)$ can be extracted from the couplings collected in Table \ref{Tab1},
\bea \tilde{g}^\mathcal{Z}_L(\nu_{iL})&=&\fr{1}{2}\sum_{k=1}^3(V^*_{\nu L})_{ki}\left(g^\mathcal{Z}_{V}(\nu_{kL})+g^\mathcal{Z}_{A}(\nu_{kL})\right)(V_{\nu L})_{ki},\label{gLneu}\\
\tilde{g}^\mathcal{Z}_{V,A}(e_i)&=&\fr{1}{2}\sum_{k=1}^3\left[(V^*_{eL})_{ki}\left(g^\mathcal{Z}_{V}(e_k)+g^\mathcal{Z}_{A}(e_k)\right)(V_{eL})_{ki}\pm (V^*_{eR})_{ki}\left(g^\mathcal{Z}_V(e_k)-g^\mathcal{Z}_A(e_k)\right)(V_{eR})_{ki}\right],\\
\tilde{g}^\mathcal{Z}_{V,A}(u_i)&=&\tilde{g}^\mathcal{Z}_{V,A}(e_i)|_{e_k\to u_k, V_{eL,R}\to V_{uL,R}},\\
\tilde{g}^\mathcal{Z}_{V,A}(d_i)&=&\tilde{g}^\mathcal{Z}_{V,A}(e_i)|_{e_k\to d_k, V_{eL,R}\to V_{dL,R}}\label{gVAd}.\eea
Hence, the total $\mathcal{Z}$ decay width predicted by our model is separated to $\Gamma_\mathcal{Z}=\Gamma^{\text{SM}}_\mathcal{Z}+\Delta\Gamma_\mathcal{Z}$, in which $\Gamma^{\text{SM}}_\mathcal{Z}$ is the SM value and the shift $\Delta\Gamma_\mathcal{Z}$ is given by 
\bea
\Delta\Gamma_\mathcal{Z}&\simeq&\frac{m^{\text{SM}}_\mathcal{Z}}{6\pi}\frac{g^2}{4c^2_W}\left\{\sum_{f}N^f_C\left[\left|\tilde{G}_{V}^\mathcal{Z}(f)\right|^2\text{Re}[\delta_V^{f}]+\left|\tilde{G}_{A}^\mathcal{Z}(f)\right|^2\text{Re}[\delta_A^{f}]\right]+2\sum_i\left|\tilde{G}_L^\mathcal{Z}(\nu_{iL})\right|^2\text{Re}[\delta^{\nu_{iL}}]\right\}\crn
&&+\frac{\Delta m_\mathcal{Z}}{12\pi}\frac{g^2}{4c^2_W}\left\{\sum_f N_C^f\left[\left|\tilde{G}_{V}^\mathcal{Z}(f)\right|^2+\left|\tilde{G}_{A}^\mathcal{Z}(f)\right|^2\right]+2\sum_i\left|\tilde{G}_L^\mathcal{Z}(\nu_{iL})\right|^2\right\},
\eea
where $m^{\text{SM}}_\mathcal{Z}$ is the SM value of the $\mathcal{Z}$-boson mass, $N_C^f$ is the color number of the fermion $f$, and the mass shift $
\Delta m_\mathcal{Z}\simeq gvt_Wg_3 \varepsilon_1/4g_{12}$. Using the total $\mathcal{Z}$ decay width measured by the experiment $\Gamma^{\text{exp}}_\mathcal{Z}=2.4952\pm0.0023$ GeV and predicted by the SM $\Gamma^{\text{SM}}_\mathcal{Z}=2.4942\pm0.0008$ GeV \cite{ParticleDataGroup:2024cfk}, we require $|\Delta\Gamma_\mathcal{Z}|<0.0041$ GeV and then obtain a bound for viable parameter space regions as determined by pink curve in Fig. \ref{plotsv23vsg3}. This bound is generally lower than that given by the two previous cases. 

\section{Collider bounds\label{Colliderbounds}}
\subsection{LEPII}
LEPII probes possess $e^+e^-\to f\bar{f}$ for $f=e,\mu,\tau$, which can be mediated by new neutral gauge bosons such as $Z_{23}$ and $Z_{12}$ \cite{ALEPH:2005ab}. Since the center-of-mass energy of the LEPII, $\sqrt{s}=209$ GeV, is much smaller than the masses of these new gauge bosons, $m_{Z_{23}}$ and $m_{Z_{12}}$, such processes can be described by effective four-fermion contact interactions, namely  
\be \mathcal{L}_{\text{eff}}=\frac{g^2}{c^2_W}\sum_I\sum_{\mathcal{A,B}=L,R}\frac{1}{m^2_I}\tilde{g}_\mathcal{A}^I(e)\tilde{g}_\mathcal{B}^I(f)(\bar{e}_\mathcal{A}\gamma_\mu e_\mathcal{A})(\bar{f}_\mathcal{B}\gamma^\mu f_\mathcal{B}),\label{chiint}\ee
where $I=Z_{23}, Z_{12}$, and $\tilde{g}^I_{L,R}(f)$ are chiral gauge couplings of boson $I$ with fermion $f$, which can be extracted from Table \ref{Tab2} in Appendix \ref{AppB} as
\bea \tilde{g}^I_L(f)&=&\fr{1}{2}\sum_{k=1}^3(V^*_{eL})_{kf}\left(g^I_{V}(e_k)+g^I_{A}(e_k)\right)(V_{eL})_{kf},\\ 
\tilde{g}^I_R(f)&=& \fr{1}{2}\sum_{k=1}^3(V^*_{eR})_{kf}\left(g^I_V(e_k)-g^I_A(e_k)\right)(V_{eR})_{kf}. \eea 

The LEPII experiment reported the lower limit of scale of these contact interaction types as $\La^\pm_{ff}$, in which $\La^+_{ff}$ for $\text{Re}[\tilde{g}_\mathcal{A}^I(e)(\tilde{g}_\mathcal{B}^I(f))^*]>0$ and $\La^-_{ff}$ for $\text{Re}[\tilde{g}_\mathcal{A}^I(e)(\tilde{g}_\mathcal{B}^I(f))^*]<0$ \cite{ALEPH:2013dgf}. Let us study a particular process $e^+e^-\to\mu^+\mu^-$, the mass of the $I$ boson is bounded by
\be m_I>\fr{g}{c_W}\sqrt{\fr{\left|\tilde{g}^I_\mathcal{A}(e)\tilde{g}^I_\mathcal{B}(\mu)\right|}{4\pi}}\La^\pm_{\mu\mu}.\label{lepc}\ee
It is checked that the strongest constraint on $m_{Z_{23}}$, and thus $v_{23}$, comes from the $VV$ model with $\La_{\mu\mu}^+=18.9$ TeV, which is displayed by orange curve in Fig. \ref{plotsv23vsg3}. In the region $g_3\gtrsim 0.42$, this bound is much lower than that from the LHC discussed below. However, for $g_Y\lesssim g_3\lesssim 0.42$, the LEPII constraint is significant, implying $m_{Z_{23}}\gtrsim 6$ TeV.

\subsection{LHC}
At the LHC, the $Z_{23}$ boson can be directly produced in proton-proton colliders by quark-antiquark annihilation through the process $\bar{q}q\to Z_{23}$. Once created, the heavy $Z_{23}$ boson subsequently decays into a pair of quarks (dijet channel) or a pair of charged leptons (dilepton channel). In the context of our model, the most promising and sensitive probe is provided by the dilepton decay modes, $Z_{23}\to ll$ with $l=e,\mu$. The production cross section for these processes can be calculated using the narrow width approximation, under the assumption that $\Gamma_{Z_{23}} \ll m_{Z_{23}}$, which yields: 
\be \sigma(pp \to Z_{23}\to l\bar{l})\simeq\fr 1 3 \sum _q \fr{d L_{q\bar{q}}}{dm^2_{Z_{23}}}\hat{\sigma}(q\bar{q}\to Z_{23})\mathrm{Br}(Z_{23}\to l\bar{l}),\label{lhcff}\ee where the luminosity $dL_{q\bar{q}}/dm^2_{Z_{23}}$ can be extracted from Ref. \cite{Martin:2009iq}. The partonic peak cross-section $\hat{\sigma}(q\bar{q}\to Z_{23})$ and the branching decay ratio $\mathrm{Br}(Z_{23}\to l\bar{l})=\Ga(Z_{23}\to l\bar{l})/\sum_{f} \Ga(Z_{23}\to f\bar{f})$ are given by 
\bea && \hat{\sigma}(q\bar{q}\to Z_{23})=\fr{\pi g^2}{12 c^2_W} \left(|\tilde{g}^{Z_{23}}_V(q)|^2+|\tilde{g}^{Z_{23}}_A(q)|^2\right),\label{lhczp}\\
&& \Ga(Z_{23}\to f\bar{f}) =\fr{g^2 m_{Z_{23}}}{48\pi c^2_W} N_C^f\left(|\tilde{g}^{Z_{23}}_V(f)|^2+|\tilde{g}^{Z_{23}}_A(f)|^2\right),\label{lhcdc}
\eea in which $f$ is all the SM fermions, and the relevant couplings are given by Eqs. (\ref{gLneu}--\ref{gVAd}). Above, we have assumed that the decay channels of $Z_{23}$ into right-handed neutrinos and new scalars negligibly contribute to the total decay width of $Z_{23}$.

The bound of parameter space induced by the process $pp\to Z'\to l\bar{l}$ with the latest ATLAS \cite{ATLAS:2019erb} (CMS \cite{CMS:2021ctt}) constraint taking width per resonance mass to be 3\% (0.6\%) is shown in Fig. \ref{plotsv23vsg3} by a black (gray) curve. This bound is strongest for $0.42\lesssim g_3\lesssim 0.79$ and implies that in gauge unification scenario $g_1=g_2=g_3=\sqrt3 g_Y$ then $v_{23}\gtrsim 12.1$ TeV and $m_{Z_{23}}\gtrsim 4.9$ TeV.

\section{\label{flavor}Flavor constraints}
Due to the non-universal charge assignments of fermion generations under the gauge group $U(1)_{Y_1}\otimes U(1)_{Y_2}\otimes U(1)_{Y_3}$, the model naturally contains tree-level flavor-changing neutral currents (FCNCs). These are mediated by the heavy gauge bosons $Z_{12}$ and $Z_{23}$, as well as the SM $Z$ boson, and contribute to $\Delta F=2$ processes, such as meson mass differences $\Delta m_{P}$ ($P=K,B_s,B_d,D$). It is important to note that the parameters $y^d_i$, which account for the observed fermion mass hierarchies and mixing, are taken to be real. Consequently, the model does not introduce new sources of CP violation in neutral meson mixing or rare decay processes. Additionally, the FCNC interactions also affect several $b\to sll$ observables and lepton flavor-violating decays. In this work, we provide a more detailed analysis of the FCNC phenomenology than in previous studies, such as Refs. \cite{FernandezNavarro:2023rhv,FernandezNavarro:2024hnv,Davighi:2023evx}. 

The effective Hamiltonian relevant for FCNCs $\Delta F=2$ processes can be expressed as 
\bea 
\mathcal{H}^{\Delta F=2}_{\text{eff}}&=&\fr{G_F^2m_W^2}{16\pi^2}\left\{(V^*_{tq}V_{tb})^2[(C^q_{\text{SM}}+C^q_{LL})(\bar{b}_{L}\ga_{\mu}q_{L})^2+C^{q}_{RR}(\bar{b}_{R}\ga_{\mu}q_{R})^2+C^{q}_{LR}(\bar{b}_{L}\ga_{\mu}q_{L})(\bar{b}_{R}\ga^{\mu}q_{R})]\right.\crn 
&&+\left.(V^*_{ts}V_{td})^2[(C^K_{\text{SM}}+C^K_{LL})(\bar{s}_{L}\ga_{\mu}d_{L})^2+C^{K}_{RR}(\bar{s}_{R}\ga_{\mu}d_{R})^2+C^{K}_{LR}(\bar{s}_{L}\ga_{\mu}d_{L})(\bar{s}_{R}\ga^{\mu}d_{R})]\right\},\label{Heff}
\eea 
where the first and second part describes for $B_q$ ($q=d,s$) and $K$ mixings, respectively. Additionally, $C^{q,K}_{\text{SM}}$ are SM Wilson coefficients for $B_q,K$ meson mixings, which read \cite{Buras:2012dp}
\bea 
C^q_{\text{SM}}&=&4S_0(x_t)\eta_B,\\ 
C^K_{\text{SM}}&=&4\eta_1(\la_c/\la_t)^2S_0(x_c)+4\eta_2S_0(x_t)+8\eta_3(\la_c/\la_t)S_0(x_c,x_t),
\eea 
where $x_{t(c)}=m_{t(c)}^2/m_W^2$, $\la_{t(c)}=V^*_{t(c)s}V_{t(c)d}$, while $S_0$ is the Inami-Lim function \cite{Inami:1980fz}, and the factors $\eta_{B,1,2,3}$ are next-to-leading order (NLO) QCD corrections \cite{Buras:2012dp}.   
The new physics (NP) Wilson coefficients $C^{q,K}_{LL,RR,LR}$ are defined at the matching scale $\mu=m_{Z_{12}}$ or $m_{Z_{23}}$, i.e.,  
\bea 
C^{q}_{LL}&=&\fr{16\pi^2}{G_F^2m_W^2(V^*_{tq}V_{tb})^2}\left[\fr{([\Ga^{Z_{12}}_{L}]_{3q})^2}{m_{Z_{12}}^2}+\fr{([\Ga^{Z_{23}}_{L}]_{3q})^2}{m_{Z_{23}}^2}+\fr{([\Ga^{Z}_{L}]_{3q})^2}{m_{Z}^2}\right], \\  
C^{q}_{RR}&=&\fr{16\pi^2}{G_F^2m_W^2(V^*_{tq}V_{tb})^2}\left[\fr{([\Ga^{Z_{12}}_{R}]_{3q})^2}{m_{Z_{12}}^2}+\fr{([\Ga^{Z_{23}}_{R}]_{3q})^2}{m_{Z_{23}}^2}+\fr{([\Ga^{Z}_{R}]_{3q})^2}{m_{Z}^2}\right],\\
  C^{q}_{LR}&=&\fr{16\pi^2}{G_F^2m_W^2(V^*_{tq}V_{tb})^2}\left[\fr{[\Ga^{Z_{12}}_{L}]_{3q}[\Ga^{Z_{12}}_{R}]_{3q}}{m_{Z_{12}}^2}+\fr{[\Ga^{Z_{23}}_{L}]_{3q}[\Ga^{Z_{23}}_{R}]_{3q}}{m_{Z_{23}}^2}+\fr{[\Ga^{Z}_{L}]_{3q}[\Ga^{Z}_{R}]_{3q}}{m_{Z}^2}\right],\\
C^{K}_{LL}&=&\fr{16\pi^2}{G_F^2m_W^2(V^*_{ts}V_{td})^2}\left[\fr{([\Ga^{Z_{12}}_{L}]_{21})^2}{m_{Z_{12}}^2}+\fr{([\Ga^{Z_{23}}_{L}]_{21})^2}{m_{Z_{23}}^2}+\fr{([\Ga^{Z}_{L}]_{21})^2}{m_{Z}^2}\right], \\  
C^{K}_{RR}&=&\fr{16\pi^2}{G_F^2m_W^2(V^*_{ts}V_{td})^2}\left[\fr{([\Ga^{Z_{12}}_{R}]_{21})^2}{m_{Z_{12}}^2}+\fr{([\Ga^{Z_{23}}_{R}]_{21})^2}{m_{Z_{23}}^2}+\fr{([\Ga^{Z}_{R}]_{21})^2}{m_{Z}^2}\right],\\
  C^{K}_{LR}&=&\fr{16\pi^2}{G_F^2m_W^2(V^*_{ts}V_{td})^2}\left[\fr{[\Ga^{Z_{12}}_{L}]_{21}[\Ga^{Z_{12}}_{R}]_{21}}{m_{Z_{12}}^2}+\fr{[\Ga^{Z_{23}}_{L}]_{21}[\Ga^{Z_{23}}_{R}]_{21}}{m_{Z_{23}}^2}+\fr{[\Ga^{Z}_{L}]_{21}[\Ga^{Z}_{R}]_{21}}{m_{Z}^2}\right],
\eea 
where the flavor-violating couplings $[\Ga^{Z_{12},Z_{23},Z}_{L(R)}]_{ij}$ induced by gauge bosons $Z_{12},Z_{23},Z$ are generally have the below forms 
\bea 
[ \Ga^{Z_{12}}_{L(R)}]_{ij} &=&-g_1Y^{L(R)}_1s_{12}(V^{*}_{q_{L(R)}})_{1i}(V_{q_{L(R)}})_{1j}+g_2Y^{L(R)}_2c_{12}(V^{*}_{q_{L(R)}})_{2i}(V_{q_{L(R)}})_{2j}, \\\relax
[\Ga^{Z_{23}}_{L(R)}]_{ij} &=&-g_{12}Y_1^{L(R)}s_{23}(V^{*}_{q_{L(R)}})_{1i}(V_{q_{L(R)}})_{1j}+g_{12}Y_2^{L(R)}s_{23}(V^{*}_{q_{L(R)}})_{2i}(V_{q_{L(R)}})_{2j}\crn &&+g_3Y_3^{L(R)}c_{23}(V^{*}_{q_{L(R)}})_{3i}(V_{q_{L(R)}})_{3j},\\\relax
[\Ga^{Z}_{L}]_{ij}&=&(T_{3L}g_Lc_W-Y_1^Ls_Wg_Y)(V^{*}_{q_{L}})_{1i}(V_{q_{L}})_{1j}+(T_{3L}g_Lc_W-Y_2^Ls_Wg_Y)(V^{*}_{q_{L}})_{2i}(V_{q_{L}})_{2j}\crn &&+(T_{3L}g_Lc_W-Y_3^Ls_Wg_Y)(V^{*}_{q_{L}})_{3i}(V_{q_{L}})_{3j},\\\relax
 [\Ga^{Z}_{R}]_{ij}&=&-Y_1^Rs_Wg_Y(V^{*}_{q_{R}})_{1i}(V_{q_{R}})_{1j}-Y_2^Rs_Wg_Y(V^{*}_{q_{R}})_{2i}(V_{q_{R}})_{2j}-Y_3^Rs_Wg_Y(V^{*}_{q_{R}})_{3i}(V_{q_{R}})_{3j},
\eea  	
in which $Y^{L,R}_{i}$ are hypercharges for $q_{iL,R}$, i.e $Y^L=\left\{1/6,1/6\right\}$ for $u_{iL},d_{iL}$, $Y^R=\left\{2/3,-1/3\right\}$ for $u_{iR},d_{iR}$. The unitary matrices $V_{q_{L,R}}$ connect the weak and mass eigenstates, which can be numerically obtained by benchmark points that successfully reproduce the fermion spectrum, as shown in Table \ref{tabquark}. It should be noted that with $Y^1_L=Y^2_L=Y^3_L$, the flavor-violating couplings of SM $Z$ boson $[\Ga^{Z}_{L(R)}]_{ij}$ vanish, due to the unitary condition of $V_{qL(R)}$, i.e $[\Ga^{Z}_{L(R)}]_{ij}\sim \sum_{k=1}^3[V^*_{qL(R)}]_{ki}[V_{qL(R)}]_{kj}=0$ with $i
\neq j$. In addition, the terms depend on $1/m^2_{Z_{12}}$ in the above WCs can be skipped due to a very high scale $m_{Z_{12}}\sim v_{12}\sim \mathcal{O}(10^3)$ TeV. This leads the WCs in $\Delta F=2$ processes depend mostly on $m_{Z_{23}}$ and couplings $g_{12},g_3$. 

With the effective Hamiltonian for $\Delta F=2$ processes in Eq. (\ref{Heff}), we can determine the ratios between SM+NP contribution with SM ones in meson mass difference $\Delta m_P$ as follows   
\bea 
\ep_K&=&\fr{\Delta m_{K}^{\text{SM}}+\Delta m_{K}^{\text{NP}}}{\Delta m_{K}^{\text{SM}}}=\fr{\text{Re}[M_{12}^{K,\text{NP}}+M_{12}^{K,\text{SM}}]}{\text{Re}[M_{12}^{K,\text{SM}}]}\crn 
&=&\fr{1}{\text{Re}[C^K_{\text{SM}}]}\text{Re}\left\{(C^{K}_{LL}+C^{K}_{RR})\eta_K^{6/23}-C^{K}_{LR}\left[\fr 3 2+ \left(\fr{m_K}{m_d+m_s}\right)^2\right]\eta_K^{3/23}\fr{B_K^{(5)}}{B_K^{(1)}}\right. \crn && + \left. C^{K}_{LR}\left[\fr 1 6+ \left(\fr{m_K}{m_d+m_s}\right)^2\right]\left(\eta_K^{3/23}-\eta_K^{-24/23}\right)\fr{B^{(4)}_{K}}{B_K^{(1)}}\right\},\label{DeltaM1} \\ 
\ep_{B_q}&=&\fr{\Delta m_{B_q}^{\text{SM+NP}} }{\Delta m_{B_q}^{\text{SM}}}=\left|1+\fr{M_{12}^{q,\text{NP}}}{M_{12}^{q,\text{SM}}}\right|\crn 
&=&\left|1+\fr{C^{q}_{LL}+C^{q}_{RR}}{C^q_{\text{SM}}}\eta_{B_q}^{6/23}-\fr{C^{q}_{LR}}{C^q_{\text{SM}}}\left[\fr 3 2+ \left(\fr{m_{B_q}}{m_q+m_b}\right)^2\right]\eta_{B_q}^{3/23}\fr{B_{B_q}^{(5)}}{B_{B_q}^{(1)}}\right. \crn && + \left. \fr{C^{q}_{LR}}{C^q_{\text{SM}}}\left[\fr 1 6+ \left(\fr{m_{B_q}}{m_q+m_b}\right)^2\right]\left(\eta_{B_q}^{3/23}-\eta_{B_q}^{-24/23}\right)\fr{B_{B_q}^{(4)}}{B_{B_q}^{(1)}}\right|, \label{DeltaM2}
\eea 
where the hadronic matrix elements are expressed in terms of non-perturbative bag parameters $B_P^{(i)}$ as follows
\bea 
\langle P|(\bar{q}_{iL}\ga_{\mu}q_{jL})^2|\bar{P}\rangle&=&\langle P|(\bar{q}_{iR}\ga_{\mu}q_{jR})^2|\bar{P}\rangle=\fr{2}{3}m_P^2f_P^2B_P^{(1)}(\mu),\\ 
\langle P|(\bar{q}_{iL}q_{jR})(\bar{q}_{iR}q_{jL})|\bar{P}\rangle&=&\fr{1}{2}\left[\fr{1}{6}+\fr{m_P^2}{(m_{q_i}+m_{q_j})^2}\right]m_P^2f_P^2B_P^{(4)}(\mu),\\ 
\langle P|(\bar{q}_{iL}\ga_{\mu}q_{jL})(\bar{q}_{iR}\ga_{\mu}q_{jR})|\bar{P}\rangle&=&-2\langle P|(\bar{q}^{\al}_{iL}q^{\beta}_{jR})(\bar{q}^{\beta}_{iR}q_{jL}^{\al})|\bar{P}\rangle\crn
&=&-\fr{1}{3}\left[\fr{3}{2}+\fr{m_P^2}{(m_{q_i}+m_{q_j})^2}\right]m_P^2f_P^2B_P^{(5)}(\mu),
\eea 
with $\al,\beta$ to be color indices. The third line appears due to the Fierz transformation of the LR operator. In addition, the coefficients  $\eta_{P}\equiv\al_s(\mu_{\text{NP}})/\al_s(\mu_P)$ present the QCD corrections at leading order (LO) approximation by using renormalization group evolution (RGE) from NP scale $\mu_{\text{NP}}=m_{Z_{23}}$ to the hadronic scale $\mu_P\sim 3$ GeV for $K$ meson and $\mu_P=4.16$ GeV for $B_{s,d}$ mesons \cite{Bagger:1997gg}. It is important to note that these running effects lead to operator mixing in non-color singlet LR operators. Thus, there exist both bag parameters $B_4$ and $B_5$ in Eqs. (\ref{DeltaM1}) and (\ref{DeltaM2}). In order to estimate the impact of NP to $\Delta m_P$, we adapt the following 2$\sigma$ constraints given in \cite{VanLoi:2024ptt}, 
\bea 
&& \fr{\Delta m_{B_s}^{\text{SM+NP}} }{\Delta m_{B_s}^{\text{SM}}}=\fr{\Delta m_{B_s}^{\text{exp}} }{\Delta m_{B_s}^{\text{SM}}}\in [0.8597,1.0332],\\ 
&& \fr{\Delta m_{B_d}^{\text{SM+NP}} }{\Delta m_{B_d}^{\text{SM}}}=\fr{\Delta m_{B_d}^{\text{exp}} }{\Delta m_{B_d}^{\text{SM}}}\in [0.8336,1.0335], \label{Bq_const} 
\eea 
for $B_s$ and $B_d$ mesons. For the $K$ meson, the uncertainty of the SM prediction $\Delta m_K^{\text{SM}}$ is considerable compared to the $B_{s,d}$ meson systems, due to the difficulty in theoretical approaches for long-distance effects. Therefore, the constraint for $\Delta m_K$ is not strong as $\Delta m_{B_{s,d}}$, and we ignore this constraint in the considering work.

The flavor-violating couplings induced by $Z_{12},Z_{23},Z$ also contribute to several $\Delta S=1$ processes such as $b\to sl^+l^-$ decays, which can be described by the following effective Hamiltonian,
\bea 
\mathcal{H}^{\Delta S=1}_{\text{eff}}&=&-\fr{4G_F(V^*_{ts}V_{tb})}{\sqrt{2}}\sum_{I=9,10}(C_I^{\text{SM}}+C_I^{\text{NP}})\mathcal{O}_I+C_I^{'\text{NP}}\mathcal{O}^{'}_I\eea 
with $C_I^{\text{SM}}$ are SM WCs which are calculated at NNLO, while $C_I^{(')\text{NP}}$ are NP contributions. The operators and their corresponding Wilson coefficients (at the scale $\mu=m_{Z_{23}}$) are 
\bea [\mathcal{O}^{(')}_9]_{ij}&=&\fr{e^2}{16\pi^2}(\bar{s}_{L(R)}\ga_{\mu}b_{L(R)})(\bar{e}_i\ga^{\mu}e_j),\\\relax
[C^{(')\text{NP}}_9]_{ij}&\simeq& \fr{16\pi^2}{e^2}\fr{g\sqrt{2}}{8G_F(V^*_{ts}V_{tb})c_W}\fr{[\Ga^{Z_{23}}_{L}]_{23}[\tilde{g}^{Z_{23}}_{V}]_{ij}}{m_{Z_{23}}^2},\\ \relax
[\mathcal{O}^{(')}_{10}]_{ij}&=&\fr{e^2}{16\pi^2}(\bar{s}_{L(R)}\ga_{\mu}b_{L(R)})(\bar{e}_i\ga^{\mu}\ga_5e_j),\\\relax
[C^{(')\text{NP}}_{10}]_{ij}&\simeq&-\fr{16\pi^2}{e^2}\fr{g\sqrt{2}}{8G_F(V^*_{ts}V_{tb})c_W}\fr{[\Ga^{Z_{23}}_{L}]_{23}[\tilde{g}^{Z_{23}}_{A}]_{ij}}{m_{Z_{23}}^2},
\eea 
where 
\be [\tilde{g}^{Z_{23}}_{V,A}]_{ij}=\fr{1}{2}\sum_{k=1}^3\left[(V^*_{eL})_{ki}\left(g^{Z_{23}}_{V}(e_k)+g^{Z_{23}}_{A}(e_k)\right)(V_{eL})_{kj}\pm (V^*_{eR})_{ki}\left(g^{Z_{23}}_V(e_k)-g^{Z_{23}}_A(e_k)\right)(V_{eR})_{kj}\right] \label{Zcouplings}\ee 
are vector and axial-vector couplings induced by new gauge boson $Z_{23}$ in mass eigenstates, while $g^{Z_{23}}_{V,A}(e_k)$ are vector and axial-vector couplings in flavor states and given explicitly in Tables \ref{Tab1} and \ref{Tab2}.  

We want to emphasize that there are no significant NP contributions to Wilson coefficients of dipole operators $C^{(')}_{7}$. This can be explained because $C^{(')}_{7}$ are induced by one-loop involving FCNCs couplings of $Z_{12},Z_{23}$, which are very suppressed by factor $m_W^2/m^2_{Z_{23}}\ll 1$ for $m_{Z_{23}}\sim \mathcal{O}(10)$ TeV. Therefore, the model provides NP contributions for four Wilson coefficients $C^{(')}_9$ and $C^{(')}_{10}$. The SM+NP contribution normalized to SM one in the branching ratio of $B_s\to \mu^+\mu^-$ is given in 
\bea 
\ep_{B_s\to \mu^+\mu^-}&=&\fr{\text{BR}(B_s\to \mu^+\mu^-)_{\text{SM+NP}}}{\text{BR}(B_s\to \mu^+\mu^-)_{\text{SM}}}=\fr{\text{BR}(B_s\to \mu^+\mu^-)_{\text{exp}}}{\text{BR}(B_s\to \mu^+\mu^-)_{\text{SM}}}\crn &=&\fr{1}{1-y_s}\fr{|C_{10}^{\text{SM}}+C^{\text{NP}}_{10}-C^{'\text{NP}}_{10}|^2}{|C^{SM}_{10}|^2}.
\eea 
In order to estimate the NP impact, we consider the predicted $\ep_{B_s\to \mu^+\mu^-}$ with corresponding $2\sigma$ range 
\bea 
\fr{\text{BR}(B_s\to \mu^+\mu^-)_{\text{exp}}}{\text{BR}(B_s\to \mu^+\mu^-)_{\text{SM}}}\in [0.7574,1.0778], \label{Bsmm_const}
\eea 
where $\text{BR}(B_s\to \mu^+\mu^-)_{\text{exp}}=3.34(27)\times 10^{-9}$ \cite{Czaja:2024the} and SM prediction including power-enhanced QED correction is $\text{BR}(B_s\to \mu^+\mu^-)_{\text{SM}}=3.64(12)\times 10^{-9}$ \cite{ParticleDataGroup:2024cfk}.

For scenarios of lepton flavor violating decaying $B_s\to l_i^+l_j^-$ ($i\neq j$), we adapt the result in \cite{Crivellin:2015era} as follows
\bea
\text{BR}(B_s\to l_i^+l_j^-)&=&\fr{\al^2G_F^2|V^*_{ts}V_{tb}|^2m_{B_s}f_{B_s}^2\tau_{B_s}\text{Max}[m_i^2,m^2_{j}]}{64\pi^3}\left(1-\fr{\text{Max}[m_i^2,m^2_{j}]}{m_{B_s}^2}\right)^2\crn 
&&\times \left(|[C^\text{NP}_9]_{ij}-[C^{'\text{NP}}_9]_{ij}|^2+|[C_{10}^\text{NP}]_{ij}-[C^{'\text{NP}}_{10}]_{ij}|^2\right).
\eea 
Among above processes with different product lepton flavors $(e\mu,e\tau,\mu\tau)$, we concentrate on the ones with decaying lepton flavors $e^+\mu^-$ since they have strongest constraint, namely BR$(B_s\to e^{\pm}\mu^{\mp})_{\text{exp}}<5.4\times 10^{-9}$  \cite{ParticleDataGroup:2024cfk}.

The flavor-violating interactions of $Z_{12}, Z_{23}$ also appear in the lepton sector, which can make the leptonic three-body decays at tree-level, such as $\tau\to 3\mu,3e$ and $\mu\to 3e$. The branching ratio of these processes is shown by 
\bea 
\text{BR}(e_i\to 3e_j)&=&\fr{m_i^5\eta^2_{\text{RGE}}}{1536\pi^3\Ga_{e_i}m^4_{Z_{23}}}\left[2(|[\tilde{g}^{Z_{23}}_{L}]_{ji}[\tilde{g}^{Z_{23}}_{L}]_{jj}|^2+|[\tilde{g}^{Z_{23}}_{R}]_{ji}[\tilde{g}^{Z_{23}}_{R}]_{jj}|^2)\right.\crn 
&&\left.+|[\tilde{g}^{Z_{23}}_{L}]_{ji}[\tilde{g}^{Z_{23}}_{R}]_{jj}|^2+|[\tilde{g}^{Z_{23}}_{R}]_{ji}[\tilde{g}^{Z_{23}}_{L}]_{jj}|^2\right],
\eea 
for $i\neq j$, $i=\mu,\tau$ and $j=e,\mu$. The factor $\eta_{\text{RGE}}$ presents for RGE running from high scale $\mu\sim m_{Z_{23}}\sim \mathcal{O}(10)$ TeV to low scale $\mu\sim 1$ GeV, which numerically is $\eta_{\text{RGE}}\sim 0.9$ \cite{FernandezNavarro:2024hnv}. In addition, the couplings $[\tilde{g}^{Z_{23}}_{L(R)}]_{ij}$ are given in Eq. (\ref{Zcouplings}). 
Besides three-body leptonic decays, there exist loop contributions involving $Z_{23}$ and SM charged lepton $e_k$ to radiative decays $e_i\to e_j\gamma$. Their branching ratios are given in the limit $m_{e_i}\gg m_{e_j}$ and  $m_{e_k}^2/m^2_{Z_{23}}\ll 1$ as follow 
\bea 
\text{BR}(e_i\to e_j\gamma)&= &\fr{m_j^3}{4\pi\Ga_{e_j}}\left(|[C^{Z_{23}}_L]_{ij}|^2+|[C^{Z_{23}}_R]_{ij}|^2\right), \eea 
with coefficients $[C^{Z_{23}}_{L(R)}]_{ij}$ read
\bea 
[C^{Z_{23}}_{L(R)}]_{ij}&\simeq&\fr{e\eta_{\text{RGE}}}{48\pi^2m^2_{Z_{23}}}\sum_{k=e,\mu,\tau}\left(m_{e_j}[\tilde{g}^{Z_{23}}_{R(L)}]_{ik}[\tilde{g}^{Z_{23}}_{R(L)}]_{kj}-3m_k[\tilde{g}^{Z_{23}}_{R(L)}]_{ik}[\tilde{g}^{Z_{23}}_{L(R)}]_{kj}\right.\crn 
&&\left.+m_i[\tilde{g}^{Z_{23}}_{L(R)}]_{ik}[\tilde{g}^{Z_{23}}_{L(R)}]_{kj}\right).
\eea 
For numerical study, we focus on the observables having the strongest experimental constraints, BR$(\mu\to 3e)_{\text{exp}}<1.0\times 10^{-12}$  and BR$(\mu \to e \ga)_{\text{exp}}<4.2\times 10^{-13}$ \cite{ParticleDataGroup:2024cfk}. 
\begin{figure}[H]
	\centering
	\includegraphics[scale=0.42]{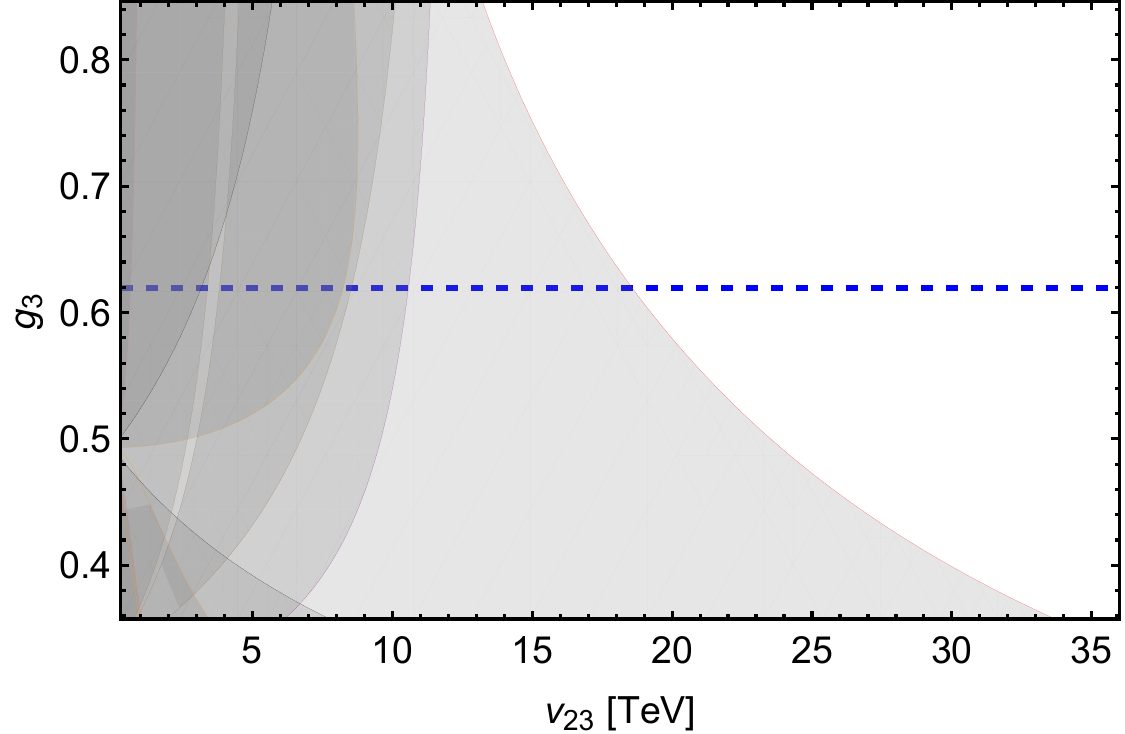}
	\includegraphics[scale=0.415]{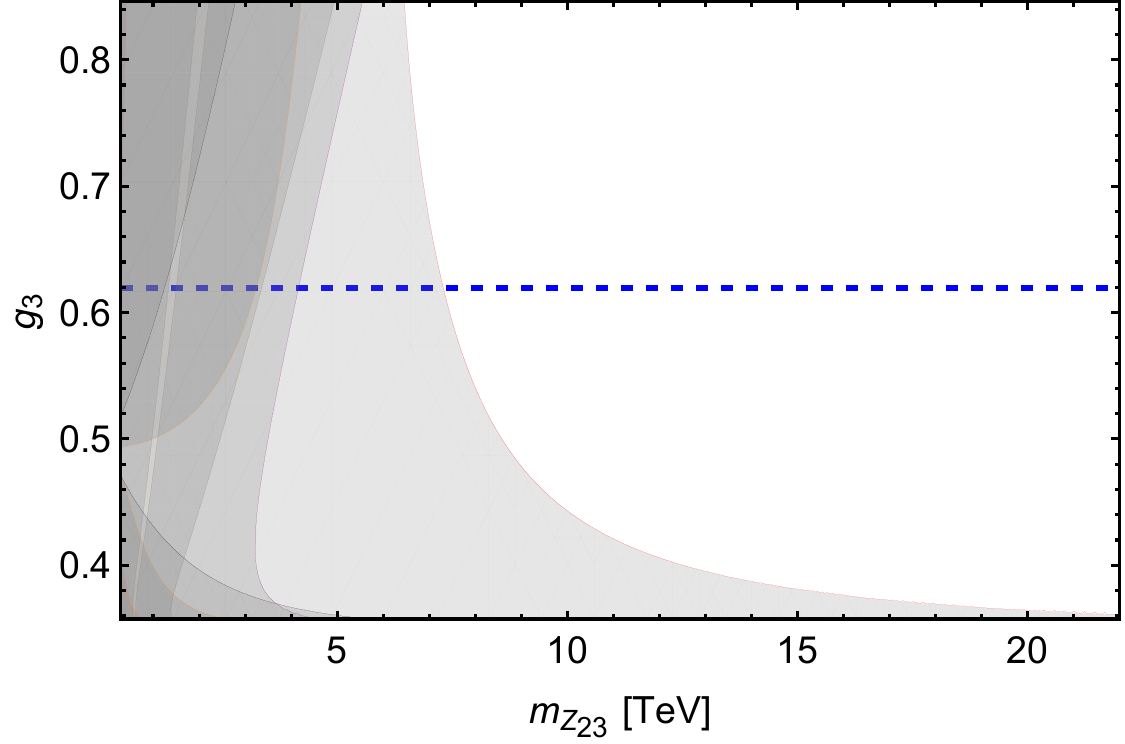}
	\caption[]{\label{plotsv23vsg3flavor}Black, brown, orange, pink, red, and purple curves denote the exclusion bounds derived from $\ep_{B_s}$, $\ep_{B_d}$, $\ep_{B_s\to\mu^+\mu^-}$, BR$(B_s\to e^{\pm}\mu^{\mp})$, BR$(\mu\to 3e)$, and BR$(\mu\to e\ga)$, respectively. The shaded regions indicate the excluded parameter space. The dashed blue line corresponds to the gauge unification scenario $g_1=g_2=g_3=\sqrt3 g_Y$.}
\end{figure}
In Fig. \ref{plotsv23vsg3flavor}, we show the correlation between the coupling $g_3$ of gauge group $U(1)_{Y_3}$ and the VEV $v_{23}$ satisfying constraints of $\ep_{B_s}$ (black), $\ep_{B_d}$ (brown), $\ep_{B_s\to\mu^+\mu^-}$ (orange), BR$(B_s\to e^{\pm}\mu^{\mp})$ (pink), BR$(\mu\to 3e)$ (red), and BR$(\mu\to e\ga)$ (purple). The parameter space, as indicated by the shaded regions, is excluded. We see that the process $\mu\to 3e$ gives the strongest constraint, i.e., $v_{23}\gtrsim 13.2$ TeV and $m_{Z_{23}}\gtrsim 6.5$ TeV. For the gauge unification case $g_1=g_2=g_3=\sqrt3 g_Y$, we have $v_{23}\gtrsim 18.6$ TeV and $m_{Z_{23}}\gtrsim 7.3$ TeV. These lower bounds are larger than the bounds given in sections \ref{EWtest} and \ref{Colliderbounds}. 

\section{\label{dark}Majorana Dark matter}
Since the dark scalars $R_{1,2}$, $I_{1,2}$ and $\eta^\pm$ are extremely heavy, with masses at the scale $v_{12}\sim\mathcal{O}(10^3)$ TeV, while the dark fermions $N_{1,2R}$ reside at a lower scale $\La\sim\mathcal{O}(1)$ TeV, our model predicts a distinctive DM candidate: the right-handed Majorana neutrino, stabilized by the conservation of dark parity. Without loss of generality, we assume that $N_{1R}$ is the lightest among the dark-sector fields responsible for DM. In this section, the mixing angles $\theta_{R, I}$ are neglected due to their strong suppression. Additionally, for simplicity, we ignore mixings between $H_2$ and $H_{3,4}$, as well as between $\nu_{1R}$ and $\nu_{2R}$, effectively taking $\mathsf{H}\simeq H_2$, $N_{1R}\simeq\nu_{1R}$, and $N_{2R}\simeq\nu_{2R}$. 

\subsection{Dark matter relic abundance}
\begin{figure}[h]
\centering
\includegraphics[scale=1.0]{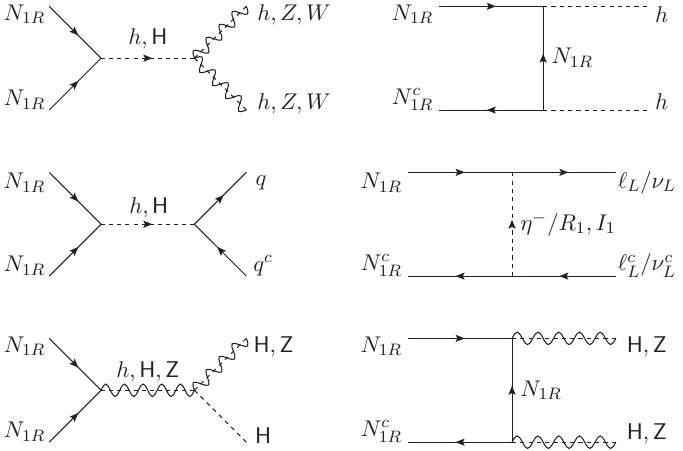}
\caption[]{\label{plotanniDM}Annihilations govern the DM relic density, where $\ell=e,\mu,\tau$ and $\nu=\nu_e,\nu_\mu,\nu_\tau$.}
\end{figure}

As discussed in Section \ref{neu}, the radiative contribution to active neutrino masses is significant and consistent with experimental observations, even for Yukawa coupling of order $\kappa\sim\mathcal{O}(1)$. This implies that the DM candidate $N_{1R}$ is appreciably coupled to the SM particles in the thermal bath of the early Universe.\footnote{The dark scalars are always in thermal equilibrium with the SM plasma through the Higgs portal with the couplings $\la_{32,38,39}$ and $\mu_9$.} Consequently, the freeze-out mechanism is operative and determines both the DM relic abundance and the DM nature as a weakly interacting massive particle. The dominant processes for DM pair annihilation involve final states consisting of the SM particle pairs, and---when kinematically allowed---pairs of the new bosons $\mathsf{H}$ and $\mathsf{Z}$, as shown in Fig. \ref{plotanniDM}. It is important to note that, due to the Majorana nature of $N_{1R}$, each $t$-channel annihilation diagram has a corresponding $u$-channel diagram, although only the former are explicitly shown in the figure. Additionally, the vector-like fermions and all fields with mass at the $v_{12,23}$ scales are significantly heavier than $N_{1R}$, and thus are kinematically inaccessible as final states in DM annihilation processes.

\begin{figure}[h]
\centering
\includegraphics[scale=0.41]{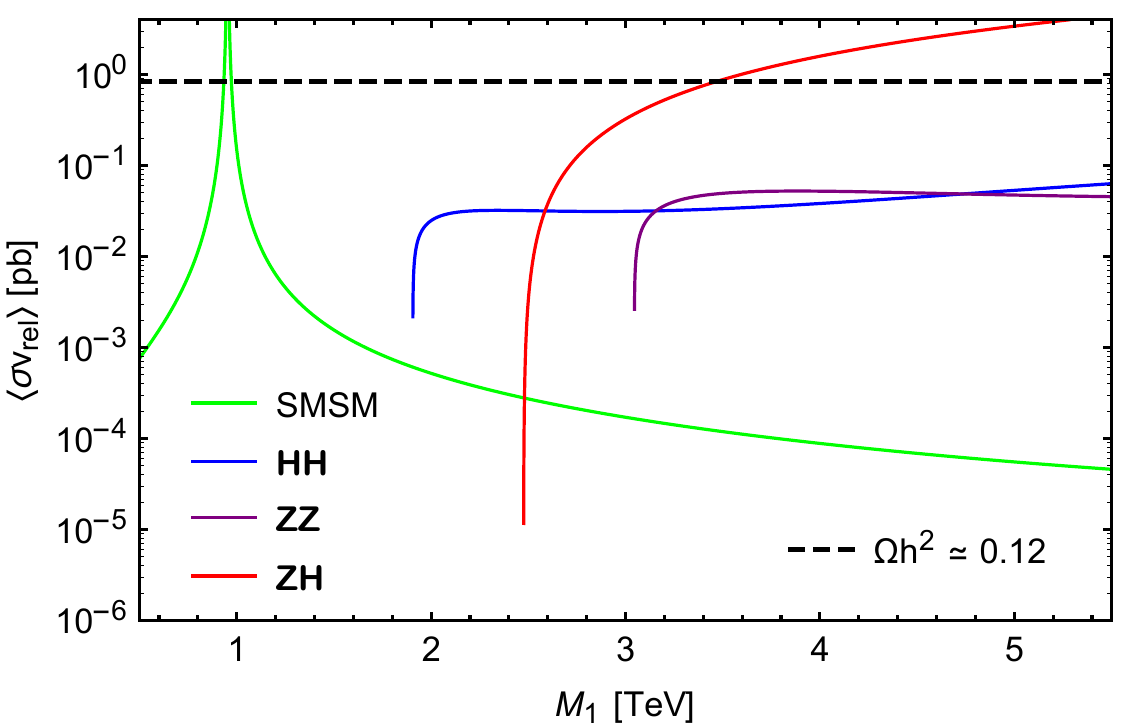}
\includegraphics[scale=0.393]{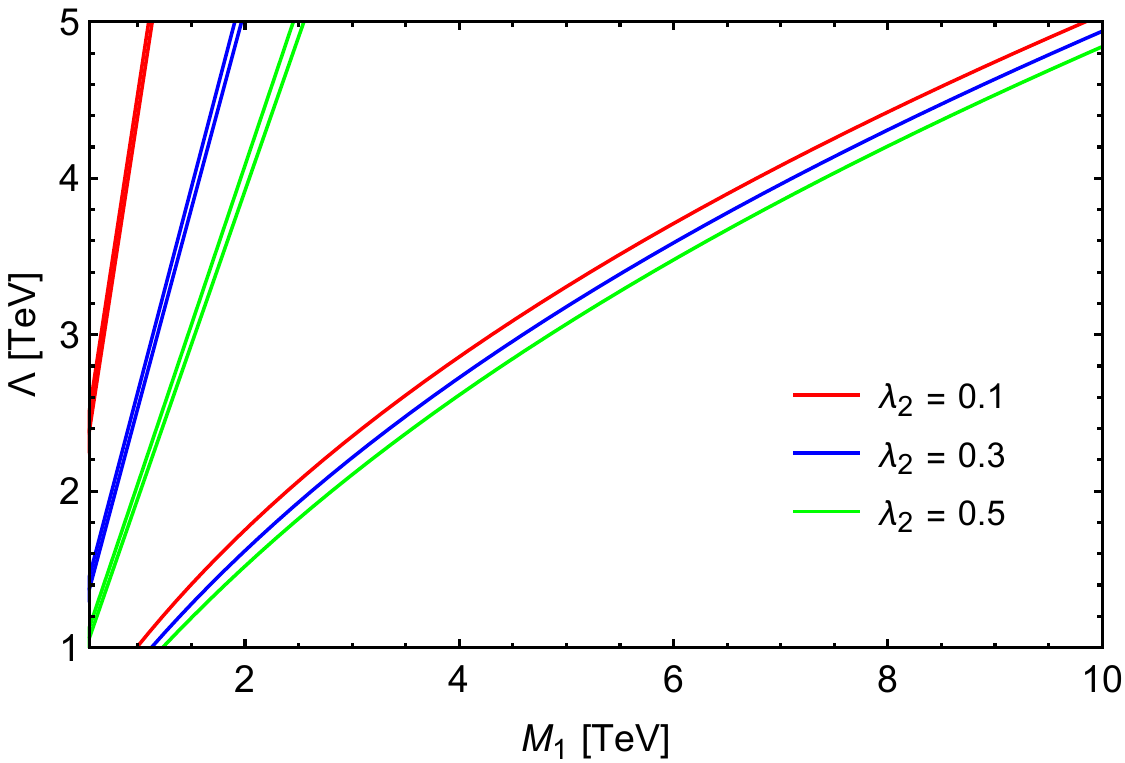}
\caption[]{\label{plotM1cross}Left panel: DM annihilation cross sections into various final states as functions of the DM mass. Right panel: Contours of the observed DM relic density in the plane of DM mass versus the new physics scale, shown for different values of the coupling $\lambda_2$.}
\end{figure}

The thermal average annihilation cross section times relative velocity for the $N_{1R}$ DM can be decomposed into four parts, namely 
\bea \langle\sigma v_\text{rel}\rangle_{N_{1R}}&=&\langle\sigma v_\text{rel}\rangle_{N_{1R}N_{1R}\to\text{SMSM}}+\langle\sigma v_\text{rel}\rangle_{N_{1R}N_{1R}\to\mathsf{HH}}\crn
&&+\langle\sigma v_\text{rel}\rangle_{N_{1R}N_{1R}\to\mathsf{ZH}}+\langle\sigma v_\text{rel}\rangle_{N_{1R}N_{1R}\to\mathsf{ZZ}},\eea
where the first part is related to the DM annihilation to the SM particles, while the remaining parts are the DM annihilation to the new bosons. In the non-relativistic approximation, it is straightforward to determine
\bea \langle\sigma v_\text{rel}\rangle_{N_{1R}}&\simeq&
\frac{\la_3^2M_1^2}{128\pi}\left[\frac{3m_h^2}{m_\mathsf{H}^2(4M_1^2-m_h^2)}-\frac{1}{4M_1^2-m_\mathsf{H}^2}\right]^2\left(1-\frac{m_h^2}{M_1^2}\right)^{\fr 1 2}\crn
&&+\frac{3\epsilon_2^2 M_1^6}{8\pi v^2\La^2}\left(\frac{1}{4M_1^2-m_\mathsf{H}^2}-\frac{1}{4M_1^2-m_h^2}\right)^2\crn
&&+\frac{\epsilon_2^2 m_t^2 M_1^4}{8\pi v^2\La^2}\left(\frac{1}{4M_1^2-m_\mathsf{H}^2}-\frac{1}{4M_1^2-m_h^2}\right)^2\left(1-\frac{m_t^2}{M_1^2}\right)^{\fr 3 2}\crn
&&+\frac{\kappa^4M_1^2}{32\pi}\left[\frac{1}{(M_1^2+m_{\eta^0}^2)^2}+\frac{1}{(M_1^2+m_{\eta^\pm}^2)^2}\right]\crn
&&+\frac{M_1^2}{128\pi \La^4}\left[\frac{9m_\mathsf{H}^4}{(4M_1^2-m_\mathsf{H}^2)^2}+\frac{32M_1^4(M_1^2-m_\mathsf{H}^2)^2}{(2M_1^2-m_\mathsf{H}^2)^4x_F}-\frac{12M_1^2m_\mathsf{H}^2(M_1^2-m_\mathsf{H}^2)}{(4M_1^2-m_\mathsf{H}^2)(2M_1^2-m_\mathsf{H}^2)^2x_F}\right]\crn
&&\times\left(1-\frac{m_\mathsf{H}^2}{M_1^2}\right)^{\fr 1 2}\Theta(M_1-m_\mathsf{H})\crn
&&+\frac{g^4_\mathsf{D}}{64\pi M_1^4 m_\mathsf{Z}^4}[m^4_\mathsf{H}-2m^2_\mathsf{H}(4M_1^2+m^2_\mathsf{Z})+(4M_1^2-m^2_\mathsf{Z})^2]^{\fr 3 2}\Theta\left(M_1-\frac{m_\mathsf{Z}+m_\mathsf{H}}{2}\right)\crn
&&+\frac{g_\mathsf{D}^4}{16\pi M_1^2}\left[1+\frac{2M_1^4(8M_1^4+m^4_\mathsf{H})}{m^4_\mathsf{Z}(4M_1^2-m^2_\mathsf{H})^2x_F}\right]\left(1-\frac{m_\mathsf{Z}^2}{M_1^2}\right)^{\fr 1 2}\Theta(M_1-m_\mathsf{Z}),\eea
for which only the dominant channels are presented. Here, $\Theta(\cdots)$ is the Heaviside step function, $x_F=M_1/T_F\simeq 25$ is given at freeze-out temperature, and $m_{\eta^0}\simeq m_{R_1,I_1}$. As shown in the left panel of Fig. \ref{plotM1cross}, the new Higgs mass resonance $M_1=m_\mathsf{H}/2$ and, when kinematically accessible, the annihilation channel $N_{1R}N_{1R}\to\mathsf{ZH}$ are crucial to set the correct DM relic density $\Om_\text{DM}h^2\simeq 0.12$ \cite{Planck:2018vyg}, where we have taken $m_{\eta^0}=m_{\eta^\pm}=100\text{ TeV}$, $\kappa=1$, $g_\mathsf{D}=\sqrt3 g_Y\simeq 0.619$, $\la_3=0.1$, $\la_2=0.3$, $\La=2.46$ TeV, and $\epsilon_2=0.01$, in addition $v=246$ GeV, $m_t=173$ GeV, and $m_h=125$ GeV. In the right panel, we present the contours corresponding to the observed DM relic density in the ($M_1,\La$) plane for various values of $\la_2$, while keeping all other parameters fixed as in the previous analysis. For each value of $\la_2$, three distinct curves emerge. Among them, two nearly parallel curves arise due to the mass resonance associated with the new Higgs boson. In contrast, the remaining curve is predominantly governed by the $\mathsf{Z}\mathsf{H}$ annihilation channel. From these results, we conclude that the dark fermion $N_{1R}$, with a mass in the TeV range, can successfully account for the observed DM relic abundance.

\subsection{Dark matter scattering off nuclei}
\begin{figure}[h]
\centering
\includegraphics[scale=1]{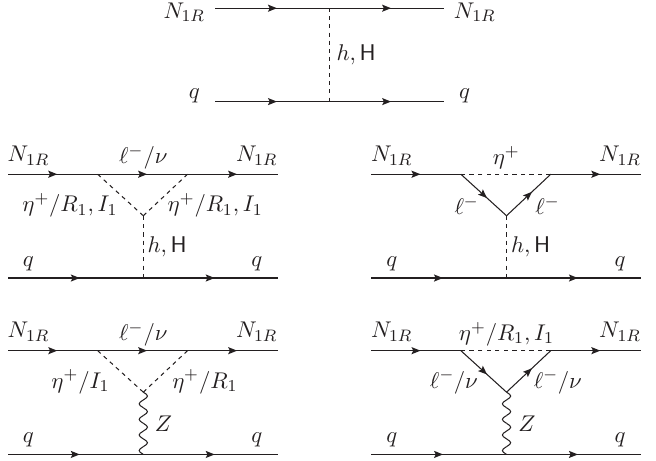}
\caption[]{\label{plotscatDM}Diagrams for DM-nucleon scattering, where $\ell=e,\mu,\tau$ and $\nu=\nu_e,\nu_\mu,\nu_\tau$.}
\end{figure}

We now turn our attention to the direct detection prospects of the dark fermion $N_{1R}$. Although $N_{1R}$ does not couple directly to the SM quarks, its scattering with nucleons can occur via Higgs mediation. In particular, due to the mixing between the SM Higgs doublet $H$ and the new scalar singlet $\Phi$, there exist tree-level elastic scattering processes mediated by the Higgs bosons $h$ and $\mathsf{H}$ through $t$-channel exchange, as illustrated in the first diagram of Fig. \ref{plotscatDM}. In addition to the tree-level contribution, one-loop processes also contribute to DM–nucleon scattering. These arise from effective couplings of $N_{1R}$ to the $h$ and $\mathsf{H}$ scalar bosons and to the SM $Z$ boson, as depicted in the remaining diagrams of the figure.  

Let us consider the first scenario, where the mixing parameter $\epsilon_2$ is quite large, i.e., $\epsilon_2\simeq 0.01$, as mentioned above. In such a case, the dominant contribution to the scattering process $N_{1R}$-nucleons results from the first diagram of Fig. \ref{plotscatDM}. The effective Lagrangian describing this scattering process is given by
\be\mathcal{L}_{\mathrm{eff}}= C_q(\bar{N}_R^cN_R)(\bar{q}q),\ee 
in which the effective coupling is
\be C_q = -\frac{\epsilon_2m_qM_1}{\La v}\left(\frac{1}{m^2_h}-\frac{1}{m^2_\mathsf{H}}\right)\simeq-\frac{\epsilon_2m_qM_1}{\La vm^2_h},  \ee
where $m_q$ denotes the $q$ quark mass. Additionally, the contribution of $\mathsf{H}$ is negligible in comparison to that of $h$, and is thus omitted. Then, the spin-independent (SI) elastic scattering cross-section of $N_{1R}$ per nucleon $N$ can be written as
\be \sigma^\mathrm{SI}_{N_{1R}}=\frac{4}{\pi}\left(\frac{M_1m_N}{M_1+m_N}\right)^2\left[\frac{\text{Z}f_p+(\text{A}-\text{Z})f_n}{\text{A}}\right]^2, \ee 
where A and Z respectively are the mass and atomic number of the target nucleus, $m_N$ is the average nucleon mass. In addition, the interaction strengths of proton $f_p$ and neutron $f_n$ with the dark fermion $N_{1R}$ is given by
\be f_{p,n}=m_{p,n}\left(\sum_{q=u,d,s}f_{Tq}^{p,n}\frac{C_q}{m_q}+\fr{2}{27}f_{TG}^{p,n}\sum_{q=c,b,t} \frac{C_q}{m_q}\right) \ee  
with $f_{TG}^{p,n}=1-\sum_{q=u,d,s}f_{Tq}^{p,n}$. Here, $m_{p,n}$ are the proton and neutron masses, respectively. Additionally, the parameters $f^{p,n}_{Tq}$ are evaluated as $f^p_{Tu}=0.020\pm 0.004$, $f^p_{Td}=0.026\pm 0.005$, $f^p_{Ts}=0.118\pm 0.062$, $f^n_{Tu}=0.014\pm 0.003$, $f^n_{Td}=0.036\pm 0.008$, and $f^n_{Ts}=0.118\pm 0.062$ \cite{Ellis:2000ds}.

\begin{figure}[h]
\centering
\includegraphics[scale=0.5]{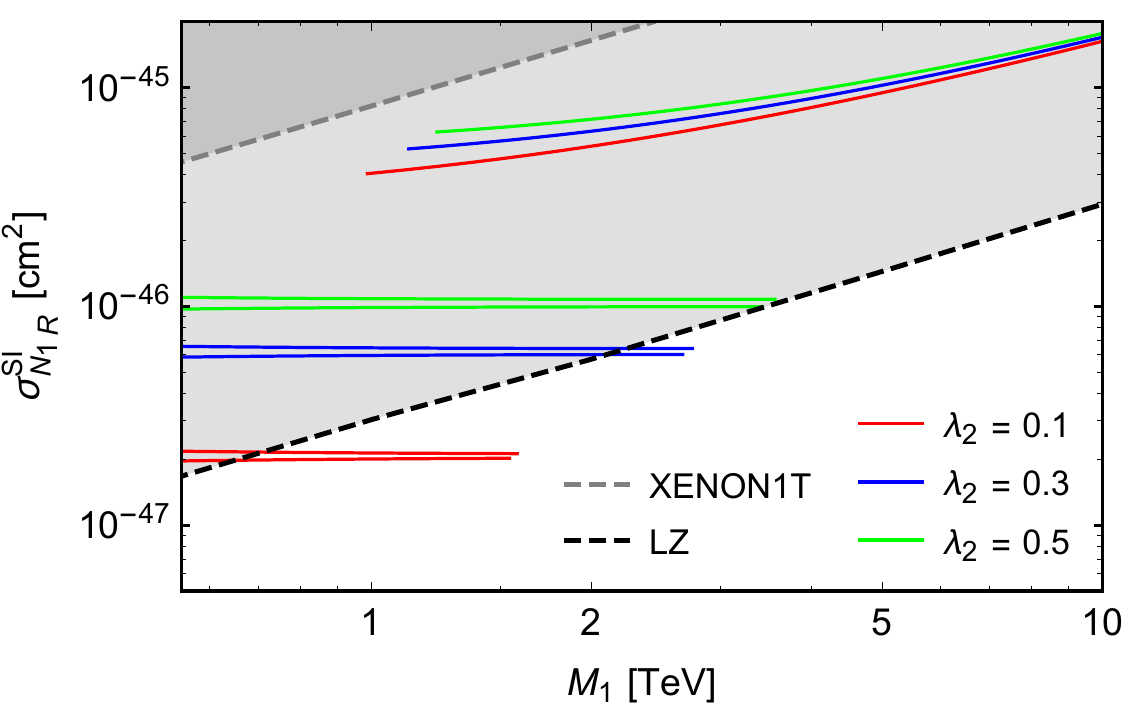}
\caption[]{\label{plotdetecDM}SI scattering cross section of the DM candidate $N_{1R}$ with nucleons as a function of the DM mass. Solid curves represent theoretical predictions. The dashed gray (black) curve shows current experimental bounds from XENON1T \cite{XENON:2018voc} (LZ \cite{LZ:2024zvo}). Shaded regions are excluded by these experiments.}
\end{figure}

Taking $\text{A}=131$, $\text{Z}=54$, $m_p\simeq m_n\simeq m_N=1$ GeV, $m_u=2.16$ MeV, $m_d=4.7$ MeV, $m_s=93.5$ MeV, $m_c=1.273$ GeV, $m_b=4.183$ GeV, together with the previously adopted parameter values that yield the correct DM relic density, we present in Fig. \ref{plotdetecDM} the contours in the ($M_1,\sigma^\text{SI}_{N_{1R}}$) plane. For comparison, we also include the most stringent current bounds on the SI scattering cross section from the XENON1T and LZ experiments \cite{XENON:2018voc,LZ:2024zvo}. As shown, all curves satisfying the relic density constraint are consistent with the XENON1T limit. However, a large portion of these curves is excluded by the LZ limit, depending on the value of $\la_2$. For instance, the LZ bound requires $2.1\text{ TeV} \lesssim M_1\lesssim 2.8$ TeV when $\la_2=0.3$. Furthermore, the projected sensitivity of upcoming direct detection experiments such as XENONnT \cite{XENON:2020kmp} and LZ \cite{LZ:2018qzl} is expected to probe deeper into the parameter space, typically favoring $\epsilon_2<0.01$.

Alternatively, we consider the scenario in which the mixing parameter $\epsilon_2$ is extremely small, i.e., $\epsilon_2\lesssim 10^{-5}$, while the quartic couplings $\la_{32,39}$ and the Yukawa coupling $\kappa$ are sizable, specifically $\la_{32,39}\simeq 2\pi$ and $\kappa\simeq 2\sqrt\pi$. In this case, the previous results for the correct DM relic density remain essentially unchanged. However, the dominant contribution to the $N_{1R}$-nucleons scattering process now arises from the one-loop diagrams shown in the second row of Fig. \ref{plotscatDM}. The corresponding effective coupling is given by

\be C_q \simeq -\frac{\kappa^2 m_q}{16\pi^2m_h^2M_1}\left[\la_{32}\mathcal{G}_1\left(\fr{M_1^2}{m_{\eta^\pm}^2}\right)+\fr{\la_{32}+\la_{39}}{2}\mathcal{G}_2\left(\fr{M_1^2}{m_{\eta^0}^2}\right)\right], \ee
where the loop function is defined as $\mathcal{G}_1(x)=[x+(1-x)\ln(1-x)]/x$. Taking the parameter values as above, we estimate the SI scattering cross-section of $N_{1R}$ per nucleon $N$ as
\be \sigma^\mathrm{SI}_{N_{1R}}\lesssim \left(\frac{M_1}{\text{TeV}}\right)^2\times 6.3\times 10^{-52}\text{ cm}^2. \ee
This result is in agreement with the current bounds provided by the XENON1T and LZ experiments \cite{XENON:2018voc,LZ:2024zvo}, even the projected bounds from upcoming direct detection experiments such as XENONnT \cite{XENON:2020kmp} and LZ \cite{LZ:2018qzl}.

Last but not least, we consider the possibility that the $N_{1R}$-nucleons scattering process is predominantly mediated by the exchange of the SM $Z$-boson, as described by the one-loop diagrams in the third row of Fig. \ref{plotscatDM}. This results in an effective axial-vector interaction of the form $A_q\bar{N}_{1R}\gamma^\mu\gamma^5 N_{1R}\bar{q}\gamma_\mu\gamma^5q$, in which 
\be A_q = \frac{\kappa^2g_A^q}{32\pi^2m_Z^2}\left[(g_V^\ell+g_A^\ell)\mathcal{G}_2\left(\frac{M_1^2}{m^2_{\eta^\pm}}\right)+(g_V^\nu+g_A^\nu)\mathcal{G}_2\left(\frac{M_1^2}{m^2_{\eta^0}}\right)\right] \ee
with the loop function $\mathcal{G}_2(x)=2[x+(1-x)\ln(1-x)]/x^2-1$ and the couplings $g_A^q=\fr 1 2 \left(-\fr 1 2 \right)$ for $q=u,c,t~(d,s,b)$, $g_V^\ell=-\frac{g}{2c_W}\left(\fr 1 2-2s_W^2\right)$, $g_A^\ell=-\frac{g}{4c_W}$, $g_V^\nu=g_A^\nu =\frac{g}{4c_W}$. Hence, the spin-dependent (SD) scattering cross section of $N_{1R}$ in the case of the proton target is given by
\be \sigma^\mathrm{SD}_{N_{1R}}=\frac{16}{\pi}\left(\frac{M_1m_p}{M_1+m_p}\right)^2J_p(J_p+1)A_p^2 \ee 
with $A_p=\sum_{q=u,d,s}\la_q^pA_q$. Here, the fractional quark-spin coefficients are $\la^p_u=0.85$, $\la^p_d=-0.42$, and $\la^p_s=-0.08$, while the angular momentum of the target nucleus in this case is $J_p=1/2$ \cite{Cheng:2012qr}. Taking $s_W^2=0.231$, $g=0.652$, $m_Z=91.188$ GeV, and the previous parameter values, we obtain
\be \sigma^\mathrm{SD}_{N_{1R}}\lesssim \left(\frac{M_1}{\text{TeV}}\right)^4\times 5.08\times 10^{-49}\text{ cm}^2, \ee
given that $\kappa\lesssim 2\sqrt\pi$. The predicted result by our model lies well below the sensitivity of present experiments \cite{LZ:2024zvo,XENON:2019rxp,PandaX-II:2018woa} and projected experiments \cite{XENON:2020kmp,LZ:2018qzl}.

\section{\label{conclusion}Conclusion}
The hierarchical structure of fermion masses suggests that the SM fermion generations may not be universal at high energy scales. This motivates the hypothesis that SM fermions could carry family-dependent hypercharges, analogous to individual lepton numbers. In contrast, right-handed neutrinos, being singlets under the SM gauge group, may instead be charged under a distinct dark gauge symmetry, decoupled from the SM fermion sector.

To explore this possibility, we have constructed a minimal and renormalizable extension of the SM—namely, the fully flipped inert doublet model—supplemented by scalar singlets, vector-like charged fermions, and right-handed Majorana neutrinos. These new fields are essential for generating the observed mass hierarchies of SM charged fermions through higher-dimensional operators, while the small active neutrino masses are realized via a hybrid mechanism combining tree-level Type-I seesaw and one-loop scotogenic contributions.

A residual dark parity, originating from the gauge structure, guarantees the stability of the lightest parity-odd particle, thus providing a viable DM candidate in the form of a dark Majorana neutrino. This symmetry also plays a central role in the radiative generation of neutrino masses. Furthermore, the decomposition of the SM hypercharge into three generation-specific hypercharges offers a natural explanation for the existence of exactly three fermion families.

In summary, the proposed framework presents a unified, minimal, and renormalizable solution to three fundamental puzzles in the SM: the origin of neutrino masses, the nature of DM, and the fermion flavor structure. The model is shown to be consistent with current experimental constraints from electroweak precision measurements, collider searches, flavor-changing processes, and DM observations, thereby offering a promising direction for physics beyond the SM.

\section*{Acknowledgement}

This research is funded by the Vietnam National Foundation for Science and Technology Development (NAFOSTED) under Grant No. 103.01-2023.50 (D.V.L., V.Q.T., and N.T.D.). A.E.C.H. is supported by ANID-Chile FONDECYT 1210378, ANID-Chile FONDECYT 1241855, ANID PIA/APOYO AFB230003 and Proyecto Milenio-ANID: ICN2019\_044.

\bibliographystyle{JHEP}

\appendix

\section{\label{decom}General hypercharge decomposition}
Because the anomalies associated with the hypercharge are canceled within each fermion generation, as in the SM, there is no fundamental reason that each generation hypercharge must correspond uniquely to a single fermion generation. In other words, multiple fermion generations may, in principle, share the same generation hypercharge. Consequently, the number of fermion generations remains theoretically unconstrained. For completeness, we consider a generalized decomposition of the SM hypercharge as $Y=\sum_{n=1}^\mathcal{N}Y_n$, with $Y_n=YC_n$, and investigate a particular form of the coefficient $C_n$ that can provide a theoretical insight into this generational puzzle, i.e., 
\be C_n = \frac{2^{(2+n-n^2)/2}}{(-1)^n}\frac{(4-n)_{\mathcal{N}-3}}{(n)_{1-2n+\mathcal{N}}}\frac{(1-a)_N}{(a-n)(N-n)!},\ee
where $a$ is a generation index, $a=1,2,\cdots,N$, for $N$ and $\mathcal{N}$ to be arbitrary integer. Additionally, we have used the Pochhammer function $(x)_y=x(x+1)\ldots (x+y-1)$. It is worthwhile that the above form of $C_n$ induces $C_n=0$ if $n\geq 4$, reducing to \be Y=Y_1+Y_2+Y_3.\ee
Additionally, $C_1=C_2=C_3=0$ if $a\geq 4$ gives a suitable argument for the existence of only three fermion generations as observed. Furthermore, for $a\leq 3$, we obtain
\bea C_1 &=&\fr 1 2 (a-2)(a-3),\crn
C_2 &=&-(a-1)(a-3),\crn
C_3 &=&\fr 1 2 (a-1)(a-2), \eea
implying $C_1=1$ and $C_2=C_3=0$ if $a=1$, $C_2=1$ and $C_1=C_3=0$ if $a=2$, $C_3=1$ and $C_1=C_2=0$ if $a=3$. These mean that each of the three fermion generations is charged under only a separate hypercharge gauge symmetry.

\section{\label{AppB}Vector and axial-vector couplings}
The interactions between the gauge bosons and the SM fermions originate from the fermion kinetic term, $\sum_F \bar{F}i\gamma^\mu D_\mu F$, where $F$ runs over all SM fermion multiplets. It is straightforward to verify that the gluon, photon, and $W$-boson interactions with the SM fermions remain identical to those in the SM. In addition, the interaction of the neutral gauge bosons $Z_I=\mathcal{Z},Z_{23},Z_{12}$ with the SM fermions takes the form
\be \mathcal{L}\supset -\frac{g}{2c_W}\bar{f}\gamma^\mu[g_V^{Z_I}(f)-g_A^{Z_I}(f)\gamma_5]fZ_{I\mu}, \ee
where $f$ denotes the SM fermions in the interaction basis. The vector ($g_V$) and axial-vector ($g_A$) couplings of the SM fermions to the $\mathcal{Z}$ boson are summarized in Table \ref{Tab1}. As expected, these couplings reduce to those of the SM $Z$ boson in the limit $\varepsilon_{1,2}\to 0$. Table \ref{Tab2} lists the corresponding couplings for the gauge bosons $Z_{23}$ and $Z_{12}$ in the simplifying limit $\varepsilon_{1,2},\zeta\to 0$. Notably, $Z_{12}$ does not couple to the third fermion generation and exhibits flavor non-universal couplings for the first and second generations. In contrast, $Z_{23}$ features flavor-universal couplings for the first two generations, while the couplings to the third generation are distinct.

\begin{table}[h]
\bc
\begin{tabular}{l|cccccccc}
\hline\hline
$f$ & $g_V^\mathcal{Z}(f)$ & $g_A^\mathcal{Z}(f)$\\ \hline
$\nu_1$ & $\fr 1 2 - \frac{g_1c_W(c_{12}s_{23}\varepsilon_1+s_{12}\varepsilon_2)}{2g}$ & $\fr 1 2 - \frac{g_1c_W(c_{12}s_{23}\varepsilon_1+s_{12}\varepsilon_2)}{2g}$\\
$\nu_2$ & $\fr 1 2 - \frac{g_2c_W(s_{12}s_{23}\varepsilon_1-c_{12}\varepsilon_2)}{2g}$ & $\fr 1 2 - \frac{g_2c_W(s_{12}s_{23}\varepsilon_1-c_{12}\varepsilon_2)}{2g}$\\
$\nu_3$ & $\fr 1 2 + \frac{g_3c_Wc_{23}\varepsilon_1}{2g}$ & $\fr 1 2 + \frac{g_3c_Wc_{23}\varepsilon_1}{2g}$\\
$e_1$ & $\fr{1-2c_{2W}}{2} - \frac{3g_1c_W(c_{12}s_{23}\varepsilon_1+s_{12}\varepsilon_2)}{2g}$ & $-\fr 1 2 + \frac{g_1c_W(c_{12}s_{23}\varepsilon_1+s_{12}\varepsilon_2)}{2g}$\\
$e_2$ & $\fr{1-2c_{2W}}{2} - \frac{3g_2c_W(s_{12}s_{23}\varepsilon_1-c_{12}\varepsilon_2)}{2g}$ & $-\fr 1 2 + \frac{g_2c_W(s_{12}s_{23}\varepsilon_1-c_{12}\varepsilon_2)}{2g}$\\
$e_3$ & $\fr{1-2c_{2W}}{2} + \frac{3g_3c_Wc_{23}\varepsilon_1}{2g}$ & $-\fr 1 2 - \frac{g_3c_Wc_{23}\varepsilon_1}{2g}$\\
$u_1$ & $\fr{4c_{2W}-1}{6} + \frac{5g_1c_W(c_{12}s_{23}\varepsilon_1+s_{12}\varepsilon_2)}{6g}$ & $\fr 1 2 - \frac{g_1c_W(c_{12}s_{23}\varepsilon_1+s_{12}\varepsilon_2)}{2g}$\\
$u_2$ & $\fr{4c_{2W}-1}{6} + \frac{5g_2c_W(s_{12}s_{23}\varepsilon_1-c_{12}\varepsilon_2)}{6g}$ & $\fr 1 2 - \frac{g_2c_W(s_{12}s_{23}\varepsilon_1-c_{12}\varepsilon_2)}{2g}$\\
$u_3$ & $\fr{4c_{2W}-1}{6} - \frac{5g_3c_Wc_{23}\varepsilon_1}{6g}$ & $\fr 1 2 + \frac{g_3c_Wc_{23}\varepsilon_1}{2g}$\\
$d_1$ & $-\fr{1+2c_{2W}}{6} - \frac{g_1c_W(c_{12}s_{23}\varepsilon_1+s_{12}\varepsilon_2)}{6g}$ & $-\fr 1 2 + \frac{g_1c_W(c_{12}s_{23}\varepsilon_1+s_{12}\varepsilon_2)}{2g}$\\
$d_2$ & $-\fr{1+2c_{2W}}{6} - \frac{g_2c_W(s_{12}s_{23}\varepsilon_1-c_{12}\varepsilon_2)}{6g}$ & $-\fr 1 2 + \frac{g_2c_W(s_{12}s_{23}\varepsilon_1-c_{12}\varepsilon_2)}{2g}$\\
$d_3$ & $-\fr{1+2c_{2W}}{6} + \frac{g_3c_Wc_{23}\varepsilon_1}{6g}$ & $-\fr 1 2 - \frac{g_3c_Wc_{23}\varepsilon_1}{2g}$\\
\hline\hline
\end{tabular}
\caption[]{\label{Tab1}The couplings of the $\mathcal{Z}$ gauge boson with the SM fermions.}
\ec
\end{table}

\begin{table}[h]
\bc
\begin{tabular}{l|cccccccc}
\hline\hline
$f$ & $g_V^{Z_{23}}(f)$ & $g_A^{Z_{23}}(f)$ & $g_V^{Z_{12}}(f)$ & $g_A^{Z_{12}}(f)$ \\ \hline
$\nu_1$ & $\frac{g_{12}c_Ws_{23}}{2g}$ & $\frac{g_{12}c_Ws_{23}}{2g}$ & $\frac{g_1c_Ws_{12}}{2g}$ & $\frac{g_1c_Ws_{12}}{2g}$\\
$\nu_2$ & $\frac{g_{12}c_Ws_{23}}{2g}$ & $\frac{g_{12}c_Ws_{23}}{2g}$ & $-\frac{g_2c_Wc_{12}}{2g}$ & $-\frac{g_2c_Wc_{12}}{2g}$\\
$\nu_3$ & $-\frac{g_3c_Wc_{23}}{2g}$ & $-\frac{g_3c_Wc_{23}}{2g}$ & $0$ & $0$\\
$e_1$ & $\frac{3g_{12}c_Ws_{23}}{2g}$ & $-\frac{g_{12}c_Ws_{23}}{2g}$ & $\frac{3g_1c_Ws_{12}}{2g}$ & $-\frac{g_1c_Ws_{12}}{2g}$\\
$e_2$ & $\frac{3g_{12}c_Ws_{23}}{2g}$ & $-\frac{g_{12}c_Ws_{23}}{2g}$ & $-\frac{3g_2c_Wc_{12}}{2g}$ & $\frac{g_2c_Wc_{12}}{2g}$\\
$e_3$ & $-\frac{3g_3c_Wc_{23}}{2g}$ & $\frac{g_3c_Wc_{23}}{2g}$ & $0$ & $0$\\
$u_1$ & $-\frac{5g_{12}c_Ws_{23}}{6g}$ & $\frac{g_{12}c_Ws_{23}}{2g}$ & $-\frac{5g_1c_Ws_{12}}{6g}$ & $\frac{g_1c_Ws_{12}}{2g}$\\
$u_2$ & $-\frac{5g_{12}c_Ws_{23}}{6g}$ & $\frac{g_{12}c_Ws_{23}}{2g}$ & $\frac{5g_2c_Wc_{12}}{6g}$ & $-\frac{g_2c_Wc_{12}}{2g}$\\
$u_3$ & $\frac{5g_3c_Wc_{23}}{6g}$ & $-\frac{g_3c_Wc_{23}}{2g}$ & $0$ & $0$\\
$d_1$ & $\frac{g_{12}c_Ws_{23}}{6g}$ & $-\frac{g_{12}c_Ws_{23}}{2g}$ & $\frac{g_1c_Ws_{12}}{6g}$ & $-\frac{g_1c_Ws_{12}}{2g}$\\
$d_2$ & $\frac{g_{12}c_Ws_{23}}{6g}$ & $-\frac{g_{12}c_Ws_{23}}{2g}$ & $-\frac{g_2c_Wc_{12}}{6g}$ & $\frac{g_2c_Wc_{12}}{2g}$\\
$d_3$ & $-\frac{g_3c_Wc_{23}}{6g}$ & $\frac{g_3c_Wc_{23}}{2g}$ & $0$ & $0$\\
\hline\hline
\end{tabular}
\caption[]{\label{Tab2}The couplings of the $Z_{23,12}$ gauge bosons with the SM fermions in the limit $\varepsilon_{1,2},\zeta\to 0$.}
\ec
\end{table}

\section{\label{AppC}One-loop Landau poles for the Abelian factors $U(1)_{Y_{1}}\otimes U(1)_{Y_{2}}\otimes U(1)_{Y_{3}}\otimes U(1)_\mathsf{D}$}
In this appendix we compute in detail the one-loop Landau poles for the four Abelian gauge groups $U(1)_{Y_{1}}\otimes U(1)_{Y_{2}}\otimes U(1)_{Y_{3}}\otimes U(1)_\mathsf{D}$ of the model under consideration. Let a $U(1)$ factor with coupling $g$ and
generator $X_{\mu }$ enter the covariant derivative as $D_{\mu }=\partial _{\mu }+i\,g\,q\,X_{\mu }$. At one loop, the $\beta $-function reads 
\begin{equation}
\mu \,\frac{\mathrm{d}g}{\mathrm{d}\mu }=\frac{b}{16\pi ^{2}}\,g^{3},\qquad
b=\frac{2}{3}\sum_{\text{Weyl }f}n_{\text{dof}}(f)\,q_{f}^{2}\;+\;\frac{1}{3}%
\sum_{\text{complex }s}n_{\text{dof}}(s)q\,_{s}^{2}.  \label{eq:betaU1}
\end{equation}%
The multiplicity $n_{\text{dof}}$ counts independent internal components
(e.g.\ color and weak isospin). For instance, an $SU(2)$ doublet
contributes $n_{\text{dof}}=2$; an $SU(3)$ color triplet
contributes $n_{\text{dof}}=3$; a field carrying both has $n_{\text{dof}%
}=2\times 3=6$. We adopt Weyl fermions and complex scalars as fundamental
degrees of freedom; a Dirac fermion counts as two Weyl fermions with
identical gauge quantum numbers.

In terms of $\alpha \equiv g^{2}/(4\pi )$ the one-loop solution is 
\begin{equation}
\frac{1}{\alpha (\mu )}=\frac{1}{\alpha (\mu _{0})}-\frac{b}{2\pi }\ln \!%
\frac{\mu }{\mu _{0}}\,,\qquad \mu _{\mathrm{LP}}=\mu
_{0}\,\exp \!\left[ \frac{2\pi }{b\,\alpha (\mu _{0})}\right] .
\label{eq:LPgeneric}
\end{equation}%
When $b>0$ the coupling grows in the UV and hits a Landau pole at finite $%
\mu _{\mathrm{LP}}$.
\paragraph{Cross-check.}
For one SM generation, summing Weyl fermions and using $n_{\text{dof}}$
explained above: 
\be
\sum_{\text{Weyl}}n_{\text{dof}}\,Y^{2}=6\left(\tfrac{1}{6}%
\right) ^{2}\;+\;3\left( \tfrac{2}{3}\right) ^{2}\;+\;3\left( -\tfrac{1}{3}\right) ^{2}\;+\;2\left( -\tfrac{1}{2}%
\right) ^{2}\;+\;1\left( -1\right) ^{2}\;=\;\frac{10}{3}.
\ee
Three generations give $10$, while one complex Higgs doublet gives $%
\sum_{s}n_{\text{dof}}Y^{2}=2(1/2)^{2}=1/2$. Then Eq.~%
\eqref{eq:betaU1} yields $b_{Y}=\tfrac{2}{3}\!\times 10+\tfrac{1}{3}\!\times 
\tfrac{1}{2}=\tfrac{41}{6}$, as expected.

For each of $U(1)_{Y_{1}}$, $U(1)_{Y_{2}}$, $U(1)_{Y_{3}}$, exactly one SM generation carries the usual hypercharges; the other two generations are neutral under that particular $U(1)$. Therefore the SM fermionic content gives rise to the following contribution: 
\begin{equation}
\sum_{\text{Weyl}}n_{\text{dof}}\,Y_{a}^{2}\Big|_{\text{SM}}=\frac{10}{3}%
\qquad (a=1,2,3),
\end{equation}%
while for $U(1)_\mathsf{D}$ only $\nu _{1R}$ and $\nu _{2R}$ are charged ($%
\pm 1$), giving 
\begin{equation}
\sum_{\text{Weyl}}n_{\text{dof}}\,\mathsf{D}^{2}=1^{2}+(-1)^{2}=2.
\end{equation}%

Each entry in Table III denotes a \emph{Dirac} fermion in $(\mathbf{3},%
\mathbf{1})$ or $(\mathbf{1},\mathbf{1})$. A Dirac field counts as two Weyl
fields with identical charges. Thus the contribution to the fermionic sum is 
$2\,n_{\text{dof}}Q^{2}$ for each charge that is nonzero. Summing over each
each of $U(1)_{Y_{1}}$, $U(1)_{Y_{2}}$, $U(1)_{Y_{3}}$ of Table III one finds 
\begin{eqnarray}
\sum_{\text{Weyl}}n_{\text{dof}}\,Y_{1}^{2}\Big|_{\text{VLF}}
&=&24 \left( \tfrac{1}{6}\right) ^{2}+6 \left( -\tfrac{1}{2}\right) ^{2}=\frac{13}{6}, \\
\sum_{\text{Weyl}}n_{\text{dof}}\,Y_{2}^{2}\Big|_{\text{VLF}}
&=&12 \left( \tfrac{1}{6}\right) ^{2}+6 \left( \tfrac{1}{2}\right) ^{2}+12 \left(- \tfrac{1}{2}\right) ^{2}=\frac{29}{6}, \\
\sum_{\text{Weyl}}n_{\text{dof}}\,Y_{3}^{2}\Big|_{\text{VLF}}
&=&16 \left( \tfrac{1}{2}\right) ^{2}+16
\left(-\tfrac{1}{2}\right) ^{2}=8, \\
\sum_{\text{Weyl}}n_{\text{dof}}\,\mathsf{D}^{2}\Big|_{\text{VLF}} &=&0.
\end{eqnarray}%
Therefore, we obtain: 
\bea
\sum_{\text{Weyl}}n_{\text{dof}}\,Y_{1}^{2}&=&\sum_{\text{Weyl}}n_{\text{dof}%
}\,Y_{1}^{2}\Big|_{\text{SM}}+\sum_{\text{Weyl}}n_{\text{dof}%
}\,Y_{1}^{2}\Big|_{\text{VLF}}=\frac{10}{3}+\frac{13}{6}=\frac{33}{6},\\
\sum_{\text{Weyl}}n_{\text{dof}}\,Y_{2}^{2}&=&\sum_{\text{Weyl}}n_{\text{dof}%
}\,Y_{2}^{2}\Big|_{\text{SM}}+\sum_{\text{Weyl}}n_{\text{dof}%
}\,Y_{2}^{2}\Big|_{\text{VLF}}=\frac{10}{3}+\frac{29}{6}=\frac{49}{6},\\
\sum_{\text{Weyl}}n_{\text{dof}}\,Y_{3}^{2}&=&\sum_{\text{Weyl}}n_{\text{dof}%
}\,Y_{3}^{2}\Big|_{\text{SM}}+\sum_{\text{Weyl}}n_{\text{dof}%
}\,Y_{3}^{2}\Big|_{\text{VLF}}=\frac{10}{3}+8=\frac{34}{3},\\
\sum_{\text{Weyl}}n_{\text{dof}}\,\mathsf{D}^{2}&=&1^{2}+1^{2}=2.
\eea
For complex scalars, summing $n_{\text{dof}}q^{2}$ from Table II we obtain 
\begin{align}
\sum_{\text{scalars}}n_{\text{dof}}\,Y_{1}^{2}& =\left( \tfrac{1}{6}\right)
^{2}+\left(-\tfrac{1}{2}\right) ^{2}=\frac{5}{18}, \\
\sum_{\text{scalars}}n_{\text{dof}}\,Y_{2}^{2}& =\left( \tfrac{1}{6}%
\right) ^{2}+\left(-\tfrac{1}{2}\right) ^{2}+\left(-\tfrac{1}{6}%
\right) ^{2}+\left( \tfrac{1}{2}\right) ^{2}=\frac{5}{9}, \\
\sum_{\text{scalars}}n_{\text{dof}}\,Y_{3}^{2}& =2\!\left( \tfrac{1}{2}%
\right) ^{2}+\left(-\tfrac{1}{6}\right) ^{2}+\left(-\tfrac{1}{2}\right) ^{2}=%
\frac{23}{18}, \\
\sum_{\text{scalars}}n_{\text{dof}}\,\mathsf{D}^{2}& =2^{2}+2\!\left(
-1\right)^{2}+1^{2} =7.
\end{align}%

Using Eq.~(\ref{eq:betaU1}) with the above sums we find 
\begin{align}
b_{Y_{1}}& =\frac{2}{3}\left( \frac{33}{6}\right) +\frac{1}{3}\left( \frac{5%
}{18}\right) =\frac{203}{54}, \\[2mm]
b_{Y_{2}}& =\frac{2}{3}\left( \frac{49}{6}\right) +\frac{1}{3}\left( \frac{5%
}{9}\right) =\frac{152}{27}, \\[2mm]
b_{Y_{3}}& =\frac{2}{3}\left( \frac{34}{3}\right) +\frac{1}{3}\left( \frac{23%
}{18}\right) =\frac{431}{54}, \\[2mm]
b_\mathsf{D}& =\frac{2}{3}\,(2)+\frac{1}{3}\,(7)=\frac{11}{3}.
\end{align}%
If a given $U(1)$ factor is unbroken up to a high scale, the Landau pole is given by
\be
\mu_{\text{LP},i} = \mu_0
	\exp\!\left[\fr{2\pi}{b_i\,\alpha_i(\mu_0)}\right], \hs i\in\{Y_1,Y_2,Y_3,\mathsf{D}\}. 
\label{eq:LPbox}
\ee
Because all $b_{i}>0$, each coupling grows with $\mu $ and hits a Landau pole at $\mu_{\mathrm{LP},i}$.
\paragraph{Determining the SM hypercharge coupling at the heaviest vector-like fermion mass scale $\mu_0 = 10^5$ TeV.}
To set realistic initial conditions, we must compute the value of the SM hypercharge coupling $\alpha_Y$ at the matching scale $\mu_0$. The one-loop $\beta$-function for $\alpha_Y^{-1}$ in the SM is:
\begin{equation}
\mu \frac{d}{d\mu} \alpha_Y^{-1} = -\frac{b_Y}{2\pi}, \quad b_Y = -\frac{41}{6}.
\end{equation}
The well-known solution is:
\begin{equation}
\alpha_Y^{-1}(\mu) = \alpha_Y^{-1}(m_Z) - \frac{b_Y}{2\pi} \ln\left(\frac{\mu}{m_Z}\right).
\end{equation}
We use the initial value at the $Z$-pole, $\alpha_Y(m_Z) = \alpha_{\text{em}}(m_Z) / \cos^2\theta_W(m_Z) \approx (1/128) / 0.768 \approx 1/98.3$. Thus, $\alpha_Y^{-1}(m_Z) \approx 98.3$.

For $\mu_0 = 10^5$ TeV, we obtain
\be
\alpha_Y^{-1}(\mu_0 = 10^5~\mathrm{TeV}) \approx 98.3 - \frac{(-41/6)}{2\pi} \ln\left(\frac{10^8}{91.2}\right) \approx 113.4.
\ee
Therefore, the hypercharge coupling at the matching scale is:
\begin{equation}
\alpha_Y(\mu_0 = 10^5~\mathrm{TeV}) \approx \frac{1}{113.4} \approx 0.00882.
\end{equation}
\paragraph{Realistic coupling values and Landau pole calculation.}
The couplings $g_a ~(a=1,2,3)$ are not independent; they are related to the SM hypercharge coupling $g_Y$ through the matching condition:
\begin{equation}
\frac{1}{g_Y^2} = \frac{1}{g_1^2} + \frac{1}{g_2^2} + \frac{1}{g_3^2}.
\end{equation}
Assuming a unified value $g_1 = g_2 = g_3 \equiv g_*$ at the matching scale $\mu_0$ implies:
\begin{equation}
g_* = \sqrt{3}\,g_Y, \quad \alpha_* \equiv \frac{g_*^2}{4\pi} = 3\,\alpha_Y.
\end{equation}
Using $\alpha_Y(\mu_0) \approx 0.00882$, we find the realistic initial value for the triad couplings:
\begin{equation}
\alpha_{Y_1}(\mu_0) = \alpha_{Y_2}(\mu_0) = \alpha_{Y_3}(\mu_0) \approx 3 \times 0.00882 = 0.02646.
\end{equation}
For $U(1)_\mathsf{D}$, the coupling $\alpha_\mathsf{D}$ is a free parameter. We choose a value comparable to the others, $\alpha_\mathsf{D}(\mu_0) = 0.03$.

Using Eq.~(\ref{eq:LPbox}), we obtain the Landau poles 
\begin{align}
\mu_{\text{LP},Y_1} &= 10^5~\mathrm{TeV} \times \exp\left[\tfrac{2\pi}{(203/54) \times 0.02646}\right] \approx 10^{32.4}~\mathrm{TeV}, \\
\mu_{\text{LP},Y_2} &= 10^5~\mathrm{TeV} \times \exp\left[\tfrac{2\pi}{(152/27) \times 0.02646}\right]  \approx 10^{23.3}~\mathrm{TeV}, \\
\mu_{\text{LP},Y_3} &= 10^5~\mathrm{TeV} \times \exp\left[\tfrac{2\pi}{(431/54) \times 0.02646}\right]  \approx 10^{17.9}~\mathrm{TeV}, \\
\mu_{\text{LP},\mathsf{D}}    &= 10^5~\mathrm{TeV} \times \exp\left[\tfrac{2\pi}{(11/3) \times 0.03}\right]  \approx 10^{29.8}~\mathrm{TeV}.
\end{align}
\paragraph{Discussion and Implications.} The Landau pole for $U(1)_{Y_3}$ is the lowest: $\mu_{\text{LP},Y_3} \approx 10^{17.9}$ TeV. This result the following implications:
\begin{itemize}
	\item Seesaw scale: A typical seesaw scale is $M_3 \sim 10^{11}$ TeV. Our calculated Landau pole is many orders of magnitude higher ($\sim 10^7$ times larger), posing no problem for neutrino mass generation via this mechanism.
	\item GUT scale: A grand unification scale is often postulated near $M_{\text{GUT}} \sim 10^{13}$ TeV. Our Landau pole is still significantly higher ($\sim 10^5$ times larger), suggesting the model could potentially be valid up to or beyond a GUT scale without encountering the pole.
	\item Coupling strength constraint: The requirement that the Landau pole remains above a desired scale (e.g., the GUT scale) places an upper bound on the individual couplings $g_a$ and $g_\mathsf{D}$. If $\alpha_*(\mu_0)$ were much larger than the unified value of $0.02646$ used here (e.g., $\alpha_* = 0.1$), the Landau pole for $Y_3$ would plummet to $\mu_{\text{LP},Y_3} / \mu_0 \sim 10^3$, placing it near $10^8$ TeV for $\mu_0 = 10^5$ TeV. This would be below the GUT scale and could be problematic. Therefore, the triad structure and the relation $g_* = \sqrt{3}g_Y$ are essential for pushing the Landau poles to safe energies.
\end{itemize}
\paragraph{Bounds on couplings from Landau pole constraints.}
The requirement that the Landau pole for $U(1)_{Y_3}$ remains above a certain scale places an upper bound on its coupling strength. We require $\mu_{\mathrm{LP},Y_3} \geq \mu_{\mathrm{min}}$, which implies:
\be
\alpha_{Y_3}(\mu_0) \leq \frac{2\pi}{b_{Y_3} \ln(\mu_{\mathrm{min}}/\mu_0)}, \quad \text{and thus} \quad g_3 \leq \sqrt{ \frac{8\pi^2}{b_{Y_3} \ln(\mu_{\mathrm{min}}/\mu_0)} }.\nn
\ee
Using $b_{Y_3} = 431/54$ and $\mu_0 = 10^5$ TeV, we obtain the following bounds for various scales:
\bea
&&\text{Seesaw scale } (\mu_{\mathrm{min}} = 10^{11}~\mathrm{TeV}):  \alpha_{Y_3} \leq 0.05698, \quad g_3 \leq 0.846, \crn
&&\text{GUT scale } (\mu_{\mathrm{min}} = 10^{13}~\mathrm{TeV}):  \alpha_{Y_3} \leq 0.04275, \quad g_3 \leq 0.733, \crn
&&\text{Planck scale } (\mu_{\mathrm{min}} = 10^{15}~\mathrm{TeV}):  \alpha_{Y_3} \leq 0.03419, \quad g_3 \leq 0.655.\nn
\eea
In the unified scenario ($g_1 = g_2 = g_3 = \sqrt{3} g_Y$), we have $g_* \approx 0.5767$ at $\mu_0 = 10^5$ TeV. This value is well within the bounds for GUT and Planck scales.

In conclusion, for the matching scale $\mu_0 = 10^5$ TeV and the consequent realistic coupling values, the Landau poles are pushed to extremely high energies. The most constraining pole, for $U(1)_{Y_3}$, lies near $10^{18}$ TeV, which is comfortably above typical scales for new physics like the seesaw mechanism or grand unification.

\bibliographystyle{JHEP}
\bibliography{combine}

\end{document}